\documentclass[a4paper,11pt]{article}
\pdfoutput=1 

\usepackage{jheppub} 

\usepackage[T1]{fontenc}

\usepackage{graphicx}
\usepackage{gensymb}
\usepackage{xcolor} 
\colorlet{lightgray}{black!10} 
\colorlet{pencil}{black!25} 
\usepackage{float}
\usepackage{cancel}
\usepackage{tkz-euclide} 
\usepackage{url}
\usepackage{pdflscape}
\usepackage{dsfont}
\usepackage[utf8]{inputenc}
\usepackage{tikz-feynman}
\usepackage{comment}

\tikzfeynmanset{compat=1.1.0}

\usetikzlibrary{
    decorations.pathmorphing,
    intersections,
    calc,
    arrows.meta
}

\usepackage{amsmath}
\usepackage{mathtools}
\usepackage{amssymb}
\usepackage{amstext}
\usepackage{amsfonts}
\usepackage{physics}
\usepackage{slashed}
\DeclareSymbolFontAlphabet{\amsmathbb}{AMSb}
\usepackage{bbold}
\usepackage{leftindex}
\usepackage{bm}

\newcommand{\AdS}{\text{AdS}}

\newcommand{\Sph}{\text{S}}

\newcommand{\SU}{\mathrm{SU}}
\newcommand{\U}{\mathrm{U}}
\newcommand{\PSU}{\mathrm{PSU}}
\newcommand{\SO}{\mathrm{SO}}

\newcommand{\id}{\mathbb 1}

\renewcommand{\Tr}{\mathrm{tr}}

\newcommand{\de}{\text{d}}
\newcommand{\ie}{\textit{i.e.}~}
\newcommand{\cf}{\textit{cf.}~}
\newcommand{\eg}{\textit{e.g.}~}

\newcommand{\ystep}{0.1cm}
\newcommand{\StepDiagram}[2]{
\begin{tikzpicture}[thick,baseline=-3pt]
\node[draw,circle,minimum size=1.0cm]
    (OL) at (0,0)
    {\scriptsize{$T^{(\alpha)}_L$}};

\node[draw,circle,minimum size=1.0cm]
    (OBar) at (2,0)
    {\scriptsize{$\overline{T}^{(\alpha)}_L$}};
\path[name path=smallcircle]
    (OL) circle[radius=0.53cm];
\path[name path=bigcircle]
    (OL) circle[radius=0.8cm];
\path[name path=upper-old]
    (OL.45) -- (OBar.135);
\path[
    name intersections={
        of=upper-old and bigcircle,
        by=upper-old-big
    }
];
\path[name path=upper-new]
    ($(OL |- OL.45)+(0,-\ystep)$)
    --
    ($(OBar |- OBar.135)+(0,-\ystep)$);
\path[
    name intersections={
        of=upper-new and smallcircle,
        by=upper-new-small
    }
];
\path[
    name intersections={
        of=upper-new and bigcircle,
        by=upper-new-big
    }
];
\path[name path=middle-old]
    (OL.0) -- (OBar.180);

\path[
    name intersections={
        of=middle-old and bigcircle,
        by=middle-old-big
    }
];
\path[name path=middle-new]
    ($(OL |- OL.0)+(0,-\ystep)$)
    --
    ($(OBar |- OBar.180)+(0,-\ystep)$);

\path[
    name intersections={
        of=middle-new and smallcircle,
        by=middle-new-small
    }
];
\path[
    name intersections={
        of=middle-new and bigcircle,
        by=middle-new-big
    }
];
\path[name path=lower-old]
    (OL.-21) -- (OBar.-159);
\path[
    name intersections={
        of=lower-old and bigcircle,
        by=lower-old-big
    }
];
\path[name path=lower-new]
    ($(OL |- OL.-21)+(0,-\ystep)$)
    --
    ($(OBar |- OBar.-159)+(0,-\ystep)$);
\path[
    name intersections={
        of=lower-new and smallcircle,
        by=lower-new-small
    }
];
\path[
    name intersections={
        of=lower-new and bigcircle,
        by=lower-new-big
    }
];
\draw[
    decorate,
    decoration={
        snake,
        aspect=1,
        amplitude=1pt,
        segment length=5.2pt
    }
]
    (upper-new-small) -- (upper-new-big);

\draw[
    decorate,
    decoration={
        snake,
        aspect=1,
        amplitude=1pt,
        segment length=5.2pt
    }
]
    (middle-new-small) -- (middle-new-big);

\draw[
    decorate,
    decoration={
        snake,
        aspect=1,
        amplitude=1pt,
        segment length=5.2pt
    }
]
    (lower-new-small) -- (lower-new-big);
\draw[
    decorate,
    decoration={
        snake,
        aspect=1,
        amplitude=1pt,
        segment length=5.2pt
    }
]
    (upper-old-big) -- (OBar.135);
\draw[
    decorate,
    decoration={
        snake,
        aspect=1,
        amplitude=1pt,
        segment length=5.2pt
    }
]
    (middle-old-big) -- (OBar.180);
\draw[
    decorate,
    decoration={
        snake,
        aspect=1,
        amplitude=1pt,
        segment length=5.2pt
    }
]
    (lower-old-big) -- (OBar.-159);
\node at (1,0.18)
    {$\vcenter{\hbox{
        \shortstack{
            $\cdot$\\[-6pt]
            $\cdot$\\[-6pt]
            $\cdot$
        }
    }}$};
\ifnum#1=1
    \draw[dashed,red]
        (OL.-45) -- (OBar.-135);
\fi
\draw[line width=1.5pt]
    (OL) circle[radius=0.53cm];

\draw[line width=1.5pt]
    (OBar) circle[radius=0.53cm];
\ifnum#2=0

    \draw[line width=1pt]
        (OL) ++(0.8cm,0)
        arc[
            start angle=0,
            end angle=360,
            radius=0.8cm
        ];
\else
    \ifnum#2=1

        \draw[line width=1pt]
            (OL) ++(0.8cm,0)
            arc[
                start angle=0,
                end angle=360,
                radius=0.8cm
            ]
            node[
                pos=0.125,
                sloped,
                allow upside down
            ]
            {
                \tikz{
                    \draw[
                        -{Triangle[length=5pt,width=5pt]}
                    ]
                    (0,0) -- (0.001,0);
                }
            };
    \else
        \draw[line width=1pt]
            (OL) ++(0.8cm,0)
            arc[
                start angle=0,
                end angle=360,
                radius=0.8cm
            ]
            node[
                pos=0.125,
                sloped,
                allow upside down
            ]
            {
                \tikz{
                    \draw[
                        -{Triangle[length=5pt,width=5pt]}
                    ]
                    (0.001,0) -- (0,0);
                }
            };

    \fi
\fi
\end{tikzpicture}
}

\title{Circling in on correlation functions \\in $\mathcal{N}=2$ orbifold theories}
\author[a,b]{Dennis le Plat,}
\author[c]{Torben Skrzypek}

\affiliation[a]{HUN-REN Wigner Research Centre for Physics, Konkoly-Thege Mikl\'os u. 29-33, 1121 Budapest,
Hungary}
\affiliation[b]{Niels Bohr International Academy, Niels Bohr Institute, Copenhagen University, Blegdamsvej 17, DK-2100 Copenhagen Ø, Denmark}
\affiliation[c]{Deutsches Elektronen-Synchrotron DESY, Notkestra{\ss}e 85, 22607 Hamburg, Germany}

\emailAdd{dennis.leplat@nbi.ku.dk}
\emailAdd{torben.skrzypek@desy.de}

\abstract{We review the harmonic superspace formalism for 4d $\mathcal{N}=2$ theories and apply it to orbifolds of $\mathcal N=4$ Super-Yang-Mills (SYM) theory. These theories have a remarkably similar description to $\mathcal N=4$ SYM, but their planar supergraphs feature topological twist lines which generate phases when intersecting bifundamental hypermultiplet propagators. These cause twisted sector BPS correlators to depend on the coupling $g$. Twisted Coulomb-branch operators of length $L$ receive a leading perturbative contribution to their two-point function at order $\mathcal{O}(g^{2L})$ from a supergraph involving a wrapped hypermultiplet circle. After performing the complete $D$-algebra, we show that the supergraph reduces to the universal conformal period $W_L=\binom{2L}{L}\zeta(2L-1)$.
We extend this analysis to multiple nested hypermultiplet circles and show that they reproduce a series of higher-order terms in the exact localisation result. A subsequent evaluation of leading contributions to BPS three-point correlators again shows complete agreement with localisation results in extremal case and generates new perturbative data in the non-extremal case. The latter involves one untwisted BPS descendant, which contains physical hypermultiplet fields and can therefore connect directly to the twist-dependent hypermultiplet circle. This reduces the suppression in $g$, resulting in a contribution $4 g^{2(\ell+1)} W_{\ell+1}$ controlled by a bridge length $\ell$ rather than by an external operator length $L\geq\ell$. These examples demonstrate the feasibility of harmonic superspace calculations in orbifold theories and point to various future applications in the context of non-BPS correlators, integrability and marginal deformations.}

\preprint{DESY-26-140}

\begin{document} 
\bibliographystyle{JHEP}
\maketitle

\section{Introduction}\label{Sec:intro}
The best understood example of the AdS/CFT correspondence relates type IIB superstring theory on $\AdS_5\times \Sph^5$ background to $\mathcal N=4$ supersymmetric Yang-Mills (SYM) theory~\cite{Maldacena:1997re,Witten:1998qj,Gubser:1998bc}. In the planar limit, where the rank $N$ of the $\SU(N)$ gauge group is taken to be large,  integrability emerges and establishes significant calculational control. 
For example, integrability methods~\cite{Minahan:2002ve,Beisert:2003tq,Beisert:2005fw,Beisert:2006ez,Arutyunov:2007tc,Bajnok:2008bm,Bombardelli:2009ns,Gromov:2009bc,Arutyunov:2009ur,Gromov:2013pga,Gromov:2014caa,Marboe:2017dmb,Gromov:2023hzc} give access to the spectrum of planar $\mathcal N=4$ SYM at any value of the effective coupling $g=g_{\text{YM}}\tfrac{\sqrt{N}}{4\pi}$\,.
It is natural to ask how much of this integrable structure survives in theories with less supersymmetry. A particularly controlled class of examples is obtained by orbifolding $\mathcal N=4$ SYM by a discrete subgroup of its $\PSU(2,2|4)$ symmetry group~\cite{Kachru:1998ys,Lawrence:1998ja}. 
In the planar limit, these theories inherit integrability from the parent theory, while the orbifold action introduces twisted sectors and modifies the Bethe equations through additional phase factors~\cite{Ideguchi:2004wm,Beisert:2005he}. We can therefore determine the spectrum of these orbifold theories with the same precise integrability technology \cite{Skrzypek:2022cgg}. In recent attempts of extending this integrable control to higher-point correlators \cite{Ferrando:2025qkr,lePlat:2025eod}, we faced the obstruction that only limited field-theory data is available for this class of theories, limiting the extend to which integrability proposals can be verified. It is the goal of this paper to overcome this obstacle by establishing a powerful framework for perturbative gauge-theory calculations.

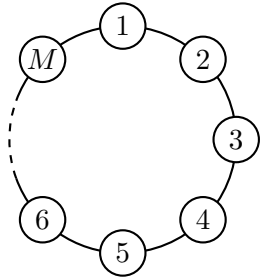
\begin{figure}[h]
    \centering
    \begin{tikzpicture}[
        node/.style={
            circle,
            draw,
            thick,
            fill=white,
            minimum size=6mm,
            inner sep=0pt
        },
        hyper/.style={
            thick
        }
    ]

        \def\R{1.5cm}

% Solid circular hypermultiplet segments
        \draw[hyper]
            (90:\R)
            arc[start angle=90,end angle=45,radius=\R];

        \draw[hyper]
            (45:\R)
            arc[start angle=45,end angle=0,radius=\R];

        \draw[hyper]
            (0:\R)
            arc[start angle=0,end angle=-45,radius=\R];

        \draw[hyper]
            (-45:\R)
            arc[start angle=-45,end angle=-90,radius=\R];

        % Continue solid circle up to the start of the omitted region
        \draw[hyper]
            (-90:\R)
            arc[start angle=-90,end angle=-160,radius=\R];

        % Dashed circular segment = arbitrary number of omitted nodes
        \draw[hyper,dashed]
            (-160:\R)
            arc[start angle=-160,end angle=-200,radius=\R];

        % Resume solid circle toward M
        \draw[hyper]
            (160:\R)
            arc[start angle=160,end angle=90,radius=\R];

        % Gauge nodes, ordered clockwise
        \node[node] (n1) at (90:\R)  {$1$};
        \node[node] (n2) at (45:\R)  {$2$};
        \node[node] (n3) at (0:\R)   {$3$};
        \node[node] (n4) at (-45:\R) {$4$};
        \node[node] (n5) at (-90:\R) {$5$};
        \node[node] (n5) at (-135:\R) {$6$};

        % Final node after the schematic omitted section
        \node[node] (nM) at (135:\R) {$M$};

    \end{tikzpicture}

    \caption{Quiver diagram of 4d $\mathcal{N}=2$ $\amsmathbb{Z}_M$
    orbifold theories. The nodes correspond to vector multiplets with
    gauge group $\SU(N)$. The links between them denote bifundamental
    hypermultiplets.}
    \label{fig:quiver}
\end{figure}

We focus on $\amsmathbb Z_M$ orbifolds preserving $\mathcal N=2$ supersymmetry.\footnote{We expect an extension to the full ADE classification of $\mathcal N=2$ orbifolds to be largely a matter of careful bookkeeping~\cite{Bath:2026gcu}.} The resulting conformal field theory (CFT) is a circular quiver gauge theory with $M$ vector multiplets and $M$ bifundamental hypermultiplets connecting neighbouring nodes, as captured by the quiver diagram in Figure~\ref{fig:quiver}. They provide a simple setting in which one can study the effect of supersymmetry reduction on  integrability. 

Furthermore, certain protected observables in these $\mathcal{N}=2$ orbifold theories are accessible via supersymmetric localisation~\cite{Pestun:2007rz}. In the planar limit in particular, exact results have been obtained for twisted two-point functions and extremal three-point functions of Coulomb-branch operators~\cite{Galvagno:2020cgq,Billo:2022gmq,Billo:2022fnb,Beccaria:2022ypy,Korchemsky:2025eyc}. These results exhibit a remarkably structured dependence on the orbifold twist and can be expressed in terms of semi-infinite matrices with Bessel kernel entries. In~\cite{Ferrando:2025qkr} the  localisation results for twisted three-point functions were matched to expectations from integrability, invoking quite non-trivial finite-size effects. 
In order to understand these better, it is desirable to identify directly in perturbative gauge theory which Feynman diagrams generate twist-sensitive contributions and how the factorised structure observed in localisation results arises. These insights would be instrumental to the outstanding investigation of non-extremal and even non-BPS correlators for which no localisation results are available and integrability has only been studied at tree-level \cite{lePlat:2025eod}. 

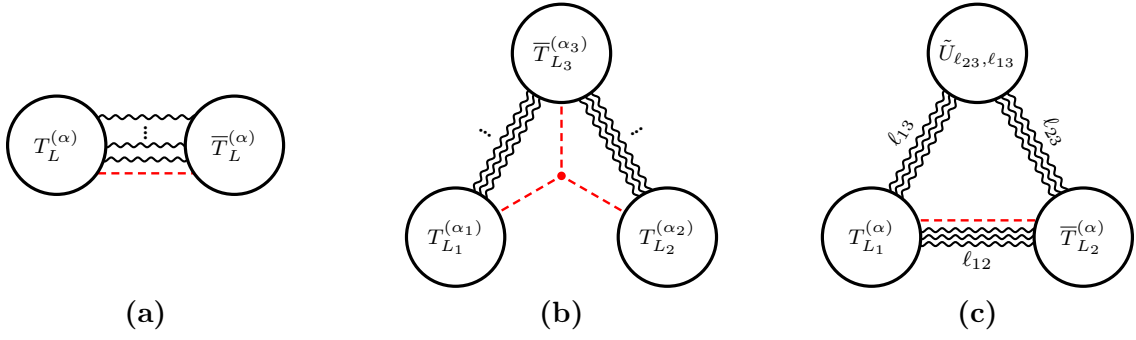
\begin{figure}[t]
\centering

\begin{tikzpicture}[
  thick,
  op/.style={
    draw=black,
    fill=white,
    circle,
    minimum size=1.3cm,
    inner sep=0pt,
    outer sep=0pt,
    line width=1.2pt,
    font=\scriptsize
  },
  curl/.style={
    black,
    line width=0.8pt,
    decorate,
    decoration={
      snake,
      aspect=1,
      amplitude=0.75pt,
      segment length=5.0pt
    }
  },
  contract/.style={
    red,
    line width=0.9pt,
    dash pattern=on 3pt off 2pt
  },
  dots/.style={
    font=\normalsize,
    inner sep=0pt
  },
  bridgelabel/.style={
    font=\small,
    inner sep=1pt,
    fill=white
  }
]

% vertical midpoint of panels (b) and (c) is about 0.405
\def\ycenter{0.405}

\begin{scope}[xshift=-1.175cm,yshift=\ycenter cm]

% operator positions
\coordinate (OL) at (0,0);
\coordinate (OR) at (2.35,0);

% three snake lines
\draw[curl]
  ($(OL)+(45:0.525cm)$) --
  ($(OR)+(135:0.525cm)$);

\draw[curl]
  ($(OL)+(0:0.525cm)$) --
  ($(OR)+(180:0.525cm)$);

\draw[curl]
  ($(OL)+(-21:0.525cm)$) --
  ($(OR)+(-159:0.525cm)$);

% red dashed contraction
\draw[contract]
  ($(OL)+(-45:0.525cm)$) --
  ($(OR)+(-135:0.525cm)$);

% dots for additional bridge lines
\node[dots] at (1.175,0.16)
  {$\vcenter{\hbox{\shortstack{
    $\cdot$\\[-6pt]
    $\cdot$\\[-6pt]
    $\cdot$
  }}}$};

\node[op] at (OL)
  {$T^{(\alpha)}_{L}$};

\node[op] at (OR)
  {$\overline{T}^{(\alpha)}_{L}$};
\end{scope}

\begin{scope}[xshift=5.5cm]

\coordinate (B3) at (0,1.62);
\coordinate (B1) at (-1.40,-0.81);
\coordinate (B2) at ( 1.40,-0.81);
\coordinate (BC) at (0,0);

% dashed contractions to common center
\draw[contract] (B3) -- (BC);
\draw[contract] (B1) -- (BC);
\draw[contract] (B2) -- (BC);

% bridge 3--1
\draw[curl]
  ([shift={(-0.095, 0.055)}]B3) --
  ([shift={(-0.095, 0.055)}]B1);
\draw[curl]
  ([shift={(-0.015, 0.009)}]B3) --
  ([shift={(-0.015, 0.009)}]B1);
\draw[curl]
  ([shift={( 0.065,-0.038)}]B3) --
  ([shift={( 0.065,-0.038)}]B1);

% bridge 3--2
\draw[curl]
  ([shift={(-0.065,-0.038)}]B3) --
  ([shift={(-0.065,-0.038)}]B2);
\draw[curl]
  ([shift={( 0.015, 0.009)}]B3) --
  ([shift={( 0.015, 0.009)}]B2);
\draw[curl]
  ([shift={( 0.095, 0.055)}]B3) --
  ([shift={( 0.095, 0.055)}]B2);

% additional lines
\node[dots,rotate=150]
  at ($(B3)!0.50!(B1)+(-0.302,0.175)$)
  {$\mathinner{\cdot\!\!\cdot\!\!\cdot}$};

\node[dots,rotate=30]
  at ($(B3)!0.50!(B2)+(0.302,0.175)$)
  {$\mathinner{\cdot\!\!\cdot\!\!\cdot}$};

% central vertex
\fill[red] (BC) circle (1.6pt);

% operators
\node[op] at (B3) {$\overline{T}^{(\alpha_3)}_{L_3}$};
\node[op] at (B1) {$T^{(\alpha_1)}_{L_1}$};
\node[op] at (B2) {$T^{(\alpha_2)}_{L_2}$};

\end{scope}

\begin{scope}[xshift=11cm]

\coordinate (C3) at (0,1.62);
\coordinate (C1) at (-1.40,-0.81);
\coordinate (C2) at ( 1.40,-0.81);

% bridge 1--3
\draw[curl]
  ([shift={(-0.095, 0.055)}]C3) --
  ([shift={(-0.095, 0.055)}]C1);
\draw[curl]
  ([shift={(-0.015, 0.009)}]C3) --
  ([shift={(-0.015, 0.009)}]C1);
\draw[curl]
  ([shift={( 0.065,-0.038)}]C3) --
  ([shift={( 0.065,-0.038)}]C1);

% bridge 2--3
\draw[curl]
  ([shift={(-0.065,-0.038)}]C3) --
  ([shift={(-0.065,-0.038)}]C2);
\draw[curl]
  ([shift={( 0.015, 0.009)}]C3) --
  ([shift={( 0.015, 0.009)}]C2);
\draw[curl]
  ([shift={( 0.095, 0.055)}]C3) --
  ([shift={( 0.095, 0.055)}]C2);

% bridge 1--2 (three horizontal snake lines)
\draw[curl]
  ([shift={(0, 0.092)}]C1) --
  ([shift={(0, 0.092)}]C2);
\draw[curl]
  (C1) -- (C2);
\draw[curl]
  ([shift={(0,-0.092)}]C1) --
  ([shift={(0,-0.092)}]C2);

% red dashed contraction ABOVE the horizontal snake lines in y-direction
\coordinate (C1u) at ($(C1)+(25:0.525cm)$);
\coordinate (C2u) at ($(C2)+(155:0.525cm)$);
\draw[contract] (C1u) -- (C2u);

% bridge length labels, replacing the previous dots
\node[rotate=60][bridgelabel]
  at ($(C3)!0.50!(C1)+(-0.322, 0.186)$)
  {\scriptsize{$\ell_{13}$}};

\node[rotate=-60][bridgelabel]
  at ($(C3)!0.50!(C2)+(0.322, 0.186)$)
  {\scriptsize{$\ell_{23}$}};

\node[bridgelabel]
  at ($(C1)!0.50!(C2)+(0,-0.31)$)
  {\scriptsize{$\ell_{12}$}};

% operators
\node[op] at (C3) {$\tilde{U}_{\ell_{23},\ell_{13}}$};
\node[op] at (C1) {$T^{(\alpha)}_{L_1}$};
\node[op] at (C2) {$\overline{T}^{(\alpha)}_{L_2}$};

\end{scope}

% panel labels, aligned horizontally
\node[anchor=north] at (0,-1.5) {\textbf{(a)}};
\node[anchor=north] at (5.5,-1.5)  {\textbf{(b)}};
\node[anchor=north] at (11,-1.5) {\textbf{(c)}};

\end{tikzpicture}

\caption{Tree-level BPS correlators, to which we will compute twist-dependent corrections. The operators $T^{(\alpha)}_L$ \eqref{eq:twisted-operator} are protected Coulomb-branch operators of length $L$ in the $\alpha$-twisted sector. As such, they source twist lines, here depicted red and dashed, which are only detected by hypermultiplet propagators crossing them. However, at tree-level the operators are connected entirely by twist-insensitive vector-multiplet propagators. The two-point function in $\mathbf{(a)}$ is subject to Section~\ref{Sec:TwoPtFcts}, the extremal three-point function in $\mathbf{(b)}$ is discussed in Section~\ref{Sec:Extremal} and the non-extremal case of $\mathbf{(c)}$ in Section~\ref{Sec:non-extremal}. }
\label{fig:tree-level}
\end{figure}

In this paper we lay the technical foundations for a perturbative field-theory analysis of twist-dependent processes at sufficiently high loop orders.
To this end, we introduce the \emph{harmonic superspace} formalism~\cite{Galperin:1984av,Galperin:1984mln,Galperin:1985bj,Galperin:1985va,Galperin_Ivanov_Ogievetsky_Sokatchev_2001} in Section~\ref{Sec:setup}, which provides a natural framework to address $\mathcal N=2$ orbifold theories. Harmonic superspace keeps $\mathcal N=2$ supersymmetry manifest and organises the fields into $\mathcal N=2$ vector prepotentials $V^{++}$ and analytic hypermultiplets $q^+$~\cite{Galperin:1984av,Galperin_Ivanov_Ogievetsky_Sokatchev_2001}. In the parent $\mathcal N=4$ theory the hypermultiplet is adjoint, whereas after the orbifold projection it becomes bifundamental and connects neighbouring quiver nodes as in Figure \ref{fig:quiver}. One may organise these fields such that the local superspace action remains largely unchanged, while the orbifold information is encoded entirely in the gauge structure. In particular, the vector and associated ghost multiplets are not charged under the orbifold group action, while bifundamental hypermultiplets detect orbifold twist. In planar supergraphs this allows us to capture the non-trivial orbifold dynamics by introducing topological twist lines. These may be moved freely through vector and ghost propagators and produce phase factors whenever they cross hypermultiplet propagators. This simplification constitutes a supergraph-level \emph{refinement of the orbifold inheritance principle} as pointed out in~\cite{ShortPaper}.

In Section~\ref{Sec:TwoPtFcts} we use this simple supergraph picture to compute the leading twist-dependent correction to the two-point functions of twisted Coulomb-branch operators depicted in Figure~\ref{fig:tree-level}(a). The first non-trivial contribution is generated by \emph{wrapped hypermultiplet circles} around one of the external operators, such as 
\begin{equation}\label{eq:wrap part}
\begin{tikzpicture}[thick,baseline=-3pt]

\node[draw, circle, minimum size=1.2cm] (OL) at (0,0)
    {\scriptsize{$T_{L}^{(\alpha)}$}};

% Vertically aligned right endpoints
\coordinate (R1) at (2, 0.44);
\coordinate (R3) at (2, 0);
\coordinate (R4) at (2,-0.215);
\coordinate (R5) at (2,-0.44);

\draw[decorate,
      decoration={snake,aspect=1, amplitude=1pt, segment length=5.2pt}]
      (OL.45) -- (R1);

\draw[dashed,red] (OL.-45) -- (R5);

\draw[decorate,
      decoration={snake,aspect=1, amplitude=1pt, segment length=5.2pt}]
      (OL.-21) -- (R4);

\draw[decorate,
      decoration={snake,aspect=1, amplitude=1pt, segment length=5.2pt}]
      (OL.0) -- (R3);

% Vertical dots indicating more lines above and below
\node at (1.2,0.2)
    {$\vcenter{\hbox{\shortstack{$\cdot$\\[-6pt]$\cdot$\\[-6pt]$\cdot$}}}$};

% Horizontal dots indicating continuation to the right
\node at (2.35,0) {$\cdots$};

% Draw thicker circle on top
\draw[line width=1.5pt] (OL) circle [radius=0.6cm];

\draw[line width=1pt] (OL) ++(0.9cm,0)
arc[start angle=0,end angle=360,radius=0.9cm]
node[pos=0.125,sloped,allow upside down]
{\tikz{\draw[-{Triangle[length=5pt,width=5pt]}] (0,0)--(0.001,0);}};

\end{tikzpicture}
\quad\sim e^{2\pi i \frac{\alpha}{M}} g^{2L}\,.
\end{equation}
In total, these generate a twist factor $s_\alpha:=\sin^2\tfrac{\alpha}{M}\pi$. We perform the complete harmonic-superspace $D$-algebra for arbitrary operator length $L$ and show that the resulting point-split integral reduces to the universal conformal period
\begin{equation}\label{eq:period}
W_L=\binom{2L}{L}\zeta(2L-1)\,,
\end{equation}
reproducing the leading localisation result. The derivation gives an explicit field-theory origin of the \emph{wheel period} $W_L$ appearing in localisation. A similar leading-order calculation has previously been performed in $\mathcal{N}=1$ superspace~\cite{Galvagno:2020cgq}.\footnote{The same wheel period $W_L$ appeared in the context of integrated correlators in \cite{Dorigoni:2021guq,Wen:2022oky}.}

We then consider the respective leading corrections for higher powers of the twist factor $s_\alpha$. These occur at order $s_\alpha^k g^{2kL}$ and are represented by $k$ nested hypermultiplet circles. 
We evaluate these supergraphs and thereby provide a direct harmonic-superspace derivation of the leading weak-coupling contribution to every level $s_\alpha^k$ in the twist expansion of the localisation expression.

In Section~\ref{Sec:ThreePtFcts} we move on to three-point functions. As first consistency check we consider extremal correlators, for which exact localisation results are available (see Figure~\ref{fig:tree-level}(b)). We again associate twist-dependent contributions exclusively to hypermultiplet circles surrounding the participating twisted operators and find agreement with localisation.

As main new application we then study non-extremal BPS three-point functions in Section~\ref{Sec:non-extremal} (see Figure~\ref{fig:tree-level}(c)). Such correlators involve an untwisted half-BPS vacuum descendant, which contains physical bifundamental hypermultiplet fields in addition to the vector-multiplet fields. We construct the descendant single-trace operator explicitly in Appendix~\ref{App:VacDescendant} and determine the relative coefficients of the hypermultiplet components. 
In contrast to the wrapped circles discussed before, the leading correction is now a \emph{punctured} hypermultiplet circle and controlled by a bridge length rather than by either external operator length $L_i$. To our knowledge we present the first perturbative result for this class of non-extremal twisted BPS three-point functions. The integrability interpretation of this result is developed in the companion paper~\cite{ShortPaper}.

The examples discussed in this paper serve as proof of concept for the applicability of harmonic superspace methods to $\mathcal{N}=2$ orbifold theories, opening up a perturbative pathway to a wide range of so far unknown observables. These will provide important benchmark data for the development of an all-loop hexagon formalism which motivated this work in the first place. We summarise our findings in Section~\ref{Sec:Conclusion} and comment on future directions.

\section{\texorpdfstring{$\bm{\mathcal N=2}$}~~orbifold theories and harmonic superspace}
\label{Sec:setup}

In this section we will review the necessary background on the $\mathcal{N}=2$ orbifold theories under consideration, the relevant localisation results and the harmonic superspace formalism. Most of the conventions introduced in this section follow previous discussions in~\cite{lePlat:2025eod,Billo:2022fnb,Korchemsky:2025eyc} and~\cite{Galperin_Ivanov_Ogievetsky_Sokatchev_2001}.

\subsection[$\mathcal N=2$ orbifold theories]{$\bm{\mathcal N=2}$ orbifold theories}
\label{Sec:quiver-conventions}

We consider planar $\mathcal N=2$ superconformal quiver theories obtained as $\amsmathbb{Z}_M$ orbifolds of $\mathcal N=4$ SYM. At the orbifold point all gauge
couplings are equal, and we denote the corresponding 't Hooft and effective coupling by
\begin{equation}
    \lambda=g_{\rm YM}^2 N\,, \qquad \text{and} \qquad g=\frac{\sqrt{\lambda}}{4\pi}\,,
    \label{eq:coupling-convention}
\end{equation}
respectively. As depicted in Figure~\ref{fig:quiver}, the quiver gauge theory features $M$ copies of the gauge group $\SU(N)_I$, labelled by $I=0,\ldots,M-1$, where we identify $I\sim I+M$. Each gauge group is mediated by an $\mathcal N=2$ vector multiplet, while neighbouring nodes are connected
by bifundamental hypermultiplets. 

In what follows, the complex scalar in the vector multiplet will be denoted by $\phi_I$, while we will encounter the physical hypermultiplet scalars $f^i_{I}$ only in Section~\ref{Sec:non-extremal}. The orbifold projection gives rise to twisted sectors of operators, which are governed by the twist phase
\begin{equation}
    \omega=e^{2\pi i/M}\,.
    \label{eq:omega-def}
\end{equation}
A particularly simple class of twisted-sector single-trace operators are the protected Coulomb-branch operators~\cite{Billo:2021rdb}
\begin{equation}
    T^{(\alpha)}_{L}(x) :=\frac{1}{\sqrt M}\sum_{I=0}^{M-1}    \omega^{\alpha I} \, \Tr \,\phi_I^L(x)\,, \qquad \alpha=0,\ldots,M-1\,,
    \label{eq:twisted-operator}
\end{equation}
whose conformal dimension $\Delta=L$ coincides with their ``Length'' $L$. The label $\alpha$ determines which twisted sector the operator belongs to. The untwisted sector corresponds to zero orbifold twist
\begin{equation}
    U_L(x):= T_{L}^{(0)}(x)\,.
    \label{eq:untwisted-operator}
\end{equation}
Complex conjugation reverses the orbifold twist, and hence we associate twist $-\alpha$ to the conjugate operator $\overline T_{L}^{(\alpha)}$. A correlator is non-vanishing only if the total orbifold twist vanishes modulo $M$.

Throughout this work we consider correlation functions of the Coulomb-branch operators defined in~\eqref{eq:twisted-operator}. Operators with the same quantum numbers may in general contain multi-trace admixtures, and in extremal three-point functions their contribution can occur at the same order in the large-$N$ expansion as the single-trace structure constant. We do not perform a full orthogonalisation here but restrict our attention to the single-trace sector. This choice agrees with the single-trace operator basis used in the localisation literature~\cite{Billo:2021rdb} to which we compare. Likewise, this choice is natural in integrability computations~\cite{Ferrando:2025qkr}, as the hexagon formalism captures only the single-trace part and does not automatically include multi-trace admixtures. These have to be added by hand~\cite{Eden:2017ozn}.

Finally, let us introduce the following twist factor
\begin{equation} \label{eq:QuiverFactor}
    s_\alpha:= \sin^2\left(\frac{\alpha \, \pi}{M}\right)\,,\quad \text{satisfying }\quad \omega^\alpha + \omega^{-\alpha} -2 =-4 s_\alpha\,, 
\end{equation}
which will play a central role in the calculations to follow.

\subsection{Results from localisation}
\label{Sec:localisation-results}

Let us quote some of the localisation results in the literature~\cite{Galvagno:2020cgq,Billo:2021rdb,Billo:2022fnb,Billo:2022xas,Billo:2022lrv,Billo:2022gmq,Korchemsky:2025eyc}. These will serve us later as benchmarks for the harmonic-superspace construction. Localisation results are available for both protected two-point functions and  extremal three-point functions. For the untwisted operators~\eqref{eq:untwisted-operator}, the respective correlation functions descend from $\mathcal{N}=4$ SYM following the \emph{inheritance principle} for untwisted observables~\cite{Bershadsky:1998mb,Bershadsky:1998cb} 
\begin{align}
    \left\langle U_{L}(x)\,\overline U_{L}(y) \right\rangle &= \frac{\mathcal{G}_L}{\abs{x-y}^{2L}}\,,\\ \left\langle U_{L_1}(x_1)\,U_{L_2}(x_2)\,\overline U_{L_3}(y) \right\rangle &=\frac{\mathcal{G}_{L_1,L_2,L_3}}{\abs{x_1-y}^{2L_1}\abs{x_2-y}^{2L_2}}\,,\qquad L_3:=L_1+L_2\,,
\end{align}
with coupling-independent constants
\begin{equation}\label{eq:untwisted factors}
    \mathcal{G}_L=L\left(\frac{N}{2}\right)^L\,,\qquad \mathcal{G}_{L_1,L_2,L_3}=\frac{L_1L_2L_3}{2\sqrt{M}}\left(\frac{N}{2}\right)^{\frac{L_1+L_2+L_3}{2}-1}\,.
\end{equation}
As such, these untwisted correlators do not capture any novel dynamics specific to orbifold theories. These instead become visible in correlators involving genuinely twisted operators $T^{(\alpha)}_L$~\eqref{eq:twisted-operator} with $\alpha\neq 0\text{ mod }M$. While twisted correlators coincide with untwisted ones in the free limit $g\to0$, \eg
\begin{equation}
    \left\langle T_{L}^{(\alpha)}(x)\,\overline T_{L}^{(\alpha)}(y) \right\rangle_{g=0}=\left\langle U_{L}(x)\,\overline U_{L}(y) \right\rangle= \frac{\mathcal{G}_L}{\abs{x-y}^{2L}} \quad \text{etc.}\,,
\end{equation}
they develop a coupling-dependence at finite values of $g$ for which localisation provides exact expressions. The order in perturbation theory at which non-trivial coupling dependence first arises is governed by the lengths of the interacting operators. We will attribute this suppression to a \emph{refined inheritance principle}~\cite{ShortPaper}.

The localisation results also allow for a strong coupling expansion and thus provide access points for the investigation of the holographic string dual~\cite{Billo:2021rdb,Billo:2022fnb,Billo:2022gmq,Korchemsky:2025eyc,Martinez:2025jjq}.

\paragraph{Twisted two-point functions}

To isolate the coupling and twist dependence of twisted two-point functions we define the ratio~\cite{Billo:2021rdb,Billo:2022fnb} 
\begin{equation}
     \mathcal{R}_{L}^{(\alpha)}(g):=\frac{\left\langle T_{L}^{(\alpha)}(x)\, \overline T_{L}^{(\alpha)}(y) \right\rangle}{\left\langle U_{L}(x)\,\overline U_{L}(y) \right\rangle} \,.
    \label{eq:R-definition}
\end{equation}
The exact planar value for this expression was obtained through localisation methods and can be written in terms of a semi-infinite matrix
$K_{L-1}(g)$ as~\cite{Korchemsky:2025eyc}
\begin{equation}
     \mathcal{R}_{L}^{(\alpha)}(g) = \left[\frac{1}{1-s_\alpha K_{L-1}(g)}\right]_{11}.
    \label{eq:R-localisation}
\end{equation}
The matrix $K_{L-1}(g)$ belongs to a one-parameter family of Bessel matrices, with individual entries given by
\begin{equation}\label{eq:Besselkernel}
    (K_L)_{ij} = (-1)^{i+j-1} \sqrt{2i+L-1} \sqrt{2j+L-1} \int_0^\infty \frac{\de t}{t} \frac{J_{2i+L-1}(\sqrt{t}) J_{2j+L-1}(\sqrt{t}) }{\sinh^2(\frac{\sqrt{t}}{4g})}\,,
\end{equation}
where $i,j\geq1$ and $J_{n}(x)$ is a Bessel function of the first kind.
Expanding~\eqref{eq:R-localisation} in the twist parameter $s_\alpha$, we get a geometric sum
\begin{equation}
     \mathcal{R}_{L}^{(\alpha)}(g) = \sum_{n=0}^{\infty} s_\alpha^n \left[K_{L-1}^n(g)\right]_{11}\,.
    \label{eq:R-kernel-expansion}
\end{equation}
We will make contact to this expansion in Section~\ref{Sec:MultipleCircles}. A further expansion in weak coupling $g$ results in the double series
\begin{equation}\label{eq:two-point structure}
    \mathcal{R}^{(\alpha)}_{L}(g)=1+\sum_{n=1}^\infty\sum_{m=0}^\infty a_{n,m}  s_\alpha^{n}g^{2Ln+2m}\,,
\end{equation}
and we recognise that the leading-order correction arises exclusively from the $n=1$ term at $m=0$
\begin{equation}
     \mathcal{R}_{L}^{(\alpha)}(g) = 1-4s_\alpha g^{2L}W_L +\order{g^{2L+2}}\,,
    \label{eq:q-def}
\end{equation}
where we introduced the conformal period
\begin{equation}
    W_L = \binom{2L}{L}\zeta(2L-1)\,.
    \label{eq:Wn}
\end{equation}
The first few values of $W_L$ are given by
\begin{equation}
    W_2=6\zeta(3)\,,
    \qquad
    W_3=20\zeta(5)\,,
    \qquad
    W_4=70\zeta(7)\,.
    \label{eq:first-W}
\end{equation}
The main result of Section~\ref{Sec:OneCircle} will be a direct harmonic-superspace derivation of this leading correction~\eqref{eq:q-def}. 

Moreover, for every power $s_\alpha^n$ in~\eqref{eq:two-point structure} we find a leading term with a matching power of $W_L$: 
\begin{equation}
    a_{n,0}=(-4W_L)^n\,. 
\end{equation}
In Section~\ref{Sec:MultipleCircles} we will show that this is quite natural from a harmonic superspace perspective.

\paragraph{Extremal three-point functions} 
As with two-point functions we will again isolate the coupling and twist dependence by considering the ratio
\begin{equation}\begin{split}
    \mathcal{C}_{L_1,L_2,L_3}^{(\alpha_1,\alpha_2,\alpha_3)}(g):&=\frac{\left\langle T_{L_1}^{(\alpha_1)}(x_1)\,T_{L_2}^{(\alpha_2)}(x_2)\, \overline T_{L_3}^{(\alpha_3)}(x_3) \right\rangle}{\left\langle U_{L_1}(x_1)\,U_{L_2}(x_2)\, \overline U_{L_3}(x_3) \right\rangle}\,.
    \label{eq:Ghat-def}
\end{split}\end{equation}
Superconformal and orbifold selection rules enforce $L_3=L_1+L_2$ and $\alpha_3=\alpha_1+\alpha_2\text{ mod }M$, where the first relation follows from $\U(1)_r$ charge conservation together with the BPS shortening condition.
Localisation predicts a remarkably simple factorised structure~\cite{Billo:2022xas,Billo:2022lrv}

\begin{equation}
    \mathcal{C}^{(\alpha_1,\alpha_2,\alpha_3)}_{L_1,L_2,L_3}(g)
    =\mathcal{V}_{L_1}^{(\alpha_1)}(g)\,
    \mathcal{V}_{L_2}^{(\alpha_2)}(g)\,
    \mathcal{V}_{L_3}^{(\alpha_3)}(g)\,,
    \label{eq:C-factorisation}
\end{equation}
with factors $\mathcal{V}_{L}^{(\alpha)}$ directly related to the two-point ratios $\mathcal{R}_{L}^{(\alpha)}$ of~\eqref{eq:R-definition}
\begin{equation}
    \mathcal{V}_{L}^{(\alpha)}(g) = \sqrt{ \mathcal{R}_{L}^{(\alpha)}(g)+\frac{g}{2L}\frac{\partial}{\partial g}  \mathcal{R}_{L}^{(\alpha)}(g)}\,.
    \label{eq:Vcal-def}
\end{equation}
If an untwisted operator is involved, it contributes a trivial factor\begin{equation}
    \mathcal{R}_{L}^{(0)}=1\quad \Rightarrow\quad \mathcal{V}_{L}^{(0)}=1\,.
    \label{eq:Vk0}
\end{equation}
Making use of the leading order expansion in~\eqref{eq:q-def} we may likewise expand
\begin{equation}
     \mathcal{V}_{L}^{(\alpha)}(g) \sim 1-4s_\alpha g^{2L}W_L +\order{g^{2L+2}}\,,
\end{equation}
or the full structure constant ratio
\begin{equation}\label{eq:compare}
    \mathcal{C}_{L_1,L_2,L_3}^{(\alpha_1,\alpha_2,\alpha_3)}(g)=1-4s_{\alpha_1} g^{2L_1}W_{L_1}-4s_{\alpha_2} g^{2L_2}W_{L_2}-4s_{\alpha_3}g^{2L_3}W_{L_3}+\order{g^{2L_{\text{min}}+2}}\,.
\end{equation}
We will match these leading contributions with direct harmonic superspace calculations in~Section~\ref{Sec:Extremal}.

\paragraph{Normalisation dependence} The two-point ratios $\mathcal{R}^{(\alpha)}_L$~\eqref{eq:R-definition} are not themselves physical observables as they may be absorbed in the relative normalisation of twisted operators~\eqref{eq:twisted-operator}. From a field-theory perspective, it is still necessary to compute these normalisations directly, since starting from a Lagrangian description, it is far from obvious that untwisted and twisted operators would have to be normalised radically different. The reappearance of $\mathcal{R}^{(\alpha)}_L$ in the three-point factors $\mathcal{V}^{(\alpha)}_L$~\eqref{eq:Vcal-def} further motivates a solid structural analysis. 

Once we move on to three-point functions we can divide out all normalisation dependence and compute physically observable structure constants ratios~\cite{Korchemsky:2025eyc}
\begin{equation}\label{eq: normed 3pt}
    C_{L_1,L_2,L_3}^{(\alpha_1,\alpha_2,\alpha_3)}(g):=\frac{\mathcal{C}_{L_1,L_2,L_3}^{(\alpha_1,\alpha_2,\alpha_3)}(g)}{\sqrt{\mathcal{R}_{L_1}^{(\alpha_1)}(g)\mathcal{R}_{L_2}^{(\alpha_2)}(g)\mathcal{R}_{L_3}^{(\alpha_3)}(g)}}=V_{L_1}^{(\alpha_1)}(g)V_{L_2}^{(\alpha_2)}(g)V_{L_3}^{(\alpha_3)}(g)\,.
\end{equation}
We again observe a decomposition into modified factors~\cite{Korchemsky:2025eyc} 
\begin{equation}
    V_{L}^{(\alpha)}(g) = \sqrt{ 1+\frac{g}{2L}\frac{\partial}{\partial g}  \log \mathcal{R}_{L}^{(\alpha)}(g)}\,.
    \label{eq:V-def}
\end{equation}
It is these normalisation-independent structure constants, that should be accessible to a suitably twisted version of the hexagon formalism in integrability~\cite{Basso:2015zoa,Ferrando:2025qkr,lePlat:2025eod}. They are slightly less natural from a gauge-theory perspective, but since harmonic superspace gives access to $\mathcal{R}_{L}^{(\alpha)}$ and $\mathcal{C}_{L_1,L_2,L_3}^{(\alpha_1,\alpha_2,\alpha_3)}$ it can of course be used to make predictions for $C_{L_1,L_2,L_3}^{(\alpha_1,\alpha_2,\alpha_3)}$ as well.\\

The localisation input for two-point and extremal three-point functions summarised here will provide valuable checks for the harmonic superspace methods presented in this work. In Section~\ref{Sec:non-extremal} we will then consider non-extremal three-point functions, which are so far inaccessible to localisation. We will compute the leading twist-sensitive contributions directly, thus generating genuinely new gauge-theory data and highlighting the predictive power of the harmonic superspace approach we are about to introduce.

\subsection{Harmonic superspace}
\label{Sec:hss-review}

Field-theory calculations in 4d $\mathcal{N}=2$ gauge theories are greatly simplified by the harmonic superspace formalism \cite{Galperin:1984av,Galperin:1984mln,Galperin:1985bj,Galperin:1985va}, which makes the underlying supersymmetry structure manifest. Here we give a brief introduction and present the conventions needed for the calculations in this paper. A thorough review of harmonic superspace can be found in~\cite{Galperin_Ivanov_Ogievetsky_Sokatchev_2001}. 

\paragraph{Superspace} In analogy to the more familiar $\mathcal{N}=1$ superspace construction, we introduce Grassmann coordinates $\theta^\alpha$, which are 4d Weyl spinors by nature. In order to capture the enlarged amount of supersymmetry, we have to add two copies, resulting in the
central superspace coordinates
\begin{equation}
    X= \left(x^\mu, \theta^\alpha_i, \bar\theta^{\dot\alpha i}\right)\,, \qquad i=1,2\,.
    \label{eq:central-coordinates}
\end{equation}
In contrast to the $\mathcal{N}=1$ construction, however, this ordinary superspace is enlarged by the (bosonic) harmonic variables $u_i^\pm$, which parametrise the coset space $\SU(2)_R/\U(1)$ and obey
\begin{equation}
    u^{+i}u_i^-=1\,, \qquad u^{+i}u_i^+ = u^{-i}u_i^- =0\,.
    \label{eq:harmonic-relations}
\end{equation}
In the following we will frequently use the shorthand notation $(u^+ v^+)=u^{+i}v_{i}^+$ for contractions of harmonic variables at different points in superspace. The $\U(1)$-quotient is taken by restricting to homogeneous functions of given $\U(1)$-charge $q$, which can be expanded as
\begin{equation}
    f^{\overbrace{++\dots+}^{q}}(u)=\sum_{n=0}^\infty f^{(i_1\dots i_{n+q} j_1\dots j_n)}u_{i_1}^+\dots u_{i_{n+q}}^+u_{j_1}^-\dots u_{j_n}^-\,,\label{eq:u-expansion}
\end{equation}
and similar for negative $q$. The indices of the coefficients are symmetrised, indicating an isospin $n+\tfrac{q}{2}$ representation at every level $n$ of the expansion. The Grassmann coordinates $\theta^\alpha_i$ transform under the $\SU(2)_R$ symmetry and can be projected to a harmonic basis
\begin{equation}
    \theta^{\pm\alpha} = u^\pm_i\theta^{i\alpha}\,,  \qquad
    \bar\theta^{\pm\dot\alpha} = u^\pm_i\bar\theta^{\dot\alpha i}\,.
    \label{eq:projected-theta}
\end{equation}
Covariant derivatives on superspace are given by
\begin{align}
    D^+_\alpha&=\frac{\partial}{\partial \theta^{-\alpha}}\,,\qquad D^-_\alpha=-\frac{\partial}{\partial \theta^{+\alpha}}+2i\sigma_{\alpha\dot\alpha}^a\bar{\theta}^{-\dot\alpha}\frac{\partial}{\partial x^a}\,, \\
    \bar D^+_{\dot\alpha}&=\frac{\partial}{\partial\bar\theta^{-\dot\alpha}}\,,\qquad \bar D^-_{\dot\alpha}=-\frac{\partial}{\partial \bar\theta^{+\dot\alpha}}-2i\theta^{-\alpha}\sigma_{\alpha\dot\alpha}^a\frac{\partial}{\partial x^a}\,,\\
    D^0&=u^{+i}\frac{\partial}{\partial u^{+i}}-u^{-i}\frac{\partial}{\partial u^{-i}}+\theta^{+\alpha}\frac{\partial}{\partial \theta^{+\alpha}}-\theta^{-\alpha}\frac{\partial}{\partial \theta^{-\alpha}}+\bar\theta^{+\dot\alpha}\frac{\partial}{\partial \bar\theta^{+\dot\alpha}}-\bar\theta^{-\dot\alpha}\frac{\partial}{\partial \bar\theta^{-\dot\alpha}}\,,\\
    D^{++}&=u^{+i}\frac{\partial}{\partial u^{-i}}-2i\theta^+\sigma^a\bar\theta^+\frac{\partial}{\partial x^a}+\theta^{+\alpha}\frac{\partial}{\partial \theta^{-\alpha}}+\bar\theta^{+\dot\alpha}\frac{\partial}{\partial \bar\theta^{-\dot\alpha}}\,,\\
    D^{--}&=u^{-i}\frac{\partial}{\partial u^{+i}}-2i\theta^-\sigma^a\bar\theta^-\frac{\partial}{\partial x^a}+\theta^{-\alpha}\frac{\partial}{\partial \theta^{+\alpha}}+\bar\theta^{-\dot\alpha}\frac{\partial}{\partial \bar\theta^{+\dot\alpha}}\,.
\end{align}
Here we use an analytic basis in which the derivatives $D^+_\alpha$ and $\bar D^+_{\dot\alpha}$ are particularly simple. In~\cite{Galperin_Ivanov_Ogievetsky_Sokatchev_2001}, this basis is specified by a subscript $A$, which we will drop for readability. The non-vanishing (anti-)commutation relations are
\begin{equation}\begin{split}
    \{D^+_\alpha,\bar{D}^-_{\dot\alpha}\}&=-\{D^-_\alpha,\bar{D}^+_{\dot\alpha}\}=-2i 
    \sigma^a_{\alpha\dot\alpha}\frac{\partial}{\partial x^a}\,,\\ [D^{++},D^{--}]&=D^0\,,\qquad [D^{0},D^{\pm\pm}]=\pm2 D^{\pm\pm}\,,\\
    [D^{\pm\pm},D^{\mp}_{\alpha,\dot{\alpha}}]&=D^{\pm}_{\alpha,\dot{\alpha}}\,,\quad\,[D^0,D^{\pm}_{\alpha,\dot{\alpha}}]=\pm D^{\pm}_{\alpha,\dot{\alpha}}\,.
\end{split}\end{equation}
We also introduce the shorthand notations
\begin{equation}\begin{split}
     \theta^2&:=\theta^\alpha\theta_\alpha\,,\qquad  \bar\theta^2:=\bar\theta_{\dot\alpha}\bar\theta^{\dot\alpha}\,,\qquad (D^\pm)^2:=D^{\pm\alpha}D^\pm_\alpha\,, \qquad
  (\bar D^\pm)^2:=\bar D^\pm_{\dot\alpha}\bar D^{\pm\dot\alpha}\,,\\
   (D^\pm)^4&:=\frac{1}{16} (D^\pm)^2(\bar D^\pm)^2\,,\quad (D)^4:=\frac{1}{16} (D^+)^2(D^-)^2\,,\quad (\bar D)^4:=\frac{1}{16} (\bar D^+)^2(\bar D^-)^2\,.
\end{split}\end{equation}
Our conventions for Berezin integration of Grassmann numbers are
\begin{equation} \label{eq:BerezinConventions}
  \int\de^2\eta\,\eta^2=1\,,\qquad 
  \int\de^2\eta\,F(\eta)=\frac14\frac{\partial}{\partial\eta^\alpha}
  \frac{\partial}{\partial\eta_\alpha}F(\eta)\bigg|_{\eta=0}\,.
\end{equation}
One also needs to introduce an appropriate notion of conjugation, which adds an $\SU(2)$-involution to the usual complex conjugation, \eg
\begin{equation}
    \tilde{f}^{i_1\dots i_n}:=\bar{f}_{i_1\dots i_n},\qquad\tilde{\theta}_{\alpha i}:=\bar{\theta}^i_{\dot{\alpha}} \qquad \text{and}\qquad \tilde{u}_i^{\pm}:=u^{\pm i}=\epsilon^{ij}u^{\pm}_j\,.
\end{equation}
Integrals of harmonic variables are normalised as 
\begin{equation}\label{eq: u-normalisation}
    \int \dd u \,1= 1\,, \qquad  \int \dd u \, u^{+i}u^-_{j}= \frac{1}{2}\delta^i_j\,, \qquad   \int \dd u \,u_{(i_1}^+\dots u_{i_{n}}^+u_{j_1}^-\dots u_{j_m)}^-=0\,.
\end{equation}
We will often encounter integrals involving several sets of harmonic variables, where the indices denote separate superspace points rather than $\SU(2)$-indices. Some useful integral relations are given by
\begin{equation}\label{eq:integralformula}
    \int \dd u_3\, (u_1^+u_3^-)(u_3^+u_2^-)=\frac{1}{2}(u_1^+u_2^-)\,,\qquad 
    \int \dd u_3\, \frac{(u_1^+u_3^+)}{(u_2^+u_3^+)}=(u_1^+u_2^-)\,.
\end{equation}

\paragraph{Superfields}
A generic superfield $\Phi(x,\theta^\pm,\bar\theta^{\pm},u^\pm)$ features an infinite number of component fields due to the expansion~\eqref{eq:expansion} in $u^\pm$. However, in parallel to the $\mathcal{N}=1$ case, such superfields are reducible and we may impose constraints to find irreducible representations. The easiest constraint to impose is \emph{analyticity} which drops dependence on $\theta^-,\bar\theta^{-}$ and may be enforced by imposing the constraints
\begin{equation}
    D^{+}_\alpha\Phi=\bar D^{+}_{\dot\alpha}\Phi=0\,.
\end{equation}
If we further assume $\U(1)$-charge +1 and impose 
\begin{equation}\label{eq:hyperconstr}
    D^{++}\Phi=0\,,
\end{equation}
we arrive at the $q^+$ superfield which captures the free Fayet-Sohnius hypermultiplet~\cite{Fayet:1975yi,Sohnius:1978fw}. This multiplet only features scalars $f^i$ and Weyl-spinors $\psi$, $\bar{\kappa}$, as can be seen from the on-shell expansion of $q^+$ after all constraints have been imposed
\begin{equation}\label{eq:q+}
    q^+(x,\theta^+,\bar\theta^+,u^\pm)=f^i(x)u_i^++\theta^{+\alpha}\psi_\alpha(x)+\bar\theta_{\dot\alpha}^+\bar\kappa^{\dot\alpha}(x)+2i\theta^+\sigma^a\bar\theta^+\partial_a f^i(x)u_i^-\,.
\end{equation}
The remaining constraints imply the free equations of motion
\begin{equation}
    \square f^i(x)=\slashed{\partial}\psi=\slashed{\partial}\bar\kappa=0\,.
\end{equation}
An action that generates the constraint~\eqref{eq:hyperconstr} is given by 
\begin{equation}
    S_q=-\int  \de\zeta^{(-4)} \,\, \tilde{q}^+ D^{++}q^+\,,
\end{equation}
where we introduced the analytic measure
\begin{equation}
    \de\zeta^{(-4)} =\de^4x\, \de^2\theta^+\, \de^2\bar\theta^+\, \de u\,.
    \label{eq:analytic-measure}
\end{equation}
We thus found an off-shell description of the free $\mathcal{N}=2$ hypermultiplet in terms of a charge-$1$ analytic superfield $q^+$. Note that the action $S_q$ has an additional rigid $\SU(2)$-symmetry called Pauli-G\"ursey (PG) symmetry, which commutes with supersymmetry and R-symmetry and acts via rotations on the doublet $(q^+,-\tilde q^+)$. One may rewrite the action in terms of these doublets $q^+_a$ to make the PG symmetry manifest
\begin{equation}
    S_q=\frac{1}{2}\int \de\zeta^{(-4)}\,\, q^+_a D^{++}q^{+a}\,.
\end{equation}
This PG symmetry will play an important role in the following. 

\paragraph{$\bm{\mathcal{N}=4}$ SYM in harmonic superspace} If we assume that the hypermultiplet we just introduced transforms in the adjoint representation of a global $\SU(N)$ symmetry, we can gauge this symmetry and arrive at an $\mathcal{N}=2$ formulation of $\mathcal{N}=4$ SYM theory. The gauging procedure introduces a covariant derivative
\begin{equation}
    \mathcal{D}^{++}q^+=D^{++}q^+ +ig_{\text{YM}} [V^{++},q^+]\,,\qquad \tilde{V}^{++}=V^{++}\,,
\end{equation}
featuring the gauge coupling constant $g_{\text{YM}}$ and the analytic prepotential $V^{++}$. We may use the gauge-freedom
\begin{equation}
    \delta V^{++}=-\mathcal{D}^{++}\lambda\,,\qquad \delta q^+=ig_{\text{YM}}[\lambda,q^+]\,.
\end{equation}
with a charge-$0$ analytic superfield $\lambda$ to annihilate pure gauge components of $V^{++}$. For example, one may choose Wess-Zumino gauge
\begin{equation}\begin{split}\label{eq:V++}
    V^{++}_{WZ}=&-2i\theta^+\sigma^a\bar\theta^+A_a(x)-i\sqrt{2}(\theta^+)^2\bar\phi(x)+i\sqrt{2}(\bar\theta^+)^2\phi(x)\\
    &+4(\bar\theta^+)^2\theta^{+\alpha}\psi_\alpha^i(x)u_i^--4(\theta^+)^2\bar\theta^+_{\dot\alpha}\bar\psi^{\dot\alpha i}(x)u_i^-\\&
    +3(\theta^+)^2(\bar\theta^+)^2D^{ij}(x)u_i^-u_j^-\,,
\end{split}\end{equation}
to identify as the physical gauge field $A_a$, scalar components $\phi$, $\bar{\phi}$ and gauginos $\psi$, $\bar{\psi}$ while $D^{ij}$ is a residual auxiliary field. In supergraph calculations it will be useful to choose a manifestly supersymmetric Fermi-Feynman gauge instead. In order to write down a kinetic term for this field, one introduces the auxiliary field $V^{--}$
\begin{equation}
    V^{--}(X,u)=\sum_{n=1}^\infty \int \dd u_1 \dots \dd u_n
\frac{(-i)^{n+1}g_{\text{YM}}^{n-1}V^{++}(X,u_1)\dots V^{++}(X,u_n)}{(u^+\cdot u^+_1)(u^+_1\cdot u^+_2)\dots (u^+_n\cdot u^+)}\,,
\end{equation}
which satisfies the differential equation
\begin{equation}
    D^{++}V^{--}-D^{--}V^{++}+ig_{\text{YM}}[V^{++},V^{--}]=0\,.
\end{equation}
We define gauge-covariant field strength tensors as
\begin{equation}
    W=-\frac{1}{4}(\bar{D}^+)^2 V^{--}\,, \qquad \bar W=\tilde W=-\frac{1}{4}(D^+)^2 V^{--}\,,
\end{equation}
and write down the action for $\mathcal{N}=2$ SYM theory
\begin{equation}
    S_{SYM}^{\mathcal{N}=2}=\frac{1}{4}\int  \dd^4 x \,\bar{D}^4\, \Tr(\bar W\bar W)\,.
\end{equation}
Combining this action with an adjoint hypermultiplet we find a supersymmetry enhancement to $\mathcal{N}=4$, which is nothing else than $\mathcal{N}=4$ SYM
\begin{equation}\label{eq:N=4 SYM}
    S_{SYM}^{\mathcal{N}=4}=S_{SYM}^{\mathcal{N}=2}+\frac{1}{2}\int  \de\zeta^{(-4)}\,\, \Tr\,(q^+_a \mathcal{D}^{++}q^{+a})\,.
\end{equation}
The additional supersymmetries act as
\begin{equation}
    \delta V^{++}=(\epsilon^{\alpha a}\theta_\alpha^++\bar\epsilon^{a}_{\dot\alpha}\bar\theta^{+\dot\alpha})q_a^+\,,\qquad \delta q_a^+=-\frac{1}{2}(D^+)^4[(\epsilon^{\alpha}_a\theta_\alpha^-+\bar\epsilon_{\dot\alpha a}\bar\theta^{-\dot\alpha})V^{--} ]\,,
\end{equation}
and the PG-symmetry of the hypermultiplet combines with the $\SU(2)$ R-symmetry to form the $\SU(4)$ R-symmetry of $\mathcal{N}=4$ SYM. 

\paragraph{Supergraphs} Perturbative field-theory calculations are organised into supergraphs, the superfield analogues of the familiar Feynman diagrams. In the large-$N$ limit, a calculation of correlators boils down to the evaluation of planar supergraphs, which involve propagators for matter $q^+$, ghost $c$ and gauge $V^{++}$ superfields, the cubic vertices  $q^{+}\tilde q^{+}V^{++}$ and $c\tilde{c}V^{++}$ as well as an infinite number of $V^{++}$ self-coupling vertices. After gauge-fixing the associated position-space Feynman rules can be read off the action~\eqref{eq:N=4 SYM} and take the respective form~\cite{Galperin_Ivanov_Ogievetsky_Sokatchev_2001} 
\begin{align}
&\begin{array}{c c}
\begin{tikzpicture}[baseline=-1pt]
\begin{feynman}
\vertex [label=left:\scriptsize{$X_1$}](a) at (0,0);
  \vertex [label=right: \scriptsize{$X_2$}](b) at (1,0);
\feynmandiagram [horizontal=a to b] {
  (a) -- [fermion] (b),
};
\end{feynman}
\end{tikzpicture}
&\qquad \sim G_q^{(1,1)}(X_1,X_2)\,,\label{eq:Feyn-first}
\end{array}\\[2pt]
&\begin{array}{c c}
\begin{tikzpicture}[baseline=-1pt]
\begin{feynman}
\vertex [label=left:\scriptsize{$X_1$}](a) at (0,0);
  \vertex [label=right: \scriptsize{$X_2$}](b) at (1,0);
\feynmandiagram [horizontal=a to b] {
  (a) -- [photon] (b),
};
\end{feynman}
\end{tikzpicture}
&\qquad\sim G_V^{(2,2)}(X_1,X_2)\,,
\end{array}\\[2pt]
&\begin{array}{c c}
\begin{tikzpicture}[baseline=-1pt]
\begin{feynman}
\vertex [label=left:\scriptsize{$X_1$}](a) at (0,0);
  \vertex [label=right: \scriptsize{$X_2$}](b) at (1,0);
\feynmandiagram [horizontal=a to b] {
  (a) -- [charged scalar] (b),
};
\end{feynman}
\end{tikzpicture}
&\qquad \sim G_c^{(0,0)}(X_1,X_2)\,,
\end{array}\\[2pt]
&\begin{array}{c c}
\begin{tikzpicture}[baseline=-1pt]
\begin{feynman}
% Define vertices manually
\vertex[label=below:\scriptsize{$X_i$}] (v) at (0.5,0);
\vertex (l) at (0,0);
\vertex (r) at (1,0);
\vertex (t) at (0.5,0.53);
% Draw edges
\feynmandiagram [horizontal= (l) to (r)]{
  (l) -- [] (v) -- [fermion] (r),
  (t) -- [photon] (v),
};
\end{feynman}
\end{tikzpicture}
&\qquad \sim -ig_{\text{YM}} \int \de\zeta^{(-4)}_i\,,
\end{array}\label{eq:analyticvertex}\\[7pt]
&\begin{array}{c c}
\begin{tikzpicture}[baseline=-1pt]
\begin{feynman}
% Define vertices manually
\vertex[label=below:\scriptsize{$X_i$}] (v) at (0.5,0);
\vertex (l) at (0,0);
\vertex (r) at (1,0);
\vertex (t) at (0.5,0.53);
% Draw edges
\feynmandiagram [horizontal= (l) to (r)]{
  (l) -- [scalar] (v) -- [charged scalar] (r),
  (t) -- [photon] (v),
};
\end{feynman}
\end{tikzpicture}
&\qquad\sim -ig_{\text{YM}} \int \de\zeta^{(-4)}_i \, D^{++}_{(\tilde{c})}\,,
\end{array}\\[7pt]
&\begin{array}{c c}
\begin{tikzpicture}[baseline=-1pt]
\begin{feynman}
\vertex[label={[yshift=3pt]above:\scriptsize{$X_i$}}] (v) at (0.5,0);
\vertex (b1) at (0,-0.5);
\vertex (b2) at (1,-0.5);
\vertex (t1) at (0,0.5);
\vertex (t2) at (1,0.5);
\feynmandiagram [horizontal= (b1) to (b2)]{
  (b1) -- [photon] (v),
  (t1) -- [photon] (v) -- [photon] (t2),
};
\end{feynman}
% Ellipsis
\fill (v) ++(0.25,-0.10) circle (0.03);
\fill (v) ++(0.20,-0.20) circle (0.03);
\fill (v) ++(0.10,-0.25) circle (0.03);
\end{tikzpicture}
&\qquad\sim-\frac{(-ig_{\text{YM}})^{n-2}}{2n}\int \dd^4 x_i\,\dd^8\theta_i \, \frac{\de u_1 \dots \de u_n}{(u_1^+u_2^+)(u_2^+u_3^+)\dots (u_n^+u_1^+)}\,.\label{eq:Feyn-last}
\end{array}
\end{align}
Here, the hypermultiplet and ghost propagators are directed from $q^+$ to $\tilde q^{+}$ and $c$ to $\tilde{c}$, respectively. We neglected the gauge structure, which, since all fields are in adjoint representations, can be restored through double-line notation~\cite{tHooft:1973alw} and colour ordering. We see that each interaction vertex comes with an integral over either analytic or full superspace and a factor of $g_{\text{YM}}$ ($g_{\text{YM}}^{n-2}$ in the case of the $n$-point self interactions), thus allowing for the usual planar large-$N$ expansion~\cite{tHooft:1973alw} in the 't Hooft coupling $\lambda$~\eqref{eq:coupling-convention}. In Fermi-Feynman gauge, the propagators are given by~\cite{Galperin_Ivanov_Ogievetsky_Sokatchev_2001} 
\begin{align}
  G_q^{(1,1)}(X_1,X_2)
  &= (u_1^+u_2^+)\exp[\mathfrak{S}_{12}]\Delta(x_1-x_2)=\frac{(u_1^+u_2^+)}{4\pi^2\,\abs{\hat{x}_{12}}^2}\,,
  \label{eq:q-propagator}\\
  G_V^{(2,2)}(X_1,X_2)
&= \Delta(x_1-x_2)
    (D_1^+)^4
 \delta^8(\theta_1-\theta_2)
\delta^{(-2,2)}(u_1,u_2)\,,
    \label{eq:vector-propagator}\\
    G_c^{(0,0)}(X_1,X_2)
  &= (u_1^-u_2^-)\, G_q^{(1,1)}(X_1,X_2)\,,
  \label{eq:c-propagator}
\end{align}
where the superscripts denote the $\U(1)$ weights at superspace points $X_1$ and $X_2$ and the bosonic propagator $\Delta(x)$ is defined as the Green's function of the Laplacian $\square_x$
\begin{equation}
    \Delta(x) := \frac{1}{4\pi^2\abs{x}^2}\,, \qquad  \square_x\Delta(x) = -\delta^{(4)}(x)\,.
    \label{eq:bosonic-propagator}
\end{equation}

The representation of the $q^+$ propagator in terms of the analytic shift operator~\cite{Howe:1999hz}
\begin{equation} \label{eq:AnalyticShift}
    \mathfrak{S}_{12} = \frac{2i}{(u_1^+ u_2^+)} \left[ (u_1^- u_2^+)\theta_1^+ \sigma^\mu \bar{\theta}_1^+ +(u_2^- u_1^+)\theta_2^+ \sigma^\mu \bar{\theta}_2^+ +\theta_1^+ \sigma^\mu \bar{\theta}_2^+ + \theta_2^+ \sigma^\mu \bar{\theta}_1^+ \right] \partial_{x_1^\mu}\,,
\end{equation}
which shifts the spacetime difference $x_{12}^\mu=x_1^\mu-x_2^\mu$ to
\begin{equation}
    \hat{x}_{12}^\mu = x_{12}^\mu + \frac{2i}{(u_1^+ u_2^+)}\bigg[(u_1^- u_2^+) \theta_1^+ \sigma^\mu \bar{\theta}_1^+ + (u_2^- u_1^+)\theta_2^+ \sigma^\mu \bar{\theta}_2^+ +\theta_1^+ \sigma^\mu \bar{\theta}_2^+ + \theta_2^+ \sigma^\mu \bar{\theta}_1^+ \bigg]\,,
\end{equation} 
will help us simplify the $D$-algebra. Superspace integrals often isolate individual Grassmann degrees, which can be extracted directly from the exponentiated analytic shift generator.

We finally note that the Grassmann and harmonic factors of the $V^{++}$ propagator constitute a Green's function w.r.t. analytic integration. Integrating the propagator against an analytic test function $F^{++}(x_2, \theta_2^+, \bar\theta_2^+, u_2)$, we observe 
\begin{equation}
\begin{aligned}
    &\int  \de^2\theta_2^+\, \de^2\bar\theta_2^+\, \de u_2\,G_V^{(2,2)}(X_1,X_2) F^{++}(x_2, \theta_2^+, \bar\theta_2^+, u_2)\\
    &\hspace{3cm} = \Delta(x_1-x_2) F^{++} \left(x_2, \theta_1^+, \bar\theta_1^+, u_1\right).
     \label{eq:analytic-delta-convolution}
\end{aligned}
\end{equation}

\paragraph{Evaluation strategy} In this article, we consider a special class of diagrams. Our external fields are the bosons $\phi$ in the vector multiplet, while the vertices are all of $q^{+}_1\tilde{q}^{+}_2V^{++}$ type. The comparatively simple structure of the above propagators suggests the following evaluation strategy:
\begin{itemize}
    \item \textbf{Write the full integrand:} 
    Use the Feynman rules given above to write the full expression. This includes appropriate projectors and point-splitting for external operators.
    \item \textbf{Carry out the $\bm{D}$-algebra:} 
    Evaluate the Grassmann integrals at the vertices, taking into account external projectors as well. Gauge propagators allow a direct integration through~\eqref{eq:analytic-delta-convolution} and for the remaining integrals, we use the Berezin integration~\eqref{eq:BerezinConventions} to extract the appropriate terms from the hypermultiplet shift operators~\eqref{eq:AnalyticShift}.
    \item \textbf{Simplify the harmonic factors:}
    We keep track of the harmonic factors arising from expanding the hypermultiplet shift operators and carry out the harmonic integrations. In the class of diagrams considered below, we will find that all explicit harmonic dependency ultimately drops out.
    \item \textbf{Evaluate the spacetime integrals:} The remaining integrations are familiar Feynman integrals over interaction loci. The integrand may be further simplified, appropriately regularised and finally evaluated. 
\end{itemize}
We will carry out an example calculation in full detail in the following Section~\ref{Sec:TwoPtFcts}.

\subsection[$\mathcal{N}=2$ orbifold theories in harmonic superspace]{$\bm{\mathcal{N}=2}$ orbifold theories in harmonic superspace} \label{Sec:HarmonicOrbifold}
With the $\mathcal{N}=2$ formulation~\eqref{eq:N=4 SYM} of $\mathcal{N}=4$ SYM in place, the orbifolding procedure is straight-forward: The orbifold projection breaks $\mathcal{N}=4$ supersymmetry down to $\mathcal{N}=2$ by quotienting by a discrete symmetry constructed out of gauge and R-symmetry. In order to remain within the harmonic superspace formalism, we have to use the PG-symmetry for the twist, as well as act with 
\begin{equation}
    \gamma=\text{diag}(\mathbb{1}_{N},\omega\cdot\,\mathbb{1}_{N},\omega^2\cdot\,\mathbb{1}_{N}\,,\dots\,,\omega^{M-1}\cdot\,\mathbb{1}_{N})\,, \label{eq:twistop}
\end{equation}
on representations of the $\SU(MN)$ gauge group. The PG-twist leaves the $\mathcal{N}=2$ vector multiplet invariant but $q^+$ and $\tilde q^+$ transform as 
\begin{equation}
    (q^+,- \tilde q^+)\to (\omega q^+,-\omega^{-1}\tilde{q}^+)\,. 
\end{equation}
One may now without further complication write down the following action for the orbifolded theory 
\begin{equation}\label{eq:action}
    S_{orb}=\sum_{I=0}^{M-1} \Tr \left(S_{SYM}^{\mathcal{N}=2}(V^{++}_I)+\frac{1}{2}\int  \de\zeta^{(-4)}\, q^+_{I} \mathcal{D}^{++}q^{+}_I \right)\,,
\end{equation}
where we define bifundamental covariant derivatives as 
\begin{align}
    \mathcal{D}^{\pm\pm}q^{+}_{I}&=D^{\pm\pm}q^{+}_{I}+ ig_{\textrm{YM}} ( V^{\pm\pm}_I q^{+}_{I} -q^{+}_{I}V^{\pm\pm}_{I+1})\,,\\
    \mathcal{D}^{\pm\pm}\tilde{q}^{+}_{I}&=D^{\pm\pm}\tilde{q}^+_{I} - ig_{\textrm{YM}} (\tilde{q}^{+} _IV^{\pm\pm}_I -V^{\pm\pm}_{I+1}\tilde{q}^{+}_I)\,.
\end{align}

Here, left- and right-action encode  fundamental and anti-fundamental action, respectively, and $V^{\pm\pm}_{M}:=V^{\pm\pm}_{0}\,.$ 
In order to unify the treatment of these $M$ copies, we introduce \emph{quiver traces} which sum over all $\SU(N)^M$ indices and combine the superfields into $M\times M$ matrices~\cite{Ideguchi:2004wm} with vector multiplets on the diagonal and hyper-multiplets on the off-diagonals
\begin{align}\label{eq:Vmatrix}
     V^{++}=\begin{pmatrix} V^{++}_0 &  0 & 0& \dots & 0\\
   0 & V^{++}_1& 0& &\\
      0 & 0& V^{++}_2& &\\[-5pt]
    \vdots &  &  & \ddots&0\\ 0&  & &0 & V_{M-1}^{++} \end{pmatrix}
    \,,
\end{align}
\begin{align}\label{eq:qmatrix}
    q^{+}=\begin{pmatrix} 0 &  q^{+}_0& 0& \dots& 0\\
   0 & 0& q^{+}_1& &\\
      0 & 0& 0& &\\[-5pt]
    \vdots & & & \ddots &q^{+}_{M-2} \\ q^{+}_{M-1} &&& 0 & 0 \end{pmatrix}
    \,,\qquad \tilde{q}^{+}=\begin{pmatrix} 0 &  0 & 0& \dots& \tilde{q}^{+}_{M-1} \\
   \tilde{q}^{+}_0 & 0& 0& &\\
    0 & \tilde{q}^{+}_1 & 0& & \\[-5pt]
   \vdots & &  &\ddots &0\\
   0& & &  \tilde{q}^{+}_{M-2} & 0\end{pmatrix}\,.
\end{align}
In this notation the orbifold action~\eqref{eq:action} takes the same form~\eqref{eq:N=4 SYM} as the $\mathcal{N}=4$ SYM action, with the orbifold deformation completely absorbed in the gauge structure. We may therefore also recycle the Feynman rules~\eqref{eq:Feyn-first}-\eqref{eq:Feyn-last}.

Single-trace operators that survive the orbifold projection are simply those on which the quiver trace closes, \ie where the field numbers balance as
\begin{equation}
    \# (q^{+} )-\# (\tilde{q}^{+})\in M\amsmathbb{Z}\,.\label{eq: condition}
\end{equation}
After introducing this matrix convention, we have to make sure that we still probe the full Hilbert space of valid single-trace operators. Just taking quiver traces of the fields in~\eqref{eq:Vmatrix} and~\eqref{eq:qmatrix} is not enough, since they will always result in a symmetric sum over all gauge groups. We can introduce relative phases by inserting powers $\gamma^\alpha$ of the twist operator $\gamma$~\eqref{eq:twistop} into the quiver trace. This generates twisted sectors which, together with the untwisted $\gamma^0$ sector, span all gauge-invariant single-trace operators. 

As an instructive example, we can now write the twisted sector Coulomb-branch operators~\eqref{eq:twisted-operator} in harmonic superspace language as gauge invariant operator
\begin{equation}\label{eq:twisted in HS}
    T^{(\alpha)}_L(x) =\frac{i^L}{\sqrt{2^LM}} \Tr\,\left(\gamma^\alpha W(X)^L\right)\Big|_{\theta=\bar\theta=0}\,.
\end{equation}
This identification is most obvious in Wess-Zumino gauge~\eqref{eq:V++}, where we may equivalently write 
\begin{equation}\label{eq:expansion}
    T^{(\alpha)}_L(x) =
  \frac{(-i)^L}{\sqrt{32^LM}} \int\dd u_1\dots\dd u_L\Tr\,\left(\gamma^\alpha (\bar D^-)^2V^{++}_{\text{WZ}}(X,u_1)\dots(\bar D^-)^2V^{++}_{\text{WZ}}(X,u_L)\right)\Big|_{\theta=\bar\theta=0}\,.
\end{equation}
In other gauge conventions this expression receives $\order{g_{\text{YM}}}$ corrections due to non-linear terms in the expansion of the field strength $W(x)$. We will neglect such $\order{g_{\text{YM}}}$ terms in the subsequent calculations, since they do not contribute to the leading twist-dependent corrections we are investigating.

\paragraph{Twist lines}
Since $\gamma$ commutes with $V^{++}$ and satisfies
\begin{equation}
    \gamma^\dagger q^+\gamma =\omega q^+\,,\qquad  \gamma^\dagger \tilde q^+\gamma =\omega^{-1} \tilde q^+  \,,
\end{equation}
we can move the twist operator through the chain of fields within the single-trace operators as long as we keep track of the resulting phases. We can furthermore split and merge twist operators $\gamma^{\alpha+\beta}\leftrightarrow\gamma^\alpha\gamma^\beta$. This makes the twist operators topological defects on the chain of fields, in keeping with their holographic origin as boundary conditions of twisted strings. 

If we now want to compute correlators involving twisted sector operators, we have to keep track of these additional twist operators and their non-trivial commutation with hypermultiplets. We therefore introduce twist-lines to the planar supergraphs, which emerge from external twist operators $\gamma$, can be moved freely through gauge and ghost propagators but cause a phase shift for intersecting hypermultiplet propagators. The corresponding Feynman rules read
\begin{align}
& \begin{array}{c c}
\begin{tikzpicture}[baseline=-1pt]
\begin{feynman}
  % Define vertices
  \vertex [label=left:$\alpha_1$](c) at (0,0);
  \vertex [label=right: $\alpha_2$](d) at (1,0);
  % Draw diagram
  \diagram* {
    (c) -- [scalar, red]  (d),
  };
\end{feynman}
\end{tikzpicture}
&\sim \delta(\alpha_1+\alpha_2\text{ mod }M)\,,
\end{array}\\[2pt]
&\begin{array}{c c}
\begin{tikzpicture}[baseline=-1pt]
\begin{feynman}
\vertex (v) at (0.5,0);
\vertex [label=left:$\alpha_1$] (b1) at (0,-0.5);
\vertex  (b2) at (1,-0.5);
\vertex [label=left:$\alpha_2$](t1) at (0,0.5);
\vertex [label=right:$\alpha_3$](t2) at (1,0.5);
\diagram* {
  (b1) -- [scalar, red] (v),
  (t1) -- [scalar, red] (v) -- [scalar, red] (t2),
};
\end{feynman}
% Ellipsis
\fill (v) ++(0.25,-0.10) circle (0.03);
\fill (v) ++(0.20,-0.20) circle (0.03);
\fill (v) ++(0.10,-0.25) circle (0.03);
\end{tikzpicture}
&\sim \delta(\sum \alpha_i \text{ mod }M)\,,
\end{array}\\[-5pt]
&\begin{array}{c c}
\begin{tikzpicture}[baseline=-1pt]
\begin{feynman}
  % Define vertices
  \vertex (a) at (0,0.5);
  \vertex (b) at (1,-0.5);
  \vertex [label=left:$\alpha_1$](c) at (0,-0.5);
  \vertex [label=right:$\alpha_2$](d) at (1,0.5);
  \vertex (v) at (0.5,0);
  % Draw diagram
  \diagram* {
    (a) -- [] (v) -- [fermion] (b),
    (c) -- [scalar, red] (v) -- [scalar, red] (d),
  };
\end{feynman}
\end{tikzpicture}
&\sim\omega^{\alpha_1} \delta(\alpha_1+\alpha_2\text{ mod }M)\,.\label{eq:twistrule}
\end{array}
\end{align}
In addition to keeping track of twist phases for hypermultiplets, the twist lines also enforce the overall twist cancellation required by orbifold invariance. Since planar diagrams mimic the worldsheet of the dual string, these twist lines should be thought of as gauge-theory analogue to the branch cuts in the worldsheet CFT which connect twisted vertex operator insertions~\cite{Hamidi:1986vh,Martinez:2025jjq}.

\paragraph{Refined inheritance principle} We have seen that supergraph computations for $\mathcal{N}=2$ orbifold theories are equivalent to those for $\mathcal{N}=4$ SYM theory, except for the appearance of additional twist lines. The orbifold inheritance principle \cite{Bershadsky:1998cb,Bershadsky:1998mb} posits that untwisted quantities, \ie those for which no twist lines appear in the first place, are inherited from $\mathcal{N}=4$ SYM theory. However, the current formalism allows for a refinement at the level of supergraphs even when considering twisted-sector correlators: 
\begin{center}
    \emph{Any planar supergraph in $\mathcal{N}=2$ orbifold theories without non-trivial intersections of twist lines and hypermultiplet propagators descends from $\mathcal{N}=4$ SYM theory.}
\end{center}
This includes cases where no intersection occurs or where a topological deformation of the twist line can get rid of the intersection such as in 
\begin{equation}
\begin{array}{c@{\qquad}c@{\qquad}c}

\begin{tikzpicture}[thick,baseline=-3pt]
\usetikzlibrary{arrows.meta}

\def\R{0.42}

% Slightly slanted red dashed line
\draw[dashed,red,line width=0.8pt]
    (-1.05,-0.08) -- (1.05,0.08);

% Decorated circle exactly in the original style
\draw[line width=1pt]
    (0,0) ++(\R cm,0)
    arc[
        start angle=0,
        end angle=360,
        radius=\R cm
    ]
    node[pos=0.125,sloped,allow upside down]
    {
        \tikz{
            \draw[-{Triangle[length=5pt,width=5pt]}]
            (0,0)--(0.001,0);
        }
    };

\end{tikzpicture}
&
\displaystyle \Rightarrow
&

\begin{tikzpicture}[thick,baseline=-3pt]
\usetikzlibrary{arrows.meta}

\def\R{0.42}
\def\RedR{0.62}

% Angle and half-length of the original red line
\pgfmathsetmacro{\redangle}{atan(0.08/1.05)}
\pgfmathsetmacro{\redlength}{sqrt(1.05*1.05 + 0.08*0.08)}

% Where the straight part starts bending
\def\shoulder{0.80}

\begin{scope}[rotate=\redangle]

\draw[dashed,red,line width=0.8pt]

    % exactly straight and aligned with original line
    (-\redlength,0)
    --
    (-\shoulder,0)

    % small rounded shoulder
    .. controls
       (-0.73,0)
       and
       (-0.66,0.02)
    ..
    (-\RedR,0)

    % exact upper semicircle
    arc[
        start angle=180,
        end angle=0,
        radius=\RedR cm
    ]

    % rounded shoulder back onto the original line
    .. controls
       (0.66,0.02)
       and
       (0.73,0)
    ..
    (\shoulder,0)

    % exactly straight and aligned with original line
    --
    (\redlength,0);

\end{scope}

\draw[line width=1pt]
    (0,0) ++(\R cm,0)
    arc[
        start angle=0,
        end angle=360,
        radius=\R cm
    ]
    node[pos=0.125,sloped,allow upside down]
    {
        \tikz{
            \draw[-{Triangle[length=5pt,width=5pt]}]
            (0,0)--(0.001,0);
        }
    };

\end{tikzpicture}
\end{array}\,.\label{eq:topo}
\end{equation}
On the other hand, if a closed hypermultiplet loop encloses a net twist source such as an operator $T^{(\alpha)}_L$ in~\eqref{eq:wrap part}, the twist factor will be picked up and an orbifold-specific modificiation is possible. We will spend the rest of this paper analysing such twist-dependent processes, for which we have now established the necessary technological tools.

\section{Two-point functions} \label{Sec:TwoPtFcts}

We begin by considering two-point functions of twisted Coulomb-branch operators $T^{(\alpha)}_L$~\eqref{eq:twisted-operator} which at tree level are depicted in Figure~\ref{fig:tree-level}(a). We will investigate which planar supergraphs contribute to the ratio $\mathcal{R}^{(\alpha)}_L(g)$~\eqref{eq:R-definition} that isolates the twist deformation and introduces a coupling-dependence to the two-point function. We compute the leading order contribution in $g$. As the exact values  are known from localisation~\cite{Galvagno:2020cgq,Beccaria:2021hvt,Billo:2021rdb}, \cf Section~\ref{Sec:localisation-results}, we can immediately check our evaluation. Having identified the elementary building block (or rather \emph{circle}), we also consider processes contributing higher powers of the orbifold twist factor $s_\alpha$.

\subsection{A single wrapped circle} \label{Sec:OneCircle}

\begin{figure}[h]
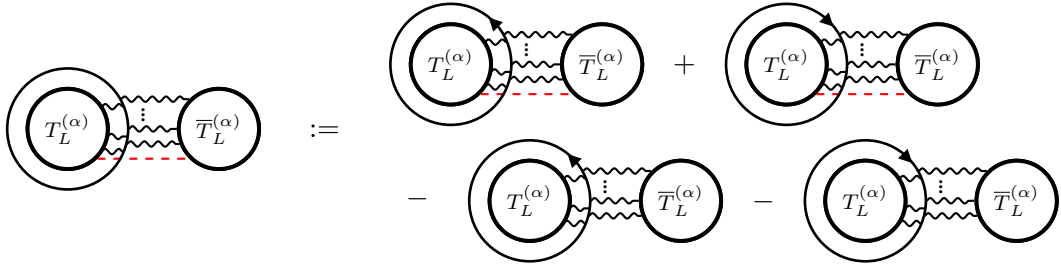

\centering
\begin{minipage}{0.33\textwidth}
\raggedleft
\StepDiagram{1}{0}
\(\quad :=\)
\end{minipage}
\begin{minipage}{0.649\textwidth}
\[ \hspace{-2.5em}
\begin{array}{l@{\quad+\quad}l}
\StepDiagram{1}{1}
&
\StepDiagram{1}{-1}
\end{array}
\]
\hspace{-1em}
\vspace{-0.5em}
\[
\quad -\quad\,
\begin{array}{c@{\quad-\quad}c}
\StepDiagram{0}{1}
&
\StepDiagram{0}{-1}
\end{array}
\]
\end{minipage}
\caption{
The leading order contribution is given by one
hypermultiplet circle wrapped around either of the twisted operators.
The twist line is indicated in red. Accounting for the different
orientations of the hypermultiplet and subtracting the untwisted
contributions, we introduce the shorthand notation on the left which is proportional to the twist factor $s_\alpha$.
}
\label{Fig:Two-pointCircle_explicit}
\end{figure}

Starting from the tree-level process in Figure~\ref{fig:tree-level}(a), we first observe that the twist line positioning is completely arbitrary since no hypermultiplet lines occur which could be affected by it. These can only appear at higher loop order as closed loops of virtual hypermultiplet fields. If these do not enclose one of the twisted operators, the twist line can be pulled out of the hypermultiplet loop as in~\eqref{eq:topo}. We therefore conclude that the twist lines only affect hypermultiplet loops that form \emph{circles wrapped around the twisted operators}. As such, they have to intersect at least $L$ gauge propagators, each of which is consequently split into two segments connected to the hypermultiplet loop through cubic $q^+\tilde q^+ V^{++}$ vertices. The process is therefore suppressed by at least a factor $g^{2L}$. The circle diagrams contributing at leading order $g^{2L}$ are shown in Figure~\ref{Fig:Two-pointCircle_explicit}. 

In computing the perturbative expansion of the ratio $\mathcal{R}^{(\alpha)}_L$~\eqref{eq:R-definition}, we compare to an untwisted correlator that does not feature any gauge-coupling-dependent corrections. 
This means that at a given order in coupling, the sum over the respective diagrams cancels exactly.
In particular, in the untwisted case the same hypermultiplet circle diagrams are cancelled exactly by further contributions, such as e.g. vector multiplet and ghost loops at the same order in perturbation theory. We can make use of this known cancellation and simply subtract the untwisted hypermultiplet circle diagrams to capture these non-descript but twist-insensitive additional contributions also in the twisted sector. We introduce an unoriented hypermultiplet circle as a shorthand notation to summarise this difference of twisted and untwisted circle diagrams, summed furthermore over both orientations of the hypermultiplet circle.
The supergraph computation for each contribution is mostly identical with the key difference provided by the twist phase that is picked up. Applying the twist-line Feynman rule~\eqref{eq:twistrule} to the supergraphs in Figure~\ref{Fig:Two-pointCircle_explicit}, we see that the twist phases add up to the linear combination
\begin{equation}\label{eq: twist contr}
    \omega^{-\alpha}+\omega^{\alpha}-1 -1 = -4s_\alpha\,,
\end{equation}
where we recognised the twist factor $s_\alpha$ defined in~\eqref{eq:QuiverFactor}. We have thus found the leading twist-dependent contribution to $\mathcal{R}^{(\alpha)}_L$ proportional to $s_\alpha g^{2L}$, which matches the expansion~\eqref{eq:q-def} of the exact localisation expression. To compute the proportionality constant, we have to evaluate the supergraphs in Figure~\ref{Fig:Two-pointCircle_explicit} explicitly, which will be the subject of the remainder of this subsection. We follow the evaluation strategy outlined in Section~\ref{Sec:hss-review}.

\subsubsection{Point-splitting and the full integrand}

We start with all external fields inserted at  distinct superspace locations
\begin{equation}
 T^{(\alpha)}_L(x)\,\rightarrow\, \Tr \gamma^\alpha W(X_1)\cdots W(X_L)\,,
  \qquad  \overline{T}^{(\alpha)}_L(y)\,\rightarrow\, 
  \Tr \gamma^{-\alpha} \bar W(Y_1)\cdots\bar W(Y_L)\,.
  \label{eq:split-external}
\end{equation}
Later we will send all $X_i \to X$ and $Y_i \to Y$ for $i=1,\dots,L$. Note, that the expansion~\eqref{eq:expansion} already introduces separate harmonic variables to each exterior $V^{++}$
field, which we will denote by $u_i$ and $v_i$ for $T^{(\alpha)}_L(x)$ and  $\overline{T}^{(\alpha)}_L(y)$, respectively. Further, we introduce analytic vertices~\eqref{eq:analyticvertex} at superspace locations $A_i=(a_i,\theta_{a_i},\bar{\theta}_{a_i},u_{a_i})$ on the hypermultiplet circle, to which we attach the external vector-multiplet lines from the $W(X_i)$ fields. In a similar fashion we attach the $\bar W(Y_i)$ legs to analytic vertices $B_i$ on the hypermultiplet circle. 

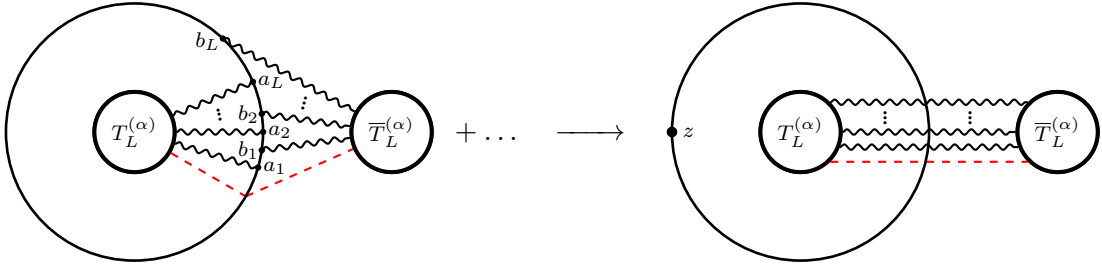
\begin{figure}[h]
\centering

\newcommand{\opsep}{3.4}
\newcommand{\Rbig}{1.7cm}
\newcommand{\raylength}{3cm}
$\begin{tikzpicture}[thick,baseline=-3pt]
\node[
    draw,
    circle,
    minimum size=1.0cm
]
(OL) at (0,0)
{\scriptsize{$T^{(\alpha)}_L$}};

\node[
    draw,
    circle,
    minimum size=1.0cm
]
(OBar) at (\opsep,0)
{\scriptsize{$\overline{T}^{(\alpha)}_L$}};
\path[name path=bigcircle]
    (OL) circle[radius=\Rbig];
\path[name path=left-a1]
    (OL.23) -- ++(23:\raylength);
\path[
    name intersections={
        of=left-a1 and bigcircle,
        by=a1
    }
];
\draw[
    decorate,
    decoration={
        snake,
        aspect=1,
        amplitude=1pt,
        segment length=5.2pt
    }
]
    (OL.23) -- (a1);
\path[name path=left-a2]
    (OL.0) -- ++(0:\raylength);

\path[
    name intersections={
        of=left-a2 and bigcircle,
        by=a2
    }
];

\draw[
    decorate,
    decoration={
        snake,
        aspect=1,
        amplitude=1pt,
        segment length=5.2pt
    }
]
    (OL.0) -- (a2);
\path[name path=left-aL]
    (OL.-16) -- ++(-16:\raylength);

\path[
    name intersections={
        of=left-aL and bigcircle,
        by=aL
    }
];

\draw[
    decorate,
    decoration={
        snake,
        aspect=1,
        amplitude=1pt,
        segment length=5.2pt
    }
]
    (OL.-16) -- (aL);
\path[name path=right-b1]
    (OBar.151) -- ++(151:\raylength);

\path[
    name intersections={
        of=right-b1 and bigcircle,
        by={b1,b1far}
    }
];

\draw[
    decorate,
    decoration={
        snake,
        aspect=1,
        amplitude=1pt,
        segment length=5.2pt
    }
]
    (b1) -- (OBar.151);

\path[name path=right-b2]
    (OBar.172) -- ++(172:\raylength);

\path[
    name intersections={
        of=right-b2 and bigcircle,
        by={b2,b2far}
    }
];

\draw[
    decorate,
    decoration={
        snake,
        aspect=1,
        amplitude=1pt,
        segment length=5.2pt
    }
]
    (b2) -- (OBar.172);

\path[name path=right-bL]
    (OBar.188) -- ++(188:\raylength);

\path[
    name intersections={
        of=right-bL and bigcircle,
        by={bL,bLfar}
    }
];

\draw[
    decorate,
    decoration={
        snake,
        aspect=1,
        amplitude=1pt,
        segment length=5.2pt
    }
]
    (bL) -- (OBar.188);

\fill (a1) circle[radius=1.2pt];
\fill (a2) circle[radius=1.2pt];
\fill (aL) circle[radius=1.2pt];

\fill (b1) circle[radius=1.2pt];
\fill (b2) circle[radius=1.2pt];
\fill (bL) circle[radius=1.2pt];

\node[
    anchor=west,
    xshift=-2pt,
]
at (a1)
{\scriptsize $a_L$};

\node[
    anchor=west,
    xshift=-2pt,
]
at (a2)
{\scriptsize $a_2$};

\node[
    anchor=west,
    xshift=-2pt
]
at (aL)
{\scriptsize $a_1$};

\node[
    anchor=east,
    xshift=3pt,
    yshift=-1pt
]
at (b1)
{\scriptsize $b_L$};

\node[
    anchor=east,
    xshift=3pt,
]
at (b2)
{\scriptsize $b_2$};

\node[
    anchor=east,
    xshift=3pt
]
at (bL)
{\scriptsize $b_1$};

\node[rotate=10]  at ({0.33*\opsep},0.22)
{
    $\vcenter{
        \hbox{
            \shortstack{
                $\cdot$\\[-6pt]
                $\cdot$\\[-6pt]
                $\cdot$
            }
        }
    }$
};

\node[rotate=-15] at ({0.66*\opsep},0.37)
{
    $\vcenter{
        \hbox{
            \shortstack{
                $\cdot$\\[-6pt]
                $\cdot$\\[-6pt]
                $\cdot$
            }
        }
    }$
};

\path[name path=twist-left]
    (OL.-30) -- ++(-30:\raylength);

\path[
    name intersections={
        of=twist-left and bigcircle,
        by=twistpoint
    }
];

\draw[dashed,red]
    (OL.-30) -- (twistpoint);

\draw[dashed,red]
    (twistpoint) -- (OBar);

\draw[line width=1.5pt]
    (OL) circle[radius=0.53cm];

\draw[line width=1.5pt]
    (OBar) circle[radius=0.53cm];

\draw[line width=1pt]
    (OL) ++(\Rbig,0)
    arc[
        start angle=0,
        end angle=360,
        radius=\Rbig
    ];
\end{tikzpicture}
\,\,\,+ \dots\quad \xrightarrow[]{\hspace{0.7cm}}
\quad
\begin{tikzpicture}[thick,baseline=-3pt]

\node[
    draw,
    circle,
    minimum size=1.0cm
]
(OL) at (0,0)
{\scriptsize{$T^{(\alpha)}_L$}};

\node[
    draw,
    circle,
    minimum size=1.0cm
]
(OBar) at (\opsep,0)
{\scriptsize{$\overline{T}^{(\alpha)}_L$}};

\draw[
    decorate,
    decoration={
        snake,
        aspect=1,
        amplitude=1pt,
        segment length=5.2pt
    }
]
(OL.45) -- (OBar.135);

\draw[
    decorate,
    decoration={
        snake,
        aspect=1,
        amplitude=1pt,
        segment length=5.2pt
    }
]
(OL.0) -- (OBar.180);

\draw[
    decorate,
    decoration={
        snake,
        aspect=1,
        amplitude=1pt,
        segment length=5.2pt
    }
]
(OL.-21) -- (OBar.-159);
\node at ({0.33*\opsep},0.18)
{
    $\vcenter{
        \hbox{
            \shortstack{
                $\cdot$\\[-6pt]
                $\cdot$\\[-6pt]
                $\cdot$
            }
        }
    }$
};
\node at ({0.66*\opsep},0.18)
{
    $\vcenter{
        \hbox{
            \shortstack{
                $\cdot$\\[-6pt]
                $\cdot$\\[-6pt]
                $\cdot$
            }
        }
    }$
};
\draw[dashed,red]
    (OL.-45) -- (OBar.-135);
\draw[line width=1.5pt]
    (OL) circle[radius=0.53cm];
\draw[line width=1.5pt]
    (OBar) circle[radius=0.53cm];
\draw[line width=1pt]
    (OL) ++(\Rbig,0)
    arc[
        start angle=0,
        end angle=360,
        radius=\Rbig
    ];
\node(z) at (-\Rbig,0){};
\node[anchor=west] at (z) {\scriptsize $z$};
\fill (z) circle[radius=2pt];    
\end{tikzpicture}
$
\caption{
Detailed view of an individual supergraph from Figure~\eqref{Fig:Two-pointCircle_explicit} with labelled interaction vertices in alternating ordering~\eqref{eq:alternating-cycle}. Following the evaluation strategy in Section~\ref{Sec:hss-review}, we will observe that the supergraph, together with the other orderings, reduces to a ladder integral~\eqref{eq:ladder-def} with an additional vertex insertion $z$ which arises from the point-splitting regularisation. We may thus interpret the right-hand picture as an accurate schematic simplification used in~\cite{ShortPaper}. \label{Fig:Two-pointCircle_simplified}}
\end{figure}

The ordering of the $A_i$ and $B_i$ vertices along the hypermultiplet circle is constrained by R-charge conservation, since, schematically, $R$-charge has to flow from $A_i$ to $B_i$ vertices. We therefore cannot place three $A_i$ vertices or three $B_i$ vertices in a row. In superspace language, the central Grassmann integral would never be saturated and the corresponding supergraph would evaluate to $0$. The $2^{L-1}+2^{L-1}-1$ nonvanishing inequivalent orderings are
\begin{align} \label{eq:non-alternating-cycle1}
  &A_1 \to B_1\to (A_2,B_2)\to\cdots\to (A_L, B_L)\to A_1\,, \quad (A_i, B_i)\in\{A_i\to B_i,B_i\to A_i\}\,,\\
  &B_L \to A_1\to (B_1,A_2)\to\cdots\to (B_{L-1}, A_L)\to B_L\,, \quad (B_i, A_{i+1})\in\{B_i\to A_{i+1},A_{i+1}\to B_i\}\,,
\label{eq:non-alternating-cycle2}
\end{align}
The strictly alternating order 
\begin{equation}
  A_1\to B_1\to A_2\to B_2\to\cdots\to A_L\to B_L\to A_1
  \label{eq:alternating-cycle}
\end{equation}
appears in both ordering schemes and should thus only be counted once.  Since it has the most complicated structure, we shall first consider this case and subsequently comment on the other orderings.  Using the configuration~\eqref{eq:alternating-cycle}, the product of analytic hypermultiplet propagators~\eqref{eq:q-propagator} is 
\begin{equation}
    \prod_{i=1}^{L}G_q^{(1,1)}(A_i,B_i)G_q^{(1,1)}(B_i,A_{i+1})=(-1)^L\prod_{i=1}^{L}G_q^{(1,1)}(A_i,B_i)G_q^{(1,1)}(A_{i+1},B_i)\,,
\end{equation}
where we identify $A_{L+1}\equiv A_1$. 
The resulting supergraph is displayed in Figure~\ref{Fig:Two-pointCircle_simplified}.

The external fields are connected to the vertices by the propagator $G_V^{(2,2)}$~\eqref{eq:vector-propagator}. Since we are interested in the scalar operators $T^{(\alpha)}_L$~\eqref{eq:twisted-operator}, we need to project on the scalar components $\phi$ and $\bar{\phi}$ of the external field strengths $W$. At leading order, we may use the expansion~\eqref{eq:expansion} which suggests we should act with derivatives $D^-_{X_i}$ and $\bar D^-_{Y_i}$, integrate the harmonic variables and ultimately set the external Grassmann numbers to zero, \ie
\begin{equation}\begin{split}
    \phi(x_i) &\sim \int \dd u_i\left[(\bar{D}^-_{X_i})^2 G_V^{(2,2)}(X_i,A_i) \right]_{\theta_i=\bar{\theta}_i=0}\,, \\ \bar{\phi}(y_i) &\sim \int \dd v_i \left[(D^-_{Y_i})^2 G_V^{(2,2)}(Y_i,B_i) \right]_{\theta_i=\bar{\theta}_i=0}\,.\label{eq:FieldComponentMap}
\end{split}\end{equation}
Since the vector-multiplet propagator acts as a Green's function, we can make use of ~\eqref{eq:analytic-delta-convolution} to carry out the Grassmann and harmonic integrations at the vertices and transfer  $(\bar{D}^-_{a_i})^2$ and $( D^-_{a_i})^2$ to the analytic remainder. 

The full integral expression takes the form
\begin{align}
  \mathcal{I}_L
  &=(-1)^L (4\pi)^{2L} g^{2L}\int\prod_{i=1}^{L}\de^4a_i\,\de^4b_i \,\de u_{a_i}\,\de u_{b_i}
  \prod_{i=1}^{L}\Delta(a_i-x_i)\Delta(b_i-y_i)
  \nonumber\\[-1mm]
  &\quad\times
  \left[
  \prod_{i=1}^{L}(\bar D^-_{A_i})^2
  \prod_{i=1}^{L}(D^-_{B_i})^2
  \prod_{i=1}^{L}G_q(a_i,b_i)G_q(a_{i+1},b_i)
  \right]_{\theta=\bar\theta=0}\,,
  \label{eq:exact-post-convolution}
\end{align} 
The derivatives $\bar D^-_{A_i}$ and $D^-_{B_i}$ act by the graded Leibniz rule on both propagators incident at each analytic vertex. The factor $(4\pi)^{2L}$ comes from the definition~\eqref{eq:coupling-convention} of the coupling~$g$.

\subsubsection{The single-circle $D$-algebra}
\label{Sec:one-circle-dalgebra}

The vector-multiplet propagators identified the external Grassmann coordinates with those at the vertices $a_i$ and $b_i$. In the end, we will project to the $\theta=\bar\theta=0$ scalar part, so the parametrisation
\begin{equation}
\theta_{a_i}^+=0\,, \qquad \bar\theta_{a_i}^{+\dot\alpha}=\bar\chi_i^{\dot\alpha}\,,\qquad \theta_{b_i}^{+\alpha}=\eta_i^\alpha\,, \qquad \bar\theta_{b_i}^+=0\,,
  \label{eq:scalar-slice}
\end{equation}
is enough to saturate the remaining Grassmann derivatives. Restricting the analytic shift generator from~\eqref{eq:AnalyticShift} to this subspace, the $q^+$-propagators~\eqref{eq:q-propagator} become finite translation operators,
\begin{equation}
  G_q(A_i,B_j)\big|_{\mathrm{subspace}}
  =(u_{a_i}^+u_{b_j}^+)
  \exp\!\left[\frac{2i}{(u_{a_i}^+u_{b_j}^+)}
(\eta_i\sigma^\mu\bar\chi_j)\partial_{a_i^\mu}\right]
  \Delta(a_i-b_j).
  \label{eq:translation-propagator}
\end{equation}
Likewise, the derivatives $(\bar D^-_{A_i})^2$ and  $(D^-_{B_i})^2$ from~\eqref{eq:exact-post-convolution} can be written as Berezin integrations 
\begin{equation}
  \left[
  \prod_i(\bar D^-_{A_i})^2\prod_i(D^-_{B_i})^2F
  \right]_{\theta=\bar\theta=0}
  =\int\prod_i\de^2\bar\chi_i\,\de^2\eta_i\,F.
  \label{eq:berezin-map}
\end{equation}

For the next step, let us introduce a short-hand notation for the bosonic propagator and the translation generator as
\begin{equation}
  f_{ij}=\Delta(a_i-b_j)\,, \qquad \lambda_{ij}=-\frac{2i}{(u_{a_i}^+u_{b_j}^+)}(\eta_i\sigma^\mu\bar\chi_j)\partial_{a_i^\mu}\,.
  \label{eq:f-ell}
\end{equation}
This allows us to write the remaining Grassmann integral as
\begin{equation}
    \mathcal{K}_L:=\mathcal{H}_L\int\prod_{i=1}^{L}\de^2\eta_i\,\de^2\bar\chi_i
  \prod_{i=1}^{L}(e^{\lambda_{ii}}f_{ii})(e^{\lambda_{i+1,i}}f_{i+1,i})\,,
  \label{eq:grassmann-kernel}
\end{equation}
where we abbreviated a Grassmann-independent harmonic factor as
\begin{equation} \label{eq:DefCalH}
    \mathcal{H}_L=\prod_{i=1}^L(u_{a_{i}}^+u_{b_j}^+)
    (u_{a_{i+1}}^+u_{b_j}^+)\,.
\end{equation}
Let us denote by $m_i\in\{0,1,2\}$ the power selected from $e^{\lambda_{ii}}$, and by $n_i\in\{0,1,2\}$ the power selected from $e^{\lambda_{i+1,i}}$. The saturation of $\dd\eta_i^2$ and $\dd\bar\chi_i^2$ gives the constraints
\begin{equation}
  m_i+n_{i-1}=2\,, \qquad m_i+n_i=2\,.
  \label{eq:incidence-equations}
\end{equation}
Subtracting the equations from each other implies $n_i=n_{i-1}$ for every $i$.

We conclude that for the strictly alternating length-$L$ wrapped circle, the complete Grassmann integral receives contributions only from  \emph{three global saturation patterns}
\begin{equation}
  (m_i,n_i)=(2,0)\,,\qquad (m_i,n_i)=(1,1)\,,\qquad (m_i,n_i)=(0,2)\,.
  \label{eq:three-sectors}
\end{equation}
with no partially mixed configurations. We thus split~\eqref{eq:grassmann-kernel} into contributions from the three patterns 
\begin{equation}
  \mathcal{K}_L=H_{+,L}K_{+,L}+H_{0,L}K_{0,L}+H_{-,L}K_{-,L}\,,
  \label{eq:complete-K}
\end{equation}
which we further factorised into the harmonic-variable-dependent part $H$ and the spacetime-dependent part $K$.
The first and third patterns are \emph{pure patterns}, only picking $\eta_i$'s and $\bar{\chi}_i$'s from either the first or the second exponent. The resulting harmonic factors are
\begin{equation}
    H_{+,L}=\prod_{i=1}^L
    \frac{(u_{a_{i+1}}^+u_{b_i}^+)}{(u_{a_{i}}^+u_{b_i}^+)}\quad \text{and} \quad  H_{-,L}=\prod_{i=1}^L
    \frac{(u_{a_{i}}^+u_{b_i}^+)}{(u_{a_{i+1}}^+u_{b_i}^+)}\,,
\end{equation}
and we may use the Grassmann identity
\begin{equation}
  \int\de^2\eta\,\de^2\bar\chi\, (\eta\sigma^\mu\bar\chi)(\eta\sigma^\nu\bar\chi) =-\frac{1}{2}\delta^{\mu\nu}\,,
  \label{eq:grassmann-two-bilinear}
\end{equation}
to evaluate the spacetime-dependent part to
\begin{equation}
  K_{+,L} =\prod_{i=1}^{L}(\Box_{a_i}f_{ii})
  \prod_{i=1}^{L}f_{(i+1)i} \quad \text{and} \quad K_{-,L} =\prod_{i=1}^{L}(\Box_{a_{i+1}}f_{(i+1)i}) \prod_{i=1}^{L}f_{ii}\,.
  \label{eq:Kplusminus}
\end{equation}
The \emph{mixed pattern} $(m_i,n_i)=(1,1)$ contains one derivative on every hypermultiplet propagator and the harmonic factors arising from this precisely cancel $\mathcal{H}_L$ in~\eqref{eq:grassmann-kernel}, resulting in a trivial factor $H_{0,L}=1$. Its Grassmann integral is the cyclic trace
\begin{equation}
  \int\prod_{i=1}^{L}\de^2\eta_i\,\de^2\bar\chi_i
  \prod_{i=1}^{L}
  (\eta_i\sigma^{\mu_i}\bar\chi_i)
  (\eta_{i+1}\sigma^{\nu_i}\bar\chi_i)
  =-\frac{1}{4^L}
  \Tr(\sigma^{\mu_1}\bar\sigma^{\nu_1}\cdots
  \sigma^{\mu_L}\bar\sigma^{\nu_L})\,.
  \label{eq:cyclic-grassmann-integral}
\end{equation}
Combining the expansion coefficients with~\eqref{eq:cyclic-grassmann-integral} gives
\begin{equation}
  K_{0,L}
  =(-1)^{L+1}
  \Tr(\sigma^{\mu_1}\bar\sigma^{\nu_1}\cdots
  \sigma^{\mu_L}\bar\sigma^{\nu_L})
  \prod_{i=1}^{L}
  (\partial_{a_i}^{\mu_i}f_{ii})
  (\partial_{a_{i+1}}^{\nu_i}f_{i+1,i})\,.
  \label{eq:Kmixed}
\end{equation}

In case of the non-alternating orderings in~\eqref{eq:non-alternating-cycle1} and~\eqref{eq:non-alternating-cycle2} the same line of argument applies but we find at least one propagator of the form 
\begin{equation}
  G_q(B_{i+1},B_{i})\big|_{\mathrm{subspace}}
  =(u_{b_{i+1}}^+u_{b_{i}}^+)
  \Delta(b_{i+1}-b_{i})\,,
\end{equation}
which does not feature any fermionic terms since the subspace restriction~\eqref{eq:scalar-slice} cancels all terms in the shift operator~\eqref{eq:AnalyticShift}. Therefore there is only one possible pure saturation pattern in each case, $(m_i,n_i)=(2,0)$ for ordering scheme~\eqref{eq:non-alternating-cycle1} and $(m_i,n_i)=(2,0)$ for ordering scheme~\eqref{eq:non-alternating-cycle2}.

\subsubsection{Simplification of the harmonic factors}\label{Sec:harmonic-one-circle-integral}

Let us first consider the harmonic factors of the strictly alternating order~\eqref{eq:alternating-cycle}. The mixed pattern is harmonic independent,
\begin{equation}
    H_{0,L}=1 \,,
\end{equation}
which integrates to unity due to~\eqref{eq: u-normalisation}. However, for either pure pattern the independent harmonic integrations do not give unity. The vector-multiplet propagators identified the exterior $(u_i,v_i)$ and interior harmonic coordinates $(u_{a_i},u_{b_i})$ through~\eqref{eq:analytic-delta-convolution}. What remains to evaluate is the integral $(\dd u_{a_i},\dd u_{b_i})$ over the harmonic factors $H_{\pm,L}$ which appeared in the pure contributions~\eqref{eq:Kplusminus}. We make use of the integral relations~\eqref{eq:integralformula} to reduce them step by step, \eg for $H_{+,L}$
\begin{equation} \label{eq:H-integral}
\begin{aligned}
    \int\prod_{i=1}^{L}\de u_{a_i}\,\de u_{b_i} \prod_{i=1}^L
    \frac{(u_{a_{i+1}}^+u_{b_i}^+)}{(u_{a_{i}}^+u_{b_i}^+)} &= \int\prod_{i=1}^{L}\de u_{a_i}\prod_{i=1}^L
  (u_{a_{i+1}}^+u_{a_i}^-) \\
  &=\frac{1}{2^{L-1}} \int\de u_{a_1}
  (u_{a_{L+1}}^+u_{1}^-)=\frac{1}{2^{L-1}}\,,
\end{aligned}
\end{equation}
where in the last step we identified $u_{a_{L+1}}^+=u_{a_1}^+$ and used~\eqref{eq:harmonic-relations} to evaluate the integrand. A similar procedure applies to $H_{-,L}$. 

The other $2^{L-1}-1$ orderings in~\eqref{eq:non-alternating-cycle1} arise from flipping $n_\text{flip}$ $(A_i,B_i)$ pairs and feature harmonic factors of the form 
\begin{equation}
(-1)^{n_\text{flip}}\prod_{i=1}^L
    \frac{1}{(u_{a_{i}}^+u_{b_i}^+)}\prod_{i\in \mathcal{A}_{aa}}
(u_{a_{i+1}}^+u_{a_i}^+) \prod_{i\in \mathcal{A}_{ba}}
    (u_{a_{i+1}}^+u_{b_i}^+)\prod_{i\in \mathcal{A}_{ab}}
    (u_{b_{i+1}}^+u_{a_i}^+)\prod_{i\in \mathcal{A}_{bb}}
(u_{b_{i+1}}^+u_{b_i}^+)\,,
\end{equation}
where the index-sets $\mathcal{A}_{cd}$ depend on the ordering in~\eqref{eq:non-alternating-cycle1} and the minus signs arise from reorienting the $G_q^{(1,1)}(B_i,A_i)$ propagators. Each $b_i$ appears in one of the factors in the numerator so we can perform the $\dd u_{b_i}$ integrals, which reduce the harmonic factor to 
\begin{equation}
\prod_{i\in \mathcal{A}_{aa}}
(u_{a_{i+1}}^+u_{a_i}^+) \prod_{i\in \mathcal{A}_{ba}}
    (u_{a_{i+1}}^+u_{a_i}^-)\prod_{i\in \mathcal{A}_{ab}}(u_{a_{i+1}}^-u_{a_{i}}^+)\prod_{i\in \mathcal{A}_{bb}}
(u_{a_{i+1}}^-u_{a_i}^-)\,.
\end{equation}
We may again use~\eqref{eq:integralformula} to sequentially integrate out this chain of $u_{a_i}$ factors with an additional factor $(-1)$ appearing whenever we integrate out a factor with the ``flipped'' polarisation $(u_{a_{i+1}}^\pm u_{a_i}^+)(u_{a_{i}}^- u_{a_{i-1}}^\pm)$. These factors precisely cancel the overall $(-1)^{n_{\text{flips}}}$ prefactor. Equal polarisations never neighbour each other in the ordering scheme~\eqref{eq:non-alternating-cycle1}. For each ordering $\rho+$ in~\eqref{eq:non-alternating-cycle1} we thus find a total factor
\begin{equation}
   \int\prod_{i=1}^{L}\de u_{a_i}\,\de u_{b_i}  H_{\rho+,L}=\frac{1}{2^{L-1}}\,,
\end{equation}
in analogy to the pure-pattern contribution in the alternating ordering. A parallel calculation for the $2^{L-1}-1$ non-alternating orderings $\rho-$ in~\eqref{eq:non-alternating-cycle2} gives the same result.

Adding to $K_{+,L}$ and $K_{-,L}$ the various spacetime-dependence factors $K_{\rho\pm,L}$ arising from the non-alternating orderings
\begin{equation}\label{eq:pure}
\begin{split}
    \mathcal{K}_{+,L}&=\frac{1}{2^{L-1}}\left(K_{+,L}+\sum_{\rho}K_{\rho+,L}\right)\sim \prod_{i=1}^{L}(\Box_{a_i}f_{ii})\\
    \mathcal{K}_{-,L}&=\frac{1}{2^{L-1}}\left(K_{-,L}+\sum_{\rho}K_{\rho-,L}\right)\sim \prod_{i=1}^{L}(\Box_{a_{i+1}}f_{(i+1)i})
\end{split}
\end{equation}
and identifying $\mathcal{K}_{0,L}:=K_{0,L}$ as the only mixed-pattern contribution, we have now reduced the supergraph evaluation to integrals over ordinary spacetime points, which we will evaluate in the next and final step:
\begin{equation}\label{eq:base case}
  \mathcal{I}_L= (-1)^L
  \frac{g^{2L}}{\pi^{2L}}
  \int
  \prod_{i=1}^{L}
  \de^4a_i\,\de^4b_i
  (\mathcal{K}_{+,L}+\mathcal{K}_{0,L}+\mathcal{K}_{-,L})
  \prod_{i=1}^{L}
  \frac{1}{|a_i-y_i|^2|b_i-x_i|^2}\,.
\end{equation}

\subsubsection{Evaluation of the spacetime integrals}\label{Subsec:one-circle-integral}

The pure contributions~\eqref{eq:pure} feature Laplace operators acting on their respective Green's function. The resulting delta distributions~\eqref{eq:bosonic-propagator} may be used to perform the integrations over the $b_i$ points resulting in the following contributions to the integral $\mathcal{I}_L$:
\begin{equation}
\begin{aligned}
    \mathcal{I}_{+,L} &= \frac{g^{2L}}{(2\pi^2)^{2L}} \int\prod_{i=1}^{L}\de^4a_i\, \frac{1}{\prod_{i=1}^{L}|a_i-y_i|^2|a_i-x_i|^2|a_i-a_{i+1}|^2}\,,\\
    \mathcal{I}_{-,L} &= \frac{g^{2L}}{(2\pi^2)^{2L}} \int\prod_{i=1}^{L}\de^4a_i\, \frac{1}{\prod_{i=1}^{L}|a_i-y_i|^2|a_{i+1}-x_i|^2|a_i-a_{i+1}|^2}\,.
\end{aligned}
\end{equation}
At this stage the difference in orderings~\eqref{eq:non-alternating-cycle1} and~\eqref{eq:non-alternating-cycle2} has become negligible and all $K_{\rho\pm,L}$ terms in~\eqref{eq:pure} become equal to $K_{\pm,L}$, cancelling the $2^{L-1}$ denominator. This justifies the schematic depiction on the right of Figure~\ref{Fig:Two-pointCircle_simplified}. The mixed contribution $\mathcal{K}_{0,L}$ can be simplified by the fermionic star--triangle identity
\begin{align}
  &\sigma^\mu\bar\sigma^\nu\partial_{B^\mu}\partial_{C^\nu}
  \int\de^4A\,
  \frac{1}{|B-A|^2|C-A|^2|D-A|^2}
  \nonumber\\
  &\hspace{35mm}
  =-4\pi^2
  \frac{\sigma\cdot(B-D)\,\bar\sigma\cdot(D-C)}
  {|B-D|^2|C-D|^2|B-C|^2}\,.
  \label{eq:fermionic-star-triangle}
\end{align}
which we use to perform the $b_i$ integrals after identifying $A=b_i$, $B=a_i$, $C=a_{i+1}$ and $D= y_i$. We furthermore introduce the abbreviations
\begin{equation}
  p_i=a_i-y_i\,,
  \qquad
  q_i=y_i-a_{i+1}\,,
  \qquad
  c_i=a_i-a_{i+1}\,.
  \label{eq:pqc}
\end{equation}
The $L$ applications of the star--triangle identity contribute a factor $(-4\pi^2)^L$, which combines with the mixed-kernel sign $(-1)^{L+1}$ in~\eqref{eq:Kmixed} and the overall $(-1)^{L}$ in~\eqref{eq:exact-post-convolution} to give a total sign $(-1)^{L+1}$. Hence the mixed contribution boils down to
\begin{equation}
  \mathcal{I}_{0,L} = \frac{(-1)^{L+1} g^{2L}}{(2\pi^2)^{2L}} \int\prod_{i=1}^{L}\de^4a_i\,  \frac{\Tr\left[\prod_{i=1}^{L} (\sigma\cdot p_i\,\bar\sigma\cdot q_i)\right]}{\prod_{i=1}^{L}|a_i-x_i|^2|p_i|^2|q_i|^2|c_i|^2}\,.
  \label{eq:I-zero}
\end{equation}

 We can now place the three contributions $\mathcal{I}_{+,L},\mathcal{I}_{0,L},\mathcal{I}_{-,L}$ over a common denominator
\begin{equation} 
\mathcal{I}_L=\mathcal{I}_{+,L}+\mathcal{I}_{0,L}+\mathcal{I}_{-,L}=\frac{g^{2L}}{(2\pi^2)^{2L}}\int\prod_{i=1}^{L}\de^4a_i\frac{\mathcal{N}_L}{  \mathcal{D}_L}\,, \quad
  \mathcal{D}_L=\prod_{i=1}^{L}|a_i-x_i|^2|p_i|^2|q_i|^2|c_i|^2\,.
  \label{eq:common-denominator}
\end{equation}
The numerator is given by  the combination
\begin{equation}
  \mathcal{N}_L =\prod_{i=1}^{L}|p_i|^2 +\prod_{i=1}^{L}|q_i|^2 +(-1)^{L+1} \Tr\!\left[\prod_{i=1}^{L}(\sigma\cdot p_i\,\bar\sigma\cdot q_i)\right]\,.
  \label{eq:global-numerator}
\end{equation}
Up to this point we kept all exterior operator locations separated. We now identify all initial points and all but one final point 
\begin{equation}
    x_1=\cdots=x_{L}=x\,,
  \qquad  y_1=\cdots=y_{L-1}=y\,,\qquad
  y_L=z\,,
  \qquad z\neq y\,.
  \label{eq:one-point-splitting}
\end{equation}
For $i=1,\ldots,L-2$, we have $q_i=-p_{i+1}$, which reduces the factors under the trace
\begin{equation}
  \bar\sigma\cdot q_i\,\sigma\cdot p_{i+1}=-|p_{i+1}|^2 \id \, \quad \text{for} \quad i=1,\ldots,L-2\,,
  \label{eq:sigma-collapse}
\end{equation}
and allows us to collapse the $2L$-sigma trace to a four-sigma trace. The two pure products factorise in the same way and the numerator can be written as
\begin{equation}
  \mathcal{N}_L = \left(\prod_{j=2}^{L-1}|p_j|^2\right)\Bigl[|p_1|^2|p_L|^2+|q_{L-1}|^2|q_L|^2 -\Tr(\sigma\cdot p_1\,\bar\sigma\cdot q_{L-1}\, \sigma\cdot p_L\,\bar\sigma\cdot q_L) \Bigr]\,.
  \label{eq:numerator-four-trace}
\end{equation}
The trace can further be simplified, given that $p_1+q_{L-1}+p_L+q_L=0$ holds by definition~\eqref{eq:pqc}. We can therefore use the identity
\begin{equation}
  |p_1|^2 |p_L|^2+|q_{L-1}|^2 |q_L|^2 -\Tr(\sigma\cdot p_1\,\bar\sigma\cdot q_{L-1}\, \sigma\cdot p_L\,\bar\sigma\cdot q_L) =|p_1+q_L|^2 |p_L+q_L|^2\,,
  \label{eq:closed-quadrilateral-identity}
\end{equation}
and obtain a simple expression for the  numerator
\begin{equation}
  \mathcal{N}_L=|z-y|^2|a_L-a_1|^2 \prod_{j=2}^{L-1}|a_j-y|^2\,.
  \label{eq:factorised-numerator}
\end{equation}
This numerator cancels the closing edge $|c_L|^2$ in the denominator $\mathcal{D}_L$~\eqref{eq:common-denominator} and  reduces the quartic powers which appear through the identification $q_i=-p_{i+1}$ to quadratic order. The resulting integral is the point-split $L$-loop ladder integral, which reads
\begin{equation}
  \mathcal{I}_L(y,z;x) =\frac{|z-y|^2 g^{2L}}{(2\pi^2)^{2L}}
  D_L(z,y,z,x)\,,
  \label{eq:point-split-ladder}
\end{equation}
where the ladder integral $D_L(z,y,z,x)$ is given by
\begin{equation}
\begin{aligned}
  D_L(z,y,z,x) =\int\prod_{i=1}^{L}\de^4a_i\,
  &\frac{1}{|z-a_1|^2|a_1-a_2|^2\cdots|a_{L-1}-a_L|^2|a_L-z|^2}\\
  &\times\prod_{i=1}^{L}\frac{1}{|a_i-x|^2|a_i-y|^2}\,.
  \label{eq:ladder-def}
\end{aligned}
\end{equation}
This coincident-endpoint limit follows from the arbitrary-rung ladder result of~\cite{Usyukina:1993ch} and evaluates to
\begin{equation}
  D_L(z,y,z,x) =\pi^{2L} \frac{|y-x|^{-2(L-1)}}{|z-x|^2|z-y|^2} \binom{2L}{L}\zeta(2L-1)\,.
  \label{eq:UD-ladder}
\end{equation}
The factor $|z-y|^2$ precisely cancels the point-splitting pole. Therefore, we obtain a finite expression when sending $z\to y$ 
\begin{equation}
  \mathcal{I}_L(x,y) = g^{2L} W_L\Delta(x-y)^{L}\,, \quad \text{with} \quad W_L=\binom{2L}{L}\zeta(2L-1) \,,
  \label{eq:one-circle-reduced-kernel}
\end{equation}
where $W_L$ is called the conformal wheel period. Multiplying the crucial twist phase $-4s_\alpha$~\eqref{eq: twist contr} we recover the leading contribution to the two-point function ratio~\eqref{eq:q-def}
\begin{equation}\label{eq:2ptresult}
\begin{aligned}
    \mathcal{R}_{L}^{(\alpha)} = 1- 4 s_\alpha \, g^{2L} \, W_L +O(g^{2L+2})\,.
\end{aligned}
\end{equation}
This result is in agreement with the leading correction~\eqref{eq:q-def} obtained by localisation~\cite{Galvagno:2020cgq, Billo:2022fnb}.
The first few conformal periods for small $L$ were given in~\eqref{eq:first-W}.

\subsection{Factorisation of multiple circles} \label{Sec:MultipleCircles}

The previous result for the leading correction at order $s_\alpha g^{2L}$ can be extended directly to the leading contribution at every order in the $s_\alpha$-expansion of $\mathcal{R}_L^{(\alpha)}$. The structure of this expansion was parametrised in~\eqref{eq:two-point structure} and we aim to compute the coefficients $a_{n,0}$.
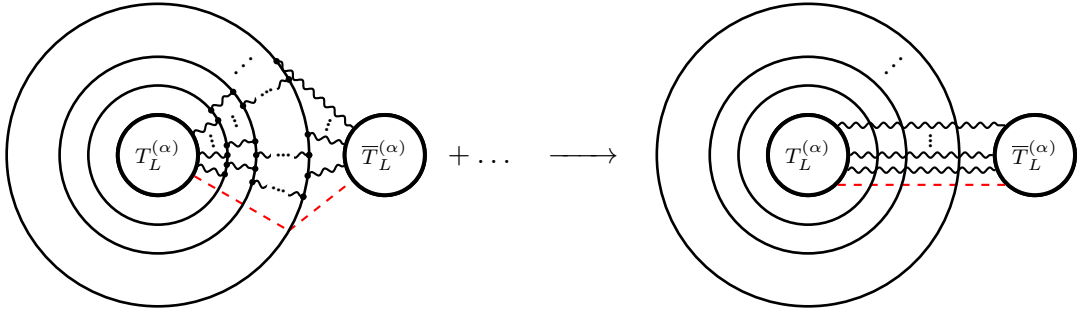
\begin{figure}[h]
\centering
\newcommand{\opsep}{3.0}
\newcommand{\Rbig}{2cm}
\newcommand{\raylength}{3cm}
\newcommand{\Rop}{0.53cm}
\pgfmathsetlengthmacro{\Rinner}{0.46*\Rbig}
\pgfmathsetlengthmacro{\Rmiddle}{0.65*\Rbig}
\pgfmathsetlengthmacro{\Rellipsis}{(\Rmiddle+\Rbig)/2}
\newcommand{\aOneInnerAngle}{30}
\newcommand{\aOneMiddleAngle}{40}
\newcommand{\aOneOuterAngle}{30}
\newcommand{\aTwoInnerAngle}{0}
\newcommand{\aTwoMiddleAngle}{8}
\newcommand{\aTwoOuterAngle}{0}
\newcommand{\aLInnerAngle}{-16}
\newcommand{\aLMiddleAngle}{-8}
\newcommand{\aLOuterAngle}{-16}
\newcommand{\DrawRadialSnake}[3]{\draw[decorate,decoration={snake,aspect=1,amplitude=1pt,segment length=5.2pt}]($(OL)+(#1:#2)$)--($(OL)+(#1:#3)$);}
$\begin{tikzpicture}[thick,baseline=-3pt]
\node[draw,circle,minimum size=1.0cm](OL) at (0,0){\scriptsize{$T^{(\alpha)}_L$}};
\node[draw,circle,minimum size=1.0cm](OBar) at (\opsep,0){\scriptsize{$\overline{T}^{(\alpha)}_L$}};
\path[name path=bigcircle](OL) circle[radius=\Rbig];
\DrawRadialSnake{\aOneInnerAngle}{\Rop}{\Rinner}
\DrawRadialSnake{\aOneMiddleAngle}{\Rinner}{\Rmiddle}
\DrawRadialSnake{\aOneOuterAngle}{\Rmiddle}{\Rbig}
\DrawRadialSnake{\aTwoInnerAngle}{\Rop}{\Rinner}
\DrawRadialSnake{\aTwoMiddleAngle}{\Rinner}{\Rmiddle}
\DrawRadialSnake{\aTwoOuterAngle}{\Rmiddle}{\Rbig}
\DrawRadialSnake{\aLInnerAngle}{\Rop}{\Rinner}
\DrawRadialSnake{\aLMiddleAngle}{\Rinner}{\Rmiddle}
\DrawRadialSnake{\aLOuterAngle}{\Rmiddle}{\Rbig}
\node[rotate={\aOneOuterAngle-90},fill=white,inner sep=0pt] at ($(OL)+(\aOneOuterAngle:\Rellipsis)$){$\vcenter{\hbox{\shortstack{$\cdot$\\[-6pt]$\cdot$\\[-6pt]$\cdot$}}}$};
\node[rotate={\aTwoOuterAngle-90},fill=white,inner sep=0pt] at ($(OL)+(\aTwoOuterAngle:\Rellipsis)$){$\vcenter{\hbox{\shortstack{$\cdot$\\[-6pt]$\cdot$\\[-6pt]$\cdot$}}}$};
\node[rotate={\aLOuterAngle-90},fill=white,inner sep=0pt] at ($(OL)+(\aLOuterAngle:\Rellipsis)$){$\vcenter{\hbox{\shortstack{$\cdot$\\[-6pt]$\cdot$\\[-6pt]$\cdot$}}}$};
\coordinate (a1-inner-in) at ($(OL)+(\aOneInnerAngle:\Rinner)$);
\coordinate (a1-inner-out) at ($(OL)+(\aOneMiddleAngle:\Rinner)$);
\coordinate (a1-middle-in) at ($(OL)+(\aOneMiddleAngle:\Rmiddle)$);
\coordinate (a1-middle-out) at ($(OL)+(\aOneOuterAngle:\Rmiddle)$);
\coordinate (a1) at ($(OL)+(\aOneOuterAngle:\Rbig)$);
\coordinate (a2-inner-in) at ($(OL)+(\aTwoInnerAngle:\Rinner)$);
\coordinate (a2-inner-out) at ($(OL)+(\aTwoMiddleAngle:\Rinner)$);
\coordinate (a2-middle-in) at ($(OL)+(\aTwoMiddleAngle:\Rmiddle)$);
\coordinate (a2-middle-out) at ($(OL)+(\aTwoOuterAngle:\Rmiddle)$);
\coordinate (a2) at ($(OL)+(\aTwoOuterAngle:\Rbig)$);
\coordinate (aL-inner-in) at ($(OL)+(\aLInnerAngle:\Rinner)$);
\coordinate (aL-inner-out) at ($(OL)+(\aLMiddleAngle:\Rinner)$);
\coordinate (aL-middle-in) at ($(OL)+(\aLMiddleAngle:\Rmiddle)$);
\coordinate (aL-middle-out) at ($(OL)+(\aLOuterAngle:\Rmiddle)$);
\coordinate (aL) at ($(OL)+(\aLOuterAngle:\Rbig)$);
\fill (a1-inner-in) circle[radius=1.2pt];
\fill (a1-inner-out) circle[radius=1.2pt];
\fill (a1-middle-in) circle[radius=1.2pt];
\fill (a1-middle-out) circle[radius=1.2pt];
\fill (a1) circle[radius=1.2pt];
\fill (a2-inner-in) circle[radius=1.2pt];
\fill (a2-inner-out) circle[radius=1.2pt];
\fill (a2-middle-in) circle[radius=1.2pt];
\fill (a2-middle-out) circle[radius=1.2pt];
\fill (a2) circle[radius=1.2pt];
\fill (aL-inner-in) circle[radius=1.2pt];
\fill (aL-inner-out) circle[radius=1.2pt];
\fill (aL-middle-in) circle[radius=1.2pt];
\fill (aL-middle-out) circle[radius=1.2pt];
\fill (aL) circle[radius=1.2pt];
\path[name path=right-b1](OBar.139)--++(139:\raylength);
\path[name intersections={of=right-b1 and bigcircle,by={b1,b1far}}];
\draw[decorate,decoration={snake,aspect=1,amplitude=1pt,segment length=5.2pt}](b1)--(OBar.139);
\path[name path=right-b2](OBar.165)--++(165:\raylength);
\path[name intersections={of=right-b2 and bigcircle,by={b2,b2far}}];
\draw[decorate,decoration={snake,aspect=1,amplitude=1pt,segment length=5.2pt}](b2)--(OBar.165);
\path[name path=right-bL](OBar.195)--++(195:\raylength);
\path[name intersections={of=right-bL and bigcircle,by={bL,bLfar}}];
\draw[decorate,decoration={snake,aspect=1,amplitude=1pt,segment length=5.2pt}](bL)--(OBar.195);
\fill (a1) circle[radius=1.2pt];
\fill (a2) circle[radius=1.2pt];
\fill (aL) circle[radius=1.2pt];
\fill (b1) circle[radius=1.2pt];
\fill (b2) circle[radius=1.2pt];
\fill (bL) circle[radius=1.2pt];
\node[rotate=15] at ({0.24*\opsep},0.18){$\vcenter{\hbox{\shortstack{$\cdot$\\[-6pt]$\cdot$\\[-6pt]$\cdot$}}}$};
\node[rotate=22] at ({0.34*\opsep},0.44){$\vcenter{\hbox{\shortstack{$\cdot$\\[-6pt]$\cdot$\\[-6pt]$\cdot$}}}$};
\node[rotate=-20] at ({0.75*\opsep},0.39){$\vcenter{\hbox{\shortstack{$\cdot$\\[-6pt]$\cdot$\\[-6pt]$\cdot$}}}$};
\node[rotate=-45] at ({0.375*\opsep},1.175){$\vcenter{\hbox{\shortstack{$\cdot$\\[-4pt]$\cdot$\\[-4pt]$\cdot$}}}$};
\path[name path=twist-left](OL.-30)--++(-30:\raylength);
\path[name intersections={of=twist-left and bigcircle,by=twistpoint}];
\draw[dashed,red](OL.-30)--(twistpoint);
\draw[dashed,red](twistpoint)--(OBar);
\draw[line width=1.5pt](OL) circle[radius=\Rop];
\draw[line width=1.5pt](OBar) circle[radius=\Rop];
\draw[line width=1pt](OL) circle[radius=\Rbig];
\draw[line width=1pt](OL) circle[radius=\Rmiddle];
\draw[line width=1pt](OL) circle[radius=\Rinner];
\end{tikzpicture}
\,\,\,+ \dots\quad \xrightarrow[]{\hspace{0.7cm}}
\quad
\begin{tikzpicture}[thick,baseline=-3pt]
\node[draw,circle,minimum size=1.0cm](OL) at (0,0){\scriptsize{$T^{(\alpha)}_L$}};
\node[draw,circle,minimum size=1.0cm](OBar) at (\opsep,0){\scriptsize{$\overline{T}^{(\alpha)}_L$}};
\draw[decorate,decoration={snake,aspect=1,amplitude=1pt,segment length=5.2pt}](OL.45)--(OBar.135);
\draw[decorate,decoration={snake,aspect=1,amplitude=1pt,segment length=5.2pt}](OL.0)--(OBar.180);
\draw[decorate,decoration={snake,aspect=1,amplitude=1pt,segment length=5.2pt}](OL.-21)--(OBar.-159);
\node[rotate=-45] at ({0.375*\opsep},1.175){$\vcenter{\hbox{\shortstack{$\cdot$\\[-4pt]$\cdot$\\[-4pt]$\cdot$}}}$};
\node at ({0.54*\opsep},0.18){$\vcenter{\hbox{\shortstack{$\cdot$\\[-6pt]$\cdot$\\[-6pt]$\cdot$}}}$};
\draw[dashed,red](OL.-45)--(OBar.-135);
\draw[line width=1.5pt](OL) circle[radius=\Rop];
\draw[line width=1.5pt](OBar) circle[radius=\Rop];
\draw[line width=1pt](OL) circle[radius=\Rbig];
\draw[line width=1pt](OL) circle[radius=\Rmiddle];
\draw[line width=1pt](OL) circle[radius=\Rinner];
\end{tikzpicture}$
\caption{A sequence of circles can be constructed inductively and the various orderings of vertices can be accounted for uniformly, thus allowing for the schematic notation to the right~\cite{ShortPaper}. Upon removing point-splitting regularisation, the circles recursively collapse, each contributing a single-circle factor.}
\label{Fig:Two-point_MultiCircle}
\end{figure}

The minimal way to obtain a twist factor $s_\alpha^k$ is by wrapping $k$ nested hypermultiplet circles around one of the operators as depicted in Figure~\ref{Fig:Two-point_MultiCircle}. Each hypermultiplet circle is twist-sensitive and combining the twist factors $\omega^{\pm \alpha}$ from the two possible orientations with the twist-independent cancellation terms in~\eqref{eq: twist contr} gives rise to one power of $s_\alpha$ per circle.\footnote{Here we made an implicit assumption that the cancellation terms factorise in the same way as we will show for the twist-dependent hypermultiplet loops. Although it would be an interesting technical challenge to show this explicitly, the assumption is not strictly necessary for the claims of this paper. We can restrict the statements of this section to the maximally twist-sensitive contribution $\omega^{k\alpha}+\omega^{-k\alpha}$ and then adjust subleading terms at lower orders in $s_\alpha$ appropriately to restore the full $s_\alpha^k$ factor.} Due to R-charge conservation, each hypermultiplet circle is connected to the consecutive one by $L$ vector-multiplet propagators which generate a factor $g^{2L}$ at their endpoints. The inner- and outermost circle couples to the external operators at $x$ and $y$, respectively, generating a final $g^{2L}$ factor. 
We thus identify the leading order in coupling as $g^{2Lk}$ in agreement with~\eqref{eq:two-point structure}. 

Let us label the different circles by $r=1,\ldots,k$. For the $r$-th circle we introduce the analytic vertices $A_i^{(r)}$ and $B_i^{(r)}$,\footnote{As in Section~\ref{Sec:OneCircle} we first discuss the strictly alternating configuration in detail and afterwards add the non-alternating orderings.} with $i=1,\ldots,L$ and
\begin{equation}
  A_{L+1}^{(r)}\equiv A_1^{(r)}\,, \qquad B_{L+1}^{(r)}\equiv B_1^{(r)}\,.
  \label{eq:multi-cyclic}
\end{equation}
We again have to sum over the different orderings~\eqref{eq:non-alternating-cycle1} and~\eqref{eq:non-alternating-cycle2}, which will become important when considering the harmonic variables.

The full point-split integral $\mathcal{I}_L^{(k)}$ corresponding to Figure~\ref{Fig:Two-point_MultiCircle}, including the external derivatives $D^-_{X_i}$ and $\bar{D}^-_{Y_i}$, integrations and projections can be written down and evaluated according to the same strategy we just performed for a single circle. Instead of going case by case, we will prove by induction that the final spacetime-integral takes the form
\begin{equation}
  \begin{aligned}
  \mathcal{I}_L^{(k)}
  ={}&
  \frac{(2g)^{2Lk}}{(2\pi)^{2L}}
  \int
  \prod_{r=1}^{k}\prod_{i=1}^{L}
  \de^4a_i^{(r)}\,\de^4b_i^{(r)}
  \prod_{r=1}^{k}K_L^{(r)}
  \\
  &\times
  \prod_{i=1}^{L}
  \frac{1}{|a_i^{(1)}-x_i|^2}
  \prod_{r=1}^{k-1}\prod_{i=1}^{L}
  \frac{1}{|b_i^{(r)}-a_i^{(r+1)}|^2}
  \prod_{i=1}^{L}
  \frac{1}{|b_i^{(k)}-y_i|^2}\,.
  \end{aligned}
  \label{eq:kcircle-master}
\end{equation}
where 
\begin{equation}
  K_L^{(r)} = K_{+,L}^{(r)}+K_{0,L}^{(r)}+K_{-,L}^{(r)}\,,
  \label{eq:multi-kernel}
\end{equation}
and the indidual terms are defined as in~\eqref{eq:Kplusminus} and $\eqref{eq:Kmixed}$ with $f_{ij}$ replaced by 
\begin{equation}
  f_{ij}^{(r)} := \Delta\left(a_i^{(r)}-b_j^{(r)}\right)\,.
  \label{eq:multi-f}
\end{equation}
We will then evaluate~\eqref{eq:kcircle-master} in Section~\ref{Sec:multi-collapse} and find that it results in powers of the conformal wheel periods $W_L$~\eqref{eq:one-circle-reduced-kernel}.

\subsubsection{Inductive construction of the spacetime integral}

We want to confirm the induction hypothesis~\eqref{eq:kcircle-master}. The base case $k=1$ is given by~\eqref{eq:base case}. We thus only need to perform the induction step from $k$ to $k+1$\,.

The induction step closely follows the previous discussion of the one-circle process. We add a $(k+1)$\textsuperscript{th} circle between the exterior operator $\overline{T}_L^{(\alpha)}$ and the $k$\textsuperscript{th} circle and evaluate it the same way as in Section~\ref{Sec:OneCircle}. However, since the vector-multiplet propagators emerging from the $a_i^{(k+1)}$ vertices now do not connect to the specific $T_L^{(\alpha)}$ operator but to the $k$\textsuperscript{th} circle, we cannot use the exterior polarisation~\eqref{eq:FieldComponentMap} to simplify the calculation. Furthermore, we have to show that after evaluating the $(k+1)$\textsuperscript{th} circle, the polarisation emerging from the exterior $\overline{T}_L^{(\alpha)}$ operator is ``pushed through'' to the $k$\textsuperscript{th} circle so that we can use the induction hypothesis. For readability we will denote
\begin{equation}
    \bm{A}_i:=A_i^{(k+1)}\,,\qquad \bm{B}_i:=B_i^{(k+1)}\,,\qquad \bm{a}_i:=a_i^{(k+1)}\,,\qquad \bm{b}_i:=b_i^{(k+1)}\,.
\end{equation}

The external vector-multiplet propagators identify the external Grassmann coordinates of the operator $\overline{T}_L^{(\alpha)}$ with those at the vertices $\bm{b}_i$ through~\eqref{eq:analytic-delta-convolution}. The projection to $\theta=\bar\theta=0$ now only reduces one Grassmann coordinate
\begin{equation}
\theta_{\bm{a}_i}^{+\alpha}=\chi_i^\alpha\,, \qquad \bar\theta_{\bm{a}_i}^{+\dot\alpha}=\bar\chi_i^{\dot\alpha}\,,\qquad \theta_{\bm{b}_i}^{+\alpha}=\eta_i^\alpha\,, \qquad \bar\theta_{\bm{b}_i}^+=0\,,
\end{equation}
which means that the analytic shift operator~\eqref{eq:AnalyticShift} in the exponential of the $q^+$-propagatorsn\eqref{eq:q-propagator} feature a second term
\begin{equation}\label{eq:subspaceshift}
  \mathfrak{S}_{12}|_{\mathrm{subspace}}
  =\frac{2i}{(u_{\bm{a}_i}^+u_{\bm{b}_j}^+)}
\left(\eta_i\sigma^\mu\bar\chi_j+(u_{\bm{a}_i}^-u_{\bm{b}_j}^+)\chi_j\sigma^\mu\bar\chi_j\right)\partial_{\bm{a}_i^\mu} \,.
\end{equation}
We only have to account for the derivatives $(D^-_{\bm{B}_i})^2$ from~$\overline{T}_L^{(\alpha)}$ which can be written as Berezin integration
\begin{equation}
  \left[\prod_i( D^-_{\bm{B}_i})^2F
  \right]_{\theta=\bar\theta=0}
  =\int\prod_i\de^2\eta_i\,F.
  \label{eq:berezin-mapF}
\end{equation}
In order to saturate these integrals, we have to pull down $2L$ factors $\eta_i\sigma^\mu\bar\chi_j$ from the exponent~\eqref{eq:subspaceshift}. Since each variable $\bar\chi_j$ can only appear with maximally two powers, we again recover the three saturation patterns~\eqref{eq:three-sectors} and crucially the additional $\chi_j\sigma^\mu\bar\chi_j$ terms in~\eqref{eq:subspaceshift} cannot appear at all. Hence, we end up with Grassmann integrals at $\bm{A}_i$ of the form
\begin{equation}
    \int \dd^2 \chi_i \dd^2 \bar\chi_i \, \bar\chi_i^2 F \quad \rightarrow  \quad D^-_{\bm{A}_i}F\vert_{\theta_{\bm{a}_i}=\bar\theta_{\bm{a}_i}=0}\,,
\end{equation}
where F is a short-hand for the remaining $k$-circle diagram. We thus see that the same Grassmann structure~\eqref{eq:FieldComponentMap} that was generated by the external $\overline{T}_L^{(\alpha)}$ operator is pushed through the $(k+1)$\textsuperscript{th} circle. This is the harmonic superspace manifestation of R-charge conservation. 

Having dealt with the Grassmann structure, we move on to the harmonic variables and we again observe that the three saturation patterns~\eqref{eq:three-sectors} generate different harmonic factors, which are either trivial for $H_{0,L}^{(k+1)}$
\begin{equation} 
    \int\prod_{i=1}^{L}\de u_{\bm{a}_i}\,\de u_{\bm{b}_i} F(u_{\bm{a}})= \int\prod_{i=1}^{L}\de u_{\bm{a}_i} F(u_{\bm{a}})\,,
\end{equation}
or require some manipulation, \eg for $H_{+,L}^{(k+1)}$
\begin{equation}\label{eq:purepatterncont}
    \int\prod_{i=1}^{L}\de u_{\bm{a}_i}\,\de u_{\bm{b}_i} \prod_{i=1}^L
    \frac{(u_{\bm{a}_{i+1}}^+u_{\bm{b}_j}^+)}{(u_{\bm{a}_{i}}^+u_{\bm{b}_j}^+)}F(u_{\bm{a}})= \int\prod_{i=1}^{L}\de u_{\bm{a}_i}\prod_{i=1}^L
  (u_{\bm{a}_{i+1}}^+u_{\bm{a}_i}^-) F(u_{\bm{a}})\,.
\end{equation}

Here we have to recall the contribution from the non-alternating orderings $\rho\pm$ in~\eqref{eq:non-alternating-cycle1} and~\eqref{eq:non-alternating-cycle2}, which contribute different spacetime and harmonic factors $K_{\rho\pm,L}^{(k+1)}$ and $H_{\rho\pm,L}^{(k+1)}$. However, in Section~\ref{Subsec:one-circle-integral} we saw that the $K_{\rho\pm,L}^{(k+1)}$ factors feature the same delta-distribution and reduce to the same expression as $K_{\pm,L}^{(k+1)}$ under integration. They are thus equivalent as distributions and we can reduce the sum over orderings to the sum over harmonic factors. After integration over $\de u_{\bm{b}_i}$ we find the expression 
\begin{equation}
\begin{split}
    \int\prod_{i=1}^{L}\de u_{\bm{b}_i} \left( H_{\pm,L}+\sum_\rho H_{\pm\rho,L}^{(k+1)}\right)&=\sum_{\xi_i = \pm1} \xi_L(u_{\bm{a}_{2}}^{\xi_{2}}u_{\bm{a}_1}^{-})(u_{\bm{a}_{1}}^{+}u_{\bm{a}_{L}}^{\xi_{L}})\prod_{i=2}^{L-1}
  \xi_i(u_{\bm{a}_{i+1}}^{\xi_{i+1}}u_{\bm{a}_i}^{-\xi_i})\\
  &=(u_{\bm{a}_{1}}^{+}u_{\bm{a}_1}^{-})=1\,,
\end{split}
\end{equation}
where we sum over all $2^{L-1}$ sign configurations $\xi_i$ in the chain with $\xi_1=1$. In the second step we used the identity
\begin{equation}
    u^{-}_iu^{+j}- u^{+}_iu^{-j}=\delta^j_i
\end{equation}
$L-1$ times to reduce the chain of harmonic factors to just a single trivial factor. Thus, the non-alternating orderings upgrade the pure pattern harmonic variable contribution~\eqref{eq:purepatterncont} to a simple integration
\begin{equation}
    \int\prod_{i=1}^{L}\de u_{\bm{a}_i}\,\de u_{\bm{b}_i}  \left( H_{\pm,L}+\sum_\rho H_{\pm\rho,L}^{(k+1)}\right)F(u_{\bm{a}})= \int\prod_{i=1}^{L}\de u_{\bm{a}_i} F(u_{\bm{a}})\,.
\end{equation}
The integration over $\dd u_{\bm{a}_i}$ is the only remaining harmonic component in each term, and again parallels exactly the projection~\eqref{eq:FieldComponentMap} imposed by the exterior $\overline{T}_L^{(\alpha)}$ operator, which is now pushed forward to the $k$\textsuperscript{th} circle. The different saturation patterns combine their spacetime-dependent factors into $K_L^{(k+1)}$ as given in~\eqref{eq:multi-kernel}.

We have thus proven that the superspace construction of $\mathcal{I}^{(k+1)}$ with separated endpoints $\{Y_i\}$ boils down to the same superspace construction for $\mathcal{I}^{(k)}$ with endpoints $\{\bm{A}_i\}$. We may then use the induction hypothesis and collect constants and spacetime factors that remain to be evaluated 
\begin{equation}
    \mathcal{I}^{(k+1)}(Y_i)=
  (2g)^{2L}
  \int\prod_{i=1}^{L}
  \de^4\bm{a}_i\,\de^4\bm{b}_i\,
  \mathcal{I}^{(k)}(\bm{A}_i) K_L^{(k+1)}
  \prod_{i=1}^{L}\frac{1}{|\bm{b}_i-y_i|^2}\,,
\end{equation}
where the last product of factors arises from the external vector-multiplet propagators. This concludes the proof of~\eqref{eq:kcircle-master} by induction.

\subsubsection{Collapse of the circles}
\label{Sec:multi-collapse}

The integral~\eqref{eq:kcircle-master} can be evaluated recursively. In order to unveil this structure, let us consider the outermost two circles in Fig~\ref{Fig:Two-point_MultiCircle} while leaving the remaining $k-2$ circles untouched. For the first circle, it is convenient to proceed similarly as for the one-circle calculation and integrate the $b_i^{(k)}$ vertices first. We point split one of the external points according to
\begin{equation}
  y_1=\cdots=y_{L-1}=y,
  \qquad
  y_L=z,
  \qquad
  z\neq y.
  \label{eq:multi-point-split}
\end{equation}
For simplicity, let us introduce $a_i:=a_i^{(k)}$, and $c_i:=b_i^{(k-1)}$.
The sum of the three Grassmann sectors of the first circle gives the one-circle numerator found in~\eqref{eq:factorised-numerator}. After cancellation against its common denominator, the outermost circle reduces to the universal block
\begin{equation}
  \begin{aligned}
  \mathcal C_L(z,y;\{c_i\})
  :={}&
  \frac{|z-y|^2g^{2L}}{(2\pi^2)^{2L}}
  \int\prod_{i=1}^{L}\dd^4a_i
  \\
  &\times
  \frac{1}
  {|z-a_1|^2
  |a_1-a_2|^2\cdots
   |a_{L-1}-a_L|^2
   |a_L-z|^2}
  \\
  &\times
  \prod_{i=1}^{L}
  \frac{1}{|a_i-y|^2|a_i-c_i|^2}\,.
  \end{aligned}
  \label{eq:first-row-block}
\end{equation}
The first circle remains coupled to the second circle at finite point splitting through
\begin{equation}
  \prod_{i=1}^{L}\frac{1}{|a_i-c_i|^2}\,.
  \label{eq:multi-coupling}
\end{equation}
The multi-circle integral therefore does not factorise before the coincidence limit $z\to y$. Instead the factorisation arises from the short-distance region selected by the point-splitting numerator. To see this set
\begin{equation}
  z^\mu=y^\mu+\delta e^\mu\,, \qquad \delta\to0\,.
  \label{eq:multi-delta}
\end{equation}
with $e^\mu$ a fixed unit vector, $\abs{e}^2=1$.
The numerator in~\eqref{eq:first-row-block} contributes $\delta^2$. The integration region, in which all $a_i-y$ remain finite, is thus suppressed. Now suppose that only $r<L$ points approach $y$ at distances of order $\delta$. If $s$ is the number of short propagators $\Delta(a_i-a_{i+1})$ belonging to the opened chain of $a_i$'s, a proper subset of the open chain satisfies
\begin{equation}
  s\leq r\,.
  \label{eq:multi-Er}
\end{equation}
In this integration region the integrand scales as
\begin{equation}
    \delta^2\,\delta^{4r}\,\delta^{-2r}\,\delta^{-2s} = \delta^{2+2(r-s)}\,,
  \label{eq:multi-partial-collapse}
\end{equation}
and hence vanishes as $\delta\to0$.
The only unsuppressed integration region is found around the complete collapse of all points
\begin{equation}
  a_i^\mu=y^\mu+\delta e_i^\mu\,, \qquad i=1,\ldots,L\,.
  \label{eq:multi-full-collapse}
\end{equation}
All of the $L$ connections to $y$ and all of the $L+1$ propagators of the point-split chain are then short, and the total scaling is
\begin{equation}
  \delta^2\,\delta^{4L}\,\delta^{-2L}\,\delta^{-2(L+1)} = \delta^0\,.
  \label{eq:multi-full-scaling}
\end{equation}
If we take the propagators between the first and second circles to remain finite in this region they can be expanded as
\begin{equation}
  \frac{1}{|a_i-c_i|^2}=\frac{1}{|y-c_i|^2}+O(\delta)\,.
  \label{eq:multi-freeze}
\end{equation}
The remaining dimensionless short-distance integral in~\eqref{eq:first-row-block} is the same completed $L$-loop ladder integral encountered for one circle~\eqref{eq:ladder-def} and consequently leads to the same wheel period
\begin{equation}
  \lim_{z\to y}\mathcal C_L(z,y;\{c_i\})= g^{2L}W_L \prod_{i=1}^{L}\Delta(c_i-y)\,.
  \label{eq:row-collapse}
\end{equation}
In this limit we excluded simultaneous collapse of vertices belonging to the next circle. Bringing an additional subset of these into the same $\delta$-region introduces four powers of $\delta$ from each integration measure, while at most two additional short propagators arise per vertex. Therefore, such mixed regions carry positive powers of $\delta$ and do not contribute to the leading residue.

Equation~\eqref{eq:row-collapse} is the crucial recursive step. The first circle has disappeared and has been replaced by the local coefficient $W_L$ together with $L$ ordinary propagators
\begin{equation}
  \prod_{i=1}^{L}\Delta(c_i-y).
  \label{eq:multi-new-spokes}
\end{equation}
These are the external propagators of the original graph with the first circle removed. The remaining integral is therefore the $(k-1)$-circle problem. We obtain the recursion relation 
\begin{equation}
  \mathcal{I}_L^{(k)}(x,y) =g^{2L}W_L\,\mathcal{I}_L^{(k-1)}(x,y)\,.
  \label{eq:kcircle-recursion}
\end{equation}
After the first collapse, the second circle becomes the new outermost circle, so the same argument can be iterated. The zeroth-order graph is the free scalar skeleton,
\begin{equation}
  \mathcal{I}_L^{(0)}(x,y) = \Delta(x-y)^L =
  \frac{1}{(2\pi)^{2L}}\frac{1}{(x-y)^{2L}}.
  \label{eq:kcircle-zero}
\end{equation}
The iteration of~\eqref{eq:kcircle-recursion} therefore gives 
\begin{equation}
  \mathcal{I}_L^{(k)}(x,y) =\bigl(g^{2L}W_L\bigr)^k\Delta(x-y)^L\,.
  \label{eq:kcircle-final}
\end{equation}

It is important to emphasise that~\eqref{eq:kcircle-final} does \emph{not} stem from a factorisation of the finite multi-circle conformal integral. At finite point splitting, consecutive circles remain coupled. The factorisation emerges only from the iterated coincidence residue~\eqref{eq:row-collapse}, which constitutes a short-distance phenomenon.

Each non-trivially wrapped bifundamental circle carries the same quiver factor $-4s_\alpha$ introduced in~\eqref{eq:QuiverFactor}.
Hence, the leading contribution from $k$ minimal circles is
\begin{equation}
  \left.\mathcal{R}_{L}^{(\alpha)}(g)\right|_{s_\alpha^k g^{2kL}}=
  (-4s_\alpha)^k\,  \frac{\mathcal{I}_L^{(k)}(x,y)}{\Delta(x-y)^L}
  =
  (-4s_\alpha)^k\,  g^{2kL} \, W_L^k \,.
  \label{eq:kcircle-localisation}
\end{equation}
This result is in perfect agreement with the localisation result~\eqref{eq:R-localisation} and its geometric interpretation in~\cite{Korchemsky:2025eyc}. Indeed, summing over the number of circles $k$ yields 
\begin{equation}\label{eq:res0}
    \mathcal{R}_{L}^{(\alpha)}(g) = \frac{1}{1+4s_\alpha \,  g^{2L} \, W_L}+\dots\,,
\end{equation}
which encodes the leading-order term at each level in $s_\alpha$.

\section{Three-point functions}
\label{Sec:ThreePtFcts}

We now want to consider three-point functions involving the twisted Coulomb branch operators~\eqref{eq:twisted-operator}. R-charge conservation and overall orbifold invariance only allow for two types of three-point functions
\begin{equation}\label{eq:4.1}
    \langle T_{L_1}^{(\alpha_1)}T_{L_2}^{(\alpha_2)}\overline{T}_{L_1+L_2}^{(\alpha_1+\alpha_2)}\rangle\,,\qquad  \langle \overline T_{L_1}^{(\alpha_1)}\overline T_{L_2}^{(\alpha_2)}T_{L_1+L_2}^{(\alpha_1+\alpha_2)}\rangle\,,
\end{equation}
each of which is extremal, \ie saturates the inequality $L_3\leq L_1+L_2$. At tree level the correlator is evaluated by a complete Wick-contraction and in extremal correlators there is no contractions between the operators 1 and 2 (see Figure~\ref{fig:tree-level}(b)). This is captured by $\ell_{12}=0$ in terms of the bridge lengths 
\begin{equation}\label{eq: bridge length}
\ell_{ij}=\tfrac{1}{2}(L_i+L_j-L_{6-i-j})\,,
\end{equation}
which count tree-level contractions between operators $i$ and $j$. 

If we want to consider also non-extremal BPS correlators we have to introduce additional BPS operators that involve a combination of $\phi$ and $\bar{\phi}$ fields. A straightforward spectral analysis (using e.g. the code in~\cite{Mantegazza:2026spd}) suggests that such BPS operators appear only in the untwisted sector and can be interpreted as vacuum descendants of $U_L$~\eqref{eq:untwisted-operator} related by an $R$-symmetry rotation in the $\mathcal{N}=4$ parent theory. We will construct the operator $\tilde{U}_{\ell_{23},\ell_{13}}$ in Appendix~\ref{App:VacDescendant}. We may then construct a wider range of twisted three-point functions
\begin{equation}\label{eq:4.2}
    \langle T_{L_1}^{(\alpha)}\overline T_{L_2}^{(\alpha)}\tilde{U}_{\ell_{23},\ell_{13}}\rangle\,,
\end{equation}
which can be non-extremal if $L_3<L_1+L_2$ (see Figure \ref{fig:tree-level}(c)).

We now cover almost all imaginable protected three-point functions involving twisted sector operators. The correlators of superconformal descendants are governed by the highest-weight correlators we just discussed. The only twisted sector BPS operators we have truly missed are special length-two operators $\mathcal{M}_3$  discussed in~\cite{Gadde:2009dj}. The protection mechanism for these operators requires an analysis of double-trace effects, so we will avoid their discussion for the time being (see however~\cite{upcoming} for further details).

In this section, we will first use the harmonic superspace construction to reproduce the leading corrections to the known extremal three-point function results for~\eqref{eq:4.1}. This shows us how the techniques from the previous Section~\ref{Sec:TwoPtFcts} can be generalised to three-point functions.
We then employ our methods to calculate and present the leading-order contributions to the non-extremal three-point functions~\eqref{eq:4.2}, which are not accessible to localisation and therefore a genuinely new result. Most notably, we find, that the leading-order correction is governed by a bridge length $\ell$ rather than by the operator lengths $L$.

\subsection{Extremal three-point functions}\label{Sec:Extremal}

\begin{figure}[t]
\centering
\begin{tikzpicture}[thick,op/.style={draw=black,fill=white,circle,minimum size=1.05cm,inner sep=0pt,outer sep=0pt,line width=1.2pt,font=\scriptsize},curl/.style={black,line width=0.8pt,decorate,decoration={snake,aspect=1,amplitude=0.75pt,segment length=5.0pt}},contract/.style={red,line width=0.9pt,dash pattern=on 3pt off 2pt},dots/.style={font=\normalsize,inner sep=0pt},highlight/.style={draw=black,line width=0.8pt}]
\def\panelSep{5.5}
\begin{scope}[xshift=0cm]
\path[use as bounding box] (-2.2,-1.65) rectangle (2.2,2.45);
\coordinate (T1) at (0,1.62);
\coordinate (L1) at (-1.40,-0.81);
\coordinate (R1) at (1.40,-0.81);
\coordinate (C1) at (0,0);
\draw[contract] (T1) -- (C1);
\draw[contract] (L1) -- (C1);
\draw[contract] (R1) -- (C1);
\draw[highlight] (L1) circle[radius=0.80cm];
\draw[curl] ([shift={(-0.095,0.055)}]T1) -- ([shift={(-0.095,0.055)}]L1);
\draw[curl] ([shift={(-0.015,0.009)}]T1) -- ([shift={(-0.015,0.009)}]L1);
\draw[curl] ([shift={(0.065,-0.038)}]T1) -- ([shift={(0.065,-0.038)}]L1);
\draw[curl] ([shift={(-0.065,-0.038)}]T1) -- ([shift={(-0.065,-0.038)}]R1);
\draw[curl] ([shift={(0.015,0.009)}]T1) -- ([shift={(0.015,0.009)}]R1);
\draw[curl] ([shift={(0.095,0.055)}]T1) -- ([shift={(0.095,0.055)}]R1);
\node[dots,rotate=150] at ($(T1)!0.50!(L1)+(-0.302,0.175)$) {$\mathinner{\cdot\!\!\cdot\!\!\cdot}$};
\node[dots,rotate=30] at ($(T1)!0.50!(R1)+(0.302,0.175)$) {$\mathinner{\cdot\!\!\cdot\!\!\cdot}$};
\fill[red] (C1) circle (1.6pt);
\node[op] at (T1) {$\overline{T}^{(\alpha_3)}_{L_3}$};
\node[op] at (L1) {$T^{(\alpha_1)}_{L_1}$};
\node[op] at (R1) {$T^{(\alpha_2)}_{L_2}$};
\node[anchor=north] at (0,-1.5) {$\sim s_{\alpha_1}g^{2L_1}$};
\end{scope}
\begin{scope}[xshift=\panelSep cm]
\path[use as bounding box] (-2.2,-1.65) rectangle (2.2,2.45);
\coordinate (T2) at (0,1.62);
\coordinate (L2) at (-1.40,-0.81);
\coordinate (R2) at (1.40,-0.81);
\coordinate (C2) at (0,0);
\draw[contract] (T2) -- (C2);
\draw[contract] (L2) -- (C2);
\draw[contract] (R2) -- (C2);
\draw[highlight] (R2) circle[radius=0.80cm];
\draw[curl] ([shift={(-0.095,0.055)}]T2) -- ([shift={(-0.095,0.055)}]L2);
\draw[curl] ([shift={(-0.015,0.009)}]T2) -- ([shift={(-0.015,0.009)}]L2);
\draw[curl] ([shift={(0.065,-0.038)}]T2) -- ([shift={(0.065,-0.038)}]L2);
\draw[curl] ([shift={(-0.065,-0.038)}]T2) -- ([shift={(-0.065,-0.038)}]R2);
\draw[curl] ([shift={(0.015,0.009)}]T2) -- ([shift={(0.015,0.009)}]R2);
\draw[curl] ([shift={(0.095,0.055)}]T2) -- ([shift={(0.095,0.055)}]R2);
\node[dots,rotate=150] at ($(T2)!0.50!(L2)+(-0.302,0.175)$) {$\mathinner{\cdot\!\!\cdot\!\!\cdot}$};
\node[dots,rotate=30] at ($(T2)!0.50!(R2)+(0.302,0.175)$) {$\mathinner{\cdot\!\!\cdot\!\!\cdot}$};
\fill[red] (C2) circle (1.6pt);
\node[op] at (T2) {$\overline{T}^{(\alpha_3)}_{L_3}$};
\node[op] at (L2) {$T^{(\alpha_1)}_{L_1}$};
\node[op] at (R2) {$T^{(\alpha_2)}_{L_2}$};
\node[anchor=north] at (0,-1.5) {$\sim s_{\alpha_2}g^{2L_2}$};
\end{scope}
\begin{scope}[xshift={2*\panelSep cm}]
\path[use as bounding box] (-2.2,-1.65) rectangle (2.2,2.45);
\coordinate (T3) at (0,1.62);
\coordinate (L3) at (-1.40,-0.81);
\coordinate (R3) at (1.40,-0.81);
\coordinate (C3) at (0,0);
\draw[contract] (T3) -- (C3);
\draw[contract] (L3) -- (C3);
\draw[contract] (R3) -- (C3);
\draw[highlight] (T3) circle[radius=0.80cm];
\draw[curl] ([shift={(-0.095,0.055)}]T3) -- ([shift={(-0.095,0.055)}]L3);
\draw[curl] ([shift={(-0.015,0.009)}]T3) -- ([shift={(-0.015,0.009)}]L3);
\draw[curl] ([shift={(0.065,-0.038)}]T3) -- ([shift={(0.065,-0.038)}]L3);
\draw[curl] ([shift={(-0.065,-0.038)}]T3) -- ([shift={(-0.065,-0.038)}]R3);
\draw[curl] ([shift={(0.015,0.009)}]T3) -- ([shift={(0.015,0.009)}]R3);
\draw[curl] ([shift={(0.095,0.055)}]T3) -- ([shift={(0.095,0.055)}]R3);
\node[dots,rotate=150] at ($(T3)!0.50!(L3)+(-0.302,0.175)$) {$\mathinner{\cdot\!\!\cdot\!\!\cdot}$};
\node[dots,rotate=30] at ($(T3)!0.50!(R3)+(0.302,0.175)$) {$\mathinner{\cdot\!\!\cdot\!\!\cdot}$};
\fill[red] (C3) circle (1.6pt);
\node[op] at (T3) {$\overline{T}^{(\alpha_3)}_{L_3}$};
\node[op] at (L3) {$T^{(\alpha_1)}_{L_1}$};
\node[op] at (R3) {$T^{(\alpha_2)}_{L_2}$};
\node[anchor=north] at (0,-1.5) {$\sim s_{\alpha_3}g^{2L_3}$};
\end{scope}
\end{tikzpicture}
\caption{We have three choices to add one hypermultiplet circle in the extremal three-point function of Figure~\ref{fig:tree-level}(b). The first two diagrams can be evaluated as for the two-point function. For the third diagram, we have to consider a \emph{split wrapped circle}.}
    \label{Fig:Three-point_Extremal}
\end{figure}
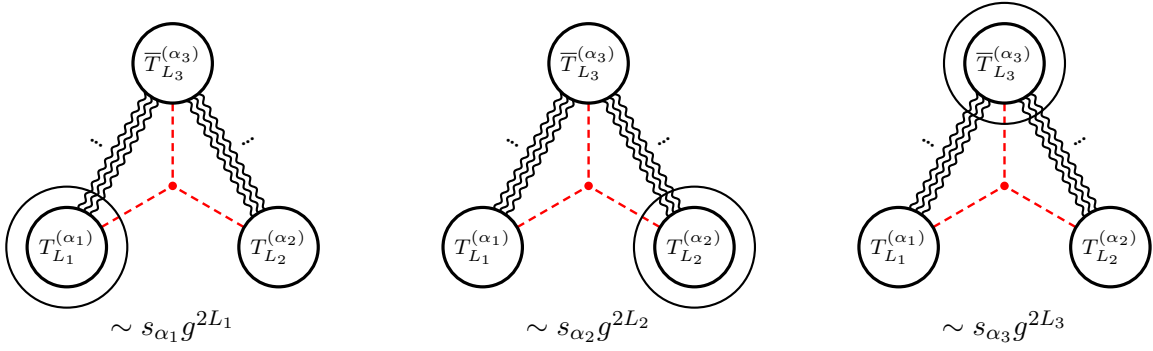
The extremal three-point correlator at tree level was depicted in Figure~\ref{fig:tree-level}(b). Here $\ell_{13}=L_1$ of the $\bar{\phi}$ scalar fields in $\overline{T}_{L_3}^{\alpha_3}(x_3)$ are contracted with $\phi$ in $T_{L_1}^{\alpha_1}(x_1)$ and $\ell_{23}=L_2$ with those in $T_{L_2}^{\alpha_2}(x_2)$, resulting in the conformal
spacetime dependence
\begin{equation}
 \Delta(x_1-x_3)^{L_1}\Delta(x_2-x_3)^{L_2}\,.
 \label{eq:three-point-tree-propagators}
\end{equation}
Following the same logic as in Section~\ref{Sec:TwoPtFcts}, we observe that twist dependence can only arise once a hypermultiplet circle wraps around one of the three operators, \cf Figure~\ref{Fig:Three-point_Extremal}. We consider these different possibilities separately.

\paragraph{A circle around one of the short operators}
A circle surrounding one of the operators $T_{L_i}^{(\alpha)}(x_i)$ with $i=1,2$, splits up the $L_i$ vector-multiplet propagators  joining $x_i$ to $x_3$. The other $L_{j\neq i}$  vector-multiplet propagators joining $x_j$ to $x_3$ are spectators and do not enter the local harmonic superspace $D$-algebra. This is depicted in the first two diagrams of Figure~\ref{Fig:Three-point_Extremal}.

The non-trivial subgraph is therefore exactly the length-$L_i$ two-point graph evaluated in Section~\ref{Sec:OneCircle}. Using~\eqref{eq:one-circle-reduced-kernel}, its reduces to the expression
\begin{equation}
 \mathcal{I}_{L_i}(x_i,x_3) = g^{2L_i}
W_{L_i}\Delta(x_i-x_3)^{L_i}\,.
 \label{eq:three-point-ell-wheel-kernel}
\end{equation}
Multiplying by the $L_j$ spectator propagators gives the complete spacetime factor~\eqref{eq:three-point-tree-propagators}. Reinstating the twist factor $-4s_{\alpha_i}$ that is picked up by the hypermultiplet circle due to its interaction with the twist line, we find a contribution to the ratio $\mathcal{C}_{L_1,L_2,L_3}^{(\alpha_1,\alpha_2,\alpha_3)}(g)$~\eqref{eq:Ghat-def} of 
\begin{equation}\label{eq: res1}
   -4s_{\alpha_1}g^{2L_1}
W_{L_1}-4s_{\alpha_2}g^{2L_2}
W_{L_2}\,.
\end{equation}

\paragraph{A wrapped circle around the split length-$\bm{L_3}$ operator} 
The circle surrounding $\overline{T}_{L_3}^{(\alpha)}(x_3)$ is the only new configuration, see the third diagram in Figure~\ref{Fig:Three-point_Extremal}. Its $L_3$ scalar propagators connect to two different points, as $L_1$ terminate at $x_1$ and $L_2$ terminate at $x_2$. The resulting scalar integral is therefore a genuine three-point integral. 

The harmonic-superspace structure of the graph is nevertheless unchanged. Starting from~\eqref{eq:exact-post-convolution} with $L$ replaced by $L_3$, the transferred derivatives still act on the complete product of all $2L_3$ hypermultiplet propagators. The Grassmann saturation equations therefore have the same three solutions as in the two-point problem, \cf~\eqref{eq:three-sectors}.

We follow a similar evaluation procedure as in Section~\ref{Sec:OneCircle}. Again we perform point-splitting and allow all but one $y_i$ coordinate to coincide
\begin{equation}  y_1=\cdots=y_{L_3-1}=x_3\,, \qquad y_{L_3}=z\,,\qquad z\neq x_3\,.
 \label{eq:three-point-p-leg-point-splitting}
\end{equation}
The superspace evaluation reduces to the spacetime integral
\begin{equation}
 \begin{aligned}
 \mathcal{I}_{L_3}^{\mathrm{split}}(z,x_3;\{x_i\}) ={}& \frac{|z-x_3|^2g^{2L_3}}{(2\pi^2)^{2L_3}} \int\prod_{i=1}^{L_3}\de^4a_i \frac{1} {|z-a_1|^2|a_1-a_{2}|^2\cdots|a_{L_3}-z|^2} \\
 &\times  \prod_{i=1}^{L_3} \frac{1}{|a_i-x_3|^2}\prod_{i=1}^{L_1} \frac{1}{|a_i-x_1|^2}\prod_{i=L_1+1}^{L_3} \frac{1}{|a_i-x_2|^2}\,.
 \end{aligned}
 \label{eq:three-point-split-point-split-integral}
\end{equation}
The derivation of~\eqref{eq:three-point-split-point-split-integral} follows the one-circle derivation presented in Section~\ref{Sec:OneCircle}. 

\subsubsection{Evaluation of the split wrapped circle}

In this section we will show that the split-circle integral~\eqref{eq:three-point-split-point-split-integral} evaluates to
\begin{equation}
 \lim_{z\to x_3} \mathcal{I}_{L_3}^{\mathrm{split}}(z,x_3;\{x_i\}) = g^{2L_3}
W_{L_3}\Delta(x_1-x_3)^{L_1}\Delta(x_2-x_3)^{L_2}\,.
 \label{eq:three-point-split-wheel-residue}
\end{equation}
upon removal of the point-splitting regularisation. In order to show this, let us set
\begin{equation}
 z^\mu=x_3^\mu+\delta e^\mu\,, \qquad \delta\to0\,,
 \label{eq:three-point-split-scaling-variable}
\end{equation}
where $e^\mu$ is once again a unit vector satisfying $|e|^2=1$.
The explicit numerator in~\eqref{eq:three-point-split-point-split-integral} contributes a factor $\delta^2$.

First consider a region in which a proper subset of $r<L_3$ integration points approaches~$x_3$. Furthermore, let $s$ be the number of propagators $\Delta(a_i-a_{i+1})$ in~\eqref{eq:three-point-split-point-split-integral} whose two endpoints are separated by a distance of order $\delta$. If a proper subset of the vertices collapses to $x_3$, possibly including $z$, we have $s\leq r$. Therefore the integration region scales as
\begin{equation}
 \delta^2\delta^{4r}\delta^{-2r}\delta^{-2s} = \delta^{2+2(r-s)}\,.
 \label{eq:three-point-proper-cluster-scaling}
\end{equation}
This suggests that any partial collapse vanishes. The only unsuppressed integration region is the complete collapse
\begin{equation}
 a_i^\mu=x_3^\mu+\delta z_i^\mu\,, \qquad i=1,\ldots,L_3\,.
 \label{eq:three-point-complete-collapse}
\end{equation}
In this region all $L_3$ propagators emerging from $x_3$ and all $L_3+1$ propagators on the point-split circle are short. The total scaling is
\begin{equation}
 \delta^2\delta^{4L_3}\delta^{-2L_3}\delta^{-2(L_3+1)} = \delta^0\,.
 \label{eq:three-point-complete-collapse-scaling}
\end{equation}
Since $x_1$ and $x_2$ remain separated from $x_3$, the propagators to the split external endpoints are finite. The leading short-distance integral is then
\begin{equation}
 \begin{aligned}
 \frac{1}{\pi^{2L_3}}\int\prod_{i=1}^{L_3}\de^4z_i\, \frac{1} {|z-z_1|^2 |z_1-z_{2}|^2\cdots |z_{L_3}-z|^2} \prod_{i=1}^{L_3}\frac{1}{z_i^2} = \,W_{L_3}\,.
 \end{aligned}
 \label{eq:three-point-affine-wheel}
\end{equation}
Adding back the remaining finite factors in~\eqref{eq:three-point-split-point-split-integral}
\begin{equation}
    g^{2L_3}\Delta(x_1-x_3)^{L_1}\Delta(x_2-x_3)^{L_2}
\end{equation}
we arrive at the expected result~\eqref{eq:three-point-split-wheel-residue}, which results in a contribution
\begin{equation}\label{eq: res2}
   -4s_{\alpha_3}g^{2L_3}
W_{L_3}
\end{equation}
to the ratio $\mathcal{C}_{L_1,L_2,L_3}^{(\alpha_1,\alpha_2,\alpha_3)}(g)$~\eqref{eq:Ghat-def}\,.

\subsubsection{Comparison with localisation results}

Having evaluated the leading-order contributions from hypermultiplet circles around each operator, we can now collect the contributions~\eqref{eq: res1} and~\eqref{eq: res2} from the diagrams in Figure~\ref{Fig:Three-point_Extremal} and find 
\begin{equation}
\mathcal{C}_{L_1,L_2,L_3}^{(\alpha_1,\alpha_2,\alpha_3)}(g)=1-4s_{\alpha_1} g^{2L_1}W_{L_1}-4s_{\alpha_2} g^{2L_2}W_{L_2}-4s_{\alpha_3}g^{2L_3}W_{L_3}+\order{g^{2L_{\text{min}}+2}}\,,
\end{equation}
in agreement with~\eqref{eq:compare}.
The normalisation-independent structure constant~\eqref{eq: normed 3pt} is obtained by dividing this correlator by the square roots of the external two-point functions~\eqref{eq:res0} 
\begin{equation}
 C_{L_1,L_2,L_3}^{(\alpha_1,\alpha_2,\alpha_3)}(g)= 1-2s_{\alpha_1} g^{2L_1}W_{L_1}-2s_{\alpha_2} g^{2L_2}W_{L_2}-2s_{\alpha_3}g^{2L_3}W_{L_3}+\order{g^{2L_{\text{min}}+2}}\,.
 \label{eq:three-point-normalised-one-circle-result}
\end{equation}
This quantity should be compared to integrability results~\cite{Ferrando:2025qkr}.

\paragraph{Several wrapped circles}
Also for three-point functions, we may consider contributions with higher powers of the twist factors $s_\alpha$. At leading order in coupling these are again due to multiple hypermultiplet circles wrapping the twisted operators. As in the two-point function, were we captured all these multi-circle diagrams in the generating function~\eqref{eq:res0}
\begin{equation}
    \mathcal{R}_{L}^{(\alpha)}(g) = \frac{1}{1+4s_\alpha \,  g^{2L} \, W_L}+\dots\,,
\end{equation}
we can construct a similar generating function for the three point function
\begin{equation}
    \mathcal{C}_{L_1,L_2,L_3}^{(\alpha_1,\alpha_2,\alpha_3)}(g)=\frac{1}{1+4s_{\alpha_1} \,  g^{2L_1} \, W_{L_1}}\frac{1}{1+4s_{\alpha_2} \,  g^{2L_2} \, W_{L_2}}\frac{1}{1+4s_{\alpha_3} \,  g^{2L_3} \, W_{L_3}}+\dots\,.
\end{equation}
We can compare this to the factorisation~\eqref{eq:C-factorisation} observed in localisation, where we may approximate the factors
\begin{equation}
 \mathcal{V}_{L}^{(\alpha)}(g)=\sqrt{ \mathcal{R}_{L}^{(\alpha)}(g) + \frac{g}{2L}  \mathcal{R}_{L}^{(\alpha)}(g)} \sim  \frac{1}{1+4s_\alpha \,  g^{2L} \, W_L}+\dots\,,
 \label{eq:three-point-minimal-derivative-identity}
\end{equation}
and thus find a precise match at leading order in coupling for each individual $s_{\alpha_1}^ks_{\alpha_2}^ls_{\alpha_3}^m$ term. In order to complete the matching at subleading orders one would have to investigate more complicated supergraphs featuring additional vector-multiplet propagator decorations. These may connect to the already present propagators and possibly connect to the external operators which have subleading terms in the expansion~\eqref{eq:expansion}. A precise study of such subleading terms is left for future research.

\subsection{Non-extremal three-point functions} \label{Sec:non-extremal}
We now turn to genuinely non-extremal three-point functions~\eqref{eq:4.2} 
\begin{equation}
    \langle T_{L_1}^{(\alpha)}(x_1)\overline T_{L_2}^{(\alpha)}(x_2)\tilde{U}_{\ell_{23},\ell_{13}}(x_3)\rangle \sim \Delta(x_1-x_2)^{\ell_{12}}\Delta(x_1-x_3)^{\ell_{13}}\Delta(x_2-x_3)^{\ell_{23}}\label{eq:nonext-tree} \,,
\end{equation}
with all bridge-lengths $\ell_{ij}>0$,  which require the introduction of an untwisted BPS \emph{vacuum descendant} $\tilde{U}_{\ell_{23},\ell_{13}}$ built upon a mixed single-trace operator of the form $\Tr\,\bar\phi^{\ell_{13}}\phi^{\ell_{23}}$. Vanishing of the anomalous dimension forces us to add various additional terms to this operator as is discussed in Appendix~\ref{App:VacDescendant}
\begin{equation}\label{eq:schematic desc}
    \tilde{U}_{\ell_{23},\ell_{13}}\sim \Tr\phi^{\ell_{23}}\bar\phi^{\ell_{13}} -\frac{1}{2}\Tr \phi^{\ell_{23}-1}f^i\bar\phi^{\ell_{13}-1}\bar{f}_i +\dots\,. 
\end{equation}
The first term is the expected term that allows for the tree-level contractions in Figure~\ref{fig:tree-level}(c). In the second term we observe an admixture of hypermultiplet scalar fields $f$ and $\bar f$ which first showed up in the expansion~\eqref{eq:q+} of the $q^+$ superfield. This component allows for a new class of twist-sensitive hypermultiplet circles whose leading perturbative order is controlled by a bridge length $(\ell_{12}+1)$~\eqref{eq: bridge length} rather than by the full length $L_i$ of one of the external operators. 

\begin{figure}[h]
    \centering
    \begin{tikzpicture}[thick,baseline=3pt,op/.style={draw=black,fill=white,circle,minimum size=1.3cm,inner sep=0pt,outer sep=0pt,line width=1.2pt,font=\scriptsize},curl/.style={black,line width=0.8pt,decorate,decoration={snake,aspect=1,amplitude=0.75pt,segment length=5.0pt}},contract/.style={red,line width=0.9pt,dash pattern=on 3pt off 2pt},highlight/.style={draw=black,line width=0.8pt},bridgelabel/.style={font=\small,inner sep=0.5pt,fill=white}]
\coordinate (T1) at (0,1.62);
\coordinate (L1) at (-1.40,-0.81);
\coordinate (R1) at (1.40,-0.81);
\coordinate (L1u) at ($(L1)+(25:0.525cm)$);
\coordinate (R1u) at ($(R1)+(155:0.525cm)$);
\draw[contract] (L1u) -- (R1u);
\node[circle,minimum size=1.05cm,inner sep=0pt,outer sep=0pt] (TopAnchor) at (T1) {};
\def\wrapR{0.95}
\def\lowerOut{0.86}
\def\entryDepth{0.94}
\coordinate (WrapUpper) at ($(L1)+(150.0524:\wrapR cm)$);
\coordinate (WrapLower) at ($(L1)+(-29.9476:\wrapR cm)$);
\coordinate (BottomEnd) at (TopAnchor.south);
\coordinate (LowerCtrl) at ($(WrapLower)+(60.0524:\lowerOut cm)$);
\coordinate (EntryCtrl) at ($(BottomEnd)+(0,-\entryDepth cm)$);
\def\topHitAngle{202}
\coordinate (TopEnd) at ($(T1)+(\topHitAngle:0.525cm)$);
\coordinate (ProjEntry) at ($(L1)!(EntryCtrl)!(T1)$);
\coordinate (TopEntryCtrl) at ($(EntryCtrl)!2!(ProjEntry)$);
\coordinate (ProjLowerCtrl) at ($(L1)!(LowerCtrl)!(T1)$);
\coordinate (UpperCtrl) at ($(LowerCtrl)!2!(ProjLowerCtrl)$);
\node[rotate=60][bridgelabel] at ($(T1)!0.50!(L1)+(-0.322, 0.186)$) {\scriptsize{$\ell_{13}-1$}};
\node[rotate=-60][bridgelabel] at ($(T1)!0.50!(R1)+(0.322, 0.186)$) {\scriptsize{$\ell_{23}-1$}};
\node[bridgelabel] at ($(L1)!0.50!(R1)+(0,-0.31)$) {\scriptsize{$\ell_{12}$}};
\draw[highlight] (TopEnd) .. controls (TopEntryCtrl) and (UpperCtrl) .. (WrapUpper) arc[start angle=150.0524,end angle=330.0524,radius=\wrapR cm] .. controls (LowerCtrl) and (EntryCtrl) .. (BottomEnd);
\draw[curl] ([shift={(-0.095,0.055)}]T1) -- ([shift={(-0.095,0.055)}]L1);
\draw[curl] ([shift={(-0.015,0.009)}]T1) -- ([shift={(-0.015,0.009)}]L1);
\draw[curl] ([shift={(0.065,-0.038)}]T1) -- ([shift={(0.065,-0.038)}]L1);
\draw[curl] ([shift={(-0.065,-0.038)}]T1) -- ([shift={(-0.065,-0.038)}]R1);
\draw[curl] ([shift={(0.015,0.009)}]T1) -- ([shift={(0.015,0.009)}]R1);
\draw[curl] ([shift={(0.095,0.055)}]T1) -- ([shift={(0.095,0.055)}]R1);
\draw[curl] ([shift={(0,0.092)}]L1) -- ([shift={(0,0.092)}]R1);
\draw[curl] (L1) -- (R1);
\draw[curl] ([shift={(0,-0.092)}]L1) -- ([shift={(0,-0.092)}]R1);
\coordinate (EL) at ($(L1)+(40:0.525cm)$);
\coordinate (ER) at ($(R1)+(140:0.525cm)$);
\def\archX{0.40}
\def\archY{0.30}
\draw[black,line width=0.8pt,decorate,decoration={snake,aspect=1,amplitude=0.75pt,segment length=5.0pt}] (EL) .. controls (-\archX,\archY) and (\archX,\archY) .. (ER);
\node[op] at (T1) {$\tilde U_{\ell_{23},\ell_{13}}$};
\node[op] at (L1) {$T^{(\alpha)}_{L_1}$};
\node[op] at (R1) {$\overline{T}^{(\alpha)}_{L_2}$};
\end{tikzpicture}$\qquad \sim \,\,s_\alpha g^{2(\ell_{12}+1)}$
    \caption{The leading twist-dependent contribution to the non-extremal three-point function in Figure~\ref{fig:tree-level}(c) forms a \emph{punctured hypermultiplet circle}.}
    \label{Fig:Three-point_NonExtremal}
\end{figure}
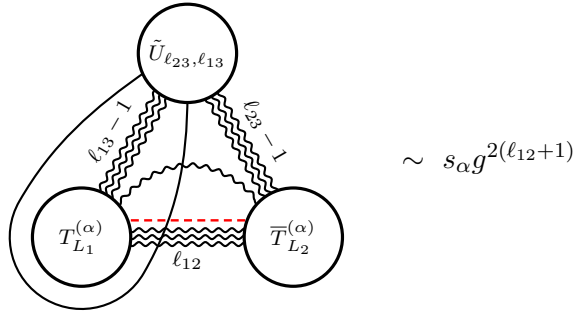

We may connect the two hypermultiplet fields by a hypermultiplet propagator that intersects $\tilde{\ell}:=(\ell_{12}+1)$ vector-multiplet propagators and the twist-line between $T_{L_1}^{(\alpha)}$ and $ T_{L_2}^{(\alpha)}$ as is depicted schematically in Figure~\ref{Fig:Three-point_NonExtremal}. This contribution, to which we will refer as a \emph{punctured hypermultiplet circle} supergraph picks up a factor $g^{2\tilde{\ell}}s_\alpha$. Note in particular that this contribution is generically of lower order in gauge coupling than the wrapped hypermultiplet circles that we discussed before.\footnote{We will comment on special cases with coincident perturbative orders in Section~\ref{Sec:LeadingCoeff_nonext3pt}.} Of course, these would also need to be taken into account eventually, alongside contributions of further terms in~\eqref{eq:schematic desc}, but the single punctured circle is the dominant contribution, to which we will restrict our attention in this section. Let us investigate the superspace evaluation of Figure~\ref{Fig:Three-point_NonExtremal}.

\subsubsection{Evaluation of the punctured circle}
As in the previous examples, we introduce interaction vertices $A_i$ and $B_i$ along the hypermultiplet circle and first consider the strictly alternating hypermultiplet ordering
\begin{equation}
X_3
\longrightarrow A_1\longrightarrow B_1
\longrightarrow A_2\longrightarrow B_2
\longrightarrow\cdots
\longrightarrow A_{\tilde{\ell}}
\longrightarrow B_{\tilde{\ell}}
\longrightarrow X_3\,.
\label{eq:punctured-chain}
\end{equation}
The $A_i$ are analytic vertices~\eqref{eq:analyticvertex} connected by a vector-multiplet propagator $G_V^{(2,2)}(X_1    ,A_i)$ \eqref{eq:vector-propagator} to $X_1$ and the $B_i$ are vertices connected to $X_2$. The internal hypermultiplet propagators~\eqref{eq:q-propagator} are
\begin{equation}\prod_{i=1}^{\tilde{\ell}}
G_q(A_i,B_i)
\prod_{i=1}^{\tilde{\ell}-1}
G_q(B_i,A_{i+1})\,.
\label{eq:punctured-propagators}
\end{equation}
In an ordinary wrapped circle from Section~\ref{Sec:OneCircle} there would be one further hypermultiplet propagator closing the circle. For the punctured circle this propagator is replaced by the physical $q^+\tilde q^+$ pair contained in $\tilde U_{\ell_{23},\ell_{13}}(x_3)$. The usual projection $\theta=\bar\theta=0$ singles out the scalar component and we consider the $\SU(2)_R$ singlet combination which arises from an identification of the harmonic variables. The endpoint factor then becomes
\begin{equation}
e_3(A_1,B_{\tilde{\ell}}) = (u_{a_1}^+u_{b_{\tilde{\ell}}}^+)\Delta(x_3-a_1)\Delta(b_{\tilde{\ell}}-x_3)\,.
\label{eq:puncture-endpoint}
\end{equation}
Thus, the two endpoint propagators are simply the scalar-component continuation of the hypermultiplet propagator chain in~\eqref{eq:punctured-propagators}.

Since the Grassmann variables at the end points vanish, the Grassmann calculation differs slightly from the closed wrapped circle. Along the open hypermultiplet chain we may again impose
\begin{equation}
\theta_{a_i}^+=0\,, \qquad \bar\theta_{a_i}^{+\dot\alpha}=\bar\chi_i^{\dot\alpha}\,,\qquad \theta_{b_i}^{+\alpha}=\eta_i^\alpha\,, \qquad \bar\theta_{b_i}^+=0\,,
  \label{eq:puncture-Grassmann-slice}
\end{equation}
and introduce $f_{ij}$ and $\lambda_{ij}$ as in~\eqref{eq:f-ell}. The Grassmann integral becomes then
\begin{equation}
\mathcal{K}_{\tilde\ell}^{\rm open}
=
\mathcal{H}_{\tilde\ell}
\int
\prod_{i=1}^{\tilde\ell}
\de^2\eta_i\,\de^2\bar\chi_i
\prod_{i=1}^{\tilde\ell}
\left(e^{\lambda_{ii}}f_{ii}\right)
\prod_{i=1}^{\tilde\ell-1}
\left(e^{\lambda_{i+1,i}}f_{i+1,i}\right)\,,
\label{eq:puncture-Grassmann-kernel}
\end{equation}
where we defined $\mathcal{H}_l$ in analogy to~\eqref{eq:DefCalH}.

Let $m_i$ denote the selected power of $\lambda_{ii}$ and $n_i$ the selected power of $\lambda_{i+1,i}$. Saturation of the integrals over $\eta_1$ and $\bar\chi_{\tilde\ell}$ uniquely fix the boundary powers
\begin{equation}
m_1=2\,, \qquad m_{\tilde\ell}=2\,.
\label{eq:puncture-endpoint-incidence}
\end{equation}
The usual saturation rules~\eqref{eq:incidence-equations} then fix a unique saturation pattern
\begin{equation}
m_i=2\,,
\qquad
i=1,\ldots,\tilde\ell\,,
\qquad
n_i=0\,,
\qquad
i=1,\ldots,\tilde\ell-1\,.
\label{eq:puncture-unique-sector}
\end{equation}
This is the essential difference with the closed hypermultiplet circle of Section~\ref{Sec:one-circle-dalgebra}, where the cyclic incidence problem admitted the two pure and one mixed saturation pattern. Opening the circle on the physical $q^+\tilde q^+\vert_{\theta=\bar\theta=0}$ pair at $x_3$ fixes its saturation pattern uniquely.

The same reduction of possibilities is visible in the non-alternating orderings, which arise from the single ordering scheme
\begin{equation}
x_3
\to (A_1,B_1)
\to\cdots
\to (A_{\tilde{\ell}},B_{\tilde{\ell}})
\to x_3\,, \qquad (A_i, B_i)\in\{A_i\to B_i,B_i\to A_i\}\,,
\label{eq:punctured-chain-disorder}
\end{equation}
and likewise only allow the saturation pattern~\eqref{eq:puncture-unique-sector}. Summing over all orderings, the harmonic variable integrals reduce to a trivial factor $1$ as before and all spacetime-dependence reduces (in the sense of equivalent distributions) to
\begin{equation}
    \mathcal{K}_{\tilde\ell}^{\rm open} = \prod_{i=1}^{\tilde\ell} \left[ \Box_{a_i}\Delta(a_i-b_i)\right]\prod_{i=1}^{\tilde\ell-1}\Delta(a_{i+1}-b_i)\,.
\label{eq:puncture-final-kernel}
\end{equation}

Using~\eqref{eq:bosonic-propagator} the $\tilde\ell$ Laplacians set $a_i=b_i$. The reduced interacting integral becomes
\begin{equation}
\begin{aligned}
    \mathcal{I}_{\tilde\ell}^{\rm punct}={}&\frac{g^{2\tilde{\ell}}}{(2\pi)^{2\tilde\ell+1}\pi^{2\tilde{\ell}}} \int \prod_{i=1}^{\tilde\ell} \de^4a_i \\
    &\times \frac{1}{|x_3-a_1|^2|a_1-a_2|^2\cdots|a_{\tilde\ell-1}-a_{\tilde\ell}|^2|a_{\tilde\ell}-x_3|^2}\prod_{i=1}^{\tilde\ell}\frac{1}{|a_i-x_1|^2|a_i-x_2|^2}\,.
\end{aligned}
\label{eq:puncture-integral}
\end{equation}
This is precisely the coincident-endpoint $\tilde\ell$-loop ladder $D_{\tilde\ell}(x_3,x_1,x_3,x_2)$ from~\eqref{eq:ladder-def} and we ultimately arrive at the expression
\begin{equation}
    \mathcal{I}_{\tilde\ell}^{\rm punct}=g^{2\tilde{\ell}}W_{\tilde{\ell}}\Delta(x_1-x_2)^{\ell_{12}}\Delta(x_1-x_3)\Delta(x_2-x_3)\,.
\end{equation}
Together with the $(\ell_{13}-1)$ spectator propagators $\Delta(x_1-x_3)$ and $(\ell_{23}-1)$ spectator propagators $\Delta(x_2-x_3)$, we recover the conformal three-point spacetime dependence~\eqref{eq:nonext-tree}.

Finally, we have to restore the twist factors. The $\SU(2)_R$ singlet $f^i\bar{f}_i$ contains two physical scalar components. For either component the bifundamental trajectory can run in the two opposite quiver orientations, giving phases $\omega^\alpha$ and $\omega^{-\alpha}$, respectively. The total sum of twist phases a marked planar skeleton is therefore $2\omega^\alpha+2\omega^{-\alpha}$, while the corresponding untwisted sum is $4$. Multiplying by the mixing coefficient in~\eqref{eq:schematic desc} gives a total factor
\begin{equation}
-\frac12 \left(2\omega^\alpha+2\omega^{-\alpha} - 4\right)
=-\left(\omega^\alpha+\omega^{-\alpha}-2\right)=+4s_\alpha\,.
\label{eq:puncture-quiver-factor}
\end{equation}
Taking the ratio with the untwisted three-point function as in~\eqref{eq:Ghat-def}, we expect a contribution 
\begin{equation}
\begin{split}
\tilde{\mathcal{C}}_{L_1,L_2,L_3}^{(\alpha)}(g):=&\frac{\langle T_{L_1}^{(\alpha)}(x_1)\overline T_{L_2}^{(\alpha)}(x_2)\tilde{U}_{\ell_{23},\ell_{13}}(x_3)\rangle}{\langle U_{L_1}(x_1)\overline U_{L_2}(x_2)\tilde{U}_{\ell_{23},\ell_{13}}(x_3)\rangle}\\=&1 + 4s_\alpha \,g^{2(\ell_{12}+1)}\, W_{\ell_{12}+1} + \mathcal{O}(g^{2(\ell_{12}+2)})+\mathcal{O}(g^{2L_i})\,,\label{eq:leading-puncture}
\end{split}\end{equation}
which is a quantity so far inaccessible to localisation and thus a genuinely new perturbative result. We will comment on the wrapping contributions at $\mathcal{O}(g^{2L_i})$ in the next section.

\subsubsection{Further contributions and normalisation}\label{Sec:LeadingCoeff_nonext3pt}
The purely adjoint component of~\eqref{eq:schematic desc},
\begin{equation}
\Tr\phi^{\frac{L+r}{2}}\bar\phi^{\frac{L-r}{2}}\,,
\end{equation}
still produces the ordinary wrapped circles considered above. Their leading contributions are
\begin{equation}
\begin{aligned}
\left.
\tilde{\mathcal{C}}_{L_1,L_2,L_3}^{(\alpha)}(g)\right|_{x_1\text{-circle}} &= -4s_\alpha g^{2L_1}W_{L_1} + O(g^{2(L_1+1)})\,, \\
\left.\tilde{\mathcal{C}}_{L_1,L_2,L_3}^{(\alpha)}(g) \right|_{x_2\text{-circle}} &= -4s_\alpha g^{2L_2}W_{L_2} + O(g^{2(L_2+1)})\,.
\label{eq:ordinary-nonext-circles}
\end{aligned}
\end{equation}
Note that there is no contribution from the untwisted operator. The wrapping effects~\eqref{eq:ordinary-nonext-circles} occur at the same order as~\eqref{eq:leading-puncture} only when
\begin{equation}
L_1=\ell_{12}+1 \quad\Longleftrightarrow\quad \ell_{13}=1\,, \qquad \text{or} \qquad
L_2=\ell_{12}+1 \quad\Longleftrightarrow\quad \ell_{23}=1\,.
\label{eq:same-order-wrap}
\end{equation}

We may furthermore consider higher components in the expansion~\eqref{eq:PJd} of $\tilde U_{\ell_{23},\ell_{13}}$ and find that their contributions are strictly subleading. A component containing $s$ hypermultiplet pairs has the form
\begin{equation}
\Tr\phi^{\frac{L+r}{2}-s}\bar\phi^{\frac{L-r}{2}-s} (f^i\bar f_i)^s\,.
\label{eq:r-pair-component}
\end{equation}
After contracting the adjoint spectators, it leaves $\tilde{\ell}_s = \ell_{12}+s$ uncontracted fields for each twisted operator. A twist-dependent matter graph involving this component therefore contributes at orders above
\begin{equation}
g^{2(\ell_{12}+s)}\,.
\label{eq:r-pair-order}
\end{equation}
All components with $s\geq2$ are thus supressed compared to the $s=1$ contribution~\eqref{eq:leading-puncture}.

Taking these possibilities into account we can slightly refine our result ~\eqref{eq:leading-puncture}
\begin{equation}
\tilde{\mathcal{C}}_{L_1,L_2,L_3}^{(\alpha)}(g) = 1 + 4s_\alpha  (1-\delta_{\ell_{23},1} - \delta_{\ell_{13},1})
g^{2(\ell_{12}+1)} \,W_{\ell_{12}+1} + \mathcal{O}(g^{2\ell_{12}+4})\,.
\label{eq:nonext-leading-general}
\end{equation}
In any case we observe that the first twist-sensitive correction of the non-extremal correlator ~\eqref{eq:nonext-tree} is always governed by the bridge length $\ell_{12}$ rather than by $\min(L_1,L_2)$.

The normalisation-independent structure constant can be constructed as in~\eqref{eq: normed 3pt} and since coupling-dependent contributions from $\mathcal{R}_{L_i}^{\alpha}$ arise at wrapping order $\mathcal{O}(g^{2L_i})$, they only contribute to the leading order in the ``near-extremal'' case~\eqref{eq:same-order-wrap}. In analogy to~\eqref{eq:three-point-normalised-one-circle-result} we find
\begin{equation}
\tilde{C}_{L_1,L_2,L_3}^{(\alpha)}(g)=1+(4-2\delta_{\ell_{23},1} -2 \delta_{\ell_{13},1})s_\alpha
g^{2(\ell_{12}+1)}W_{\ell_{12}+1}+ \mathcal{O}(g^{2\ell_{12}+4})\,.
\label{eq:nonext-normalised}
\end{equation}

Equation~\eqref{eq:nonext-normalised} determines the first non-trivial correction to the twisted structure constant of~\eqref{eq:nonext-tree}. The next-to-leading terms at order $g^{2(\ell_{12}+2)}$ will arise both from dressings by gauge propagators and potentially from contributions of further terms~\eqref{eq:r-pair-component} in $\tilde{U}_{\ell_{23},\ell_{13}}$. However, since these ``local'' dressings do not generate additional crossings through the twist line, we expect that a twist factor $s_\alpha^2$ shows up only at order $g^{4(\ell_{12}+1)}$ or higher. Investigating these subleading contributions would be a technical challenge that should first be tested in the extremal correlator cases where localisation data provides a consistency check. We comment on such possibilities in the discussion in Section~\ref{Sec:Conclusion} and may return to this challenge in future publications. At the current stage, the leading result~\eqref{eq:nonext-normalised} already provides an interesting field-theory realisation of a bridge-controlled finite-size correction. In the companion paper~\cite{ShortPaper} we investigate how these effects are captured by integrability.

\paragraph{Zero-bridge limit} Finally, let us stress that the derivation above assumes $\ell_{12}\geq1$, or equivalently $\tilde\ell\geq2$. The extremal zero-bridge case $L_3=L_1+L_2$ is qualitatively different. Due to $\tilde\ell=1$ the reduced punctured graph contains the logarithmically divergent contact integral 
\begin{equation}\label{eq:log}
    \int \de^4a\, \frac{1}{|a-x_3|^4|a-x_1|^2|a-x_2|^2}\sim\frac{1}{|x_3-x_1|^2|x_3-x_2|^2}\int \de^4b\, \frac{1}{|b|^4} \,, 
\end{equation}
for which the formal continuation of~\eqref{eq:UD-ladder} to $W_1=2\zeta(1)$ should not be interpreted as a physical conformal period. The divergence is local at the descendant insertion and signals that the zero-bridge problem is not a closed renormalised object but requires a careful renormalisation and potentially mixing with other protected operators. In particular, at extremality the single-trace descendant is degenerate with multi-trace protected operators, whose admixture can contribute at planar order to the three-point function. We therefore restrict the result~\eqref{eq:leading-puncture} to $\ell_{12}\geq1$ and leave the zero-bridge mixing problem for a separate analysis.

\section{Conclusion}\label{Sec:Conclusion}

In this work we have developed a harmonic-superspace description of planar correlation functions in $\mathcal N=2$ $\amsmathbb{Z}_M$ orbifolds of $\mathcal N=4$ SYM, placing particular emphasis on those involving twisted sector operators. Their treatment in this formalism is particularly natural, since only bifundamental hypermultiplets detect the orbifold twist. Twisted-sector insertions can therefore be represented by topological twist lines in planar supergraphs, which generate phases whenever they intersect a hypermultiplet propagator. Diagrams without such intersections descend directly from $\mathcal{N}=4$, establishing a refined version of the inheritance principle.

This observation leads to a simple field-theory picture for the leading twist-sensitive corrections to correlators of Coulomb-branch operators $T^{(\alpha)}_L$~\eqref{eq:twisted-operator}. Their two-point functions receive a first non-trivial contribution from a hypermultiplet circle encircling one of the twisted operators and intersecting the emerging twist-line (see Figure~\ref{Fig:Two-pointCircle_explicit}). We performed the complete harmonic-superspace $D$-algebra for arbitrary operator length $L$ and showed that the resulting point-split supergraph reduces to the conformal period $W_L$~\eqref{eq:period}.
Restoring the coupling dependence $g^{2L}$ and twist factor $-4s_\alpha$ the result is in agreement with the leading weak-coupling expansion of the exact localisation result~\cite{Billo:2021rdb,Billo:2022fnb,Korchemsky:2025eyc}.

We extended the construction to several nested wrapped circles (see Figure~\ref{Fig:Two-point_MultiCircle}). At every order in twist $s_\alpha^k$, the leading contribution is generated by $k$ radially ordered hypermultiplet circles. The corresponding point-split conformal integral does not factorise, immediately. Instead, factorisation emerges in the coincidence limit, where each circle collapses onto the same local wheel residue and the reduction can be performed recursively. We thus obtain an associated contribution $(-4s_\alpha)^k g^{2kL}W_L^k$, again in agreement with localisation. This provides a direct harmonic-superspace realisation of the leading terms of the localisation kernel expansion.

The same mechanism was used for the evaluation of extremal three-point-functions. For these correlators, circles surrounding either of the two short operators reduce directly to the corresponding two-point problem. A circle around the third operator is more subtle, as it gives a genuine three-point conformal integral. However, the coincidence residue is again the ordinary period $W_L$. Combining the three possible circles (see Figure~\ref{Fig:Three-point_Extremal}) reproduces the factorised localisation result to leading order.

We then presented an application of the formalism to genuinely non-extremal BPS three-point functions, for which no previous results are available. In this case the untwisted BPS operator $\tilde U_{\ell_{23},\ell_{13}}$~\eqref{eq:schematic desc} contains physical hypermultiplet admixtures. We constructed the vacuum descendant $\tilde U_{\ell_{23},\ell_{13}}$ explicitly in Appendix~\ref{App:VacDescendant}.  The appearance of a  physical hypermultiplet pair in an external operator changes the topology of the twist-sensitive graph. The matter trajectory can terminate on the composite operator and becomes a punctured circle (see Figure~\ref{Fig:Three-point_NonExtremal}). After contracting the spectator fields, the interacting graph has effective length
\begin{equation}
\tilde\ell=\ell_{12}+1\,,
\end{equation}
where $\ell_{12}\geq1$ is the tree-level bridge between the two twisted operators. The punctured harmonic-superspace $D$-algebra is correspondingly different from that of a closed circle. The opened circle removes the three-fold global Grassmann ambiguity and leaves a unique saturation sector. Nevertheless, the remaining bosonic integral reduces to the same family of coincident-endpoint ladder integrals, whose coincidence values are the wheel periods $W_{\tilde{\ell}}$. For generic positive bridges we obtain~\eqref{eq:nonext-normalised}
\begin{equation}
\tilde{C}_{L_1,L_2,L_3}^{(\alpha)}(g)=1+(4-2\delta_{\ell_{23},1} -2 \delta_{\ell_{13},1})s_\alpha
g^{2(\ell_{12}+1)}W_{\ell_{12}+1}+ \mathcal{O}(g^{2\ell_{12}+4})\,.
\end{equation}
Thus, in contrast to ordinary wrapping, the first twist-sensitive correction is governed by a bridge length $\ell_{12}$ rather than by the length $L_i$ of an external operator. This provides a perturbative prediction for a class of non-extremal twisted BPS three-point functions so far inaccessible to localisation. Its interpretation in terms of integrability is discussed in the companion paper~\cite{ShortPaper}.

The current paper serves as a proof of concept that $\mathcal{N}=2$ orbifold correlators can be studied perturbatively with relative ease thanks to the harmonic superspace formalism.  This opens up a pathway to a plethora of other observables which could be computed in the near future, such as subleading corrections, non-BPS correlators and higher-point functions. We furthermore expect that the full ADE class of $\mathcal{N}=2$ orbifold theories should be accessible to these methods by a suitable modification of the underlying gauge structure~\cite{Bath:2026gcu}. 

The localisation results \eqref{eq:R-kernel-expansion} and \eqref{eq:C-factorisation} suggest an intriguing structure of subleading corrections, which are organised by twist-factors $s_\alpha$ and the Bessel kernel elements $(K_L)_{ij}$ \eqref{eq:Besselkernel}. In this paper we matched the leading contribution of the element $(K_L)_{11}$ to the hypermultiplet circle \eqref{eq:wrap part} and it stands to reason that the other elements should admit a similar diagrammatic interpretation through (appropriately dressed) hypermultiplet circles, as we  conjectured in \cite{ShortPaper}. In order to establish the conjecture, a systematic study of subleading processes would be necessary. A subsequent puzzle is the superspace interpretation of the factorisation~\eqref{eq:C-factorisation} observed in the localisation expression for extremal three-point functions.

For the non-extremal correlators, the structure of subleading corrections is even richer. The punctured graph itself admits vector-multiplet propagator dressings, higher $(f\bar f)^s$ components of the BPS descendant enter at subsequent orders, and ordinary wrapped circles can compete with the punctured contribution for sufficiently short bridges. A complete NLO calculation will therefore require the simultaneous treatment of several classes of supergraphs rather than a simple dressing of the leading period.

Finally, the zero-bridge limit deserves a separate analysis. In that case the formal continuation of the punctured-circle result would lead to $W_1=2\zeta(1)$, reflecting the logarithmically divergent contact integral~\eqref{eq:log}. This divergence signals operator mixing and renormalisation at the descendant insertion, where single- and multi-trace protected operators become degenerate.

In conclusion our results show that harmonic superspace provides a systematic perturbative framework for twisted correlation functions in $\mathcal N=2$ orbifold theories. It reproduces the known localisation results in the sectors where exact data is available and, at the same time, extends naturally to non-extremal observables beyond the present reach of localisation. As explained in Section \ref{Sec:intro} and~\cite{ShortPaper}, we expect this to come in useful for the establishment of integrability technology. One could also try to adopt this framework for the investigation of marginally deformed orbifold theories~\cite{Gadde:2009dj,Mantegazza:2026spd}.

\section*{Acknowledgements}
We thank Zoltan Bajnok, Burkhard Eden, and all participants of the workshop \emph{Let's twist again}, held in June 2026 at DESY Hamburg, for discussions on many points related to this work. 
Further, the authors acknowledge the use of ChatGPT 5.6 Sol to simplify the $D$-algebra. All steps were carefully checked by the authors and they take full responsibility for the results presented.
This work was supported by the German Research Foundation DFG under Germany’s Excellence Strategy – EXC 2121 ‘Quantum Universe’ (390833306) through the 2025 QU Postdoc Grant. DlP is grateful for funding through a Feodor Lynen Fellowship by the Alexander von Humboldt Foundation and further acknowledges the grant NKKP Advanced 152467. 

\appendix 

\section{Vacuum descendants} \label{App:VacDescendant}

In this appendix we explicitly construct the untwisted sector BPS \emph{vacuum descendants}~$\tilde{U}_{K,L}$ which descend from the half-BPS multiplets in the parent $\mathcal{N}=4$ SYM theory that span the symmetric traceless $[0,K+L,0]$ representation of the $\SO(6)$ R-symmetry group and have conformal dimension $\Delta=K+L$. The orbifold projection breaks the R-symmetry group down to $\SU(2)_R\times U(1)_r$. The vector-multiplet scalar fields $\phi, \bar{\phi}$ carry $\U(1)_r$ charge of $\pm1$, respectively, while the hypermultiplet scalars $f^i$ live in a fundamental representation of $\SU(2)_R$. An expansion of the $\tilde{q}^+$ superfield instead yields the conjugate scalars $\bar{f}_i$ in an antifundamental representation of $\SU(2)_R$. 

We may now extract the component at fixed $\U(1)_r$-charge $K-L$ that survives the orbifold projection. It is fixed uniquely by symmetrisation and tracelessness. For simplicity we introduce the symmetrised trace
\begin{equation}
\Tr_{\rm sym}\left(\Phi_1\cdots\Phi_n\right) \equiv \frac{1}{n!} \sum_{\sigma\in S_n} \Tr\left( \Phi_{\sigma(1)}\cdots\Phi_{\sigma(n)} \right)\,.
\label{eq:symmetrised-trace}
\end{equation}
In this notation the component in question can be parameterised as\footnote{Note, in particular that the chiral untwisted operators~\eqref{eq:untwisted-operator} fall in this class of operators as $U_L=\tilde{U}_{L,0}$ and $\overline{U}_L=\tilde{U}_{0,L}$.}
\begin{equation}
\tilde U_{K,L} =\mathcal N_{K,L}\sum_{s=0}^{\min(K,L)}c_s\,\Tr_{\rm sym}\left(\phi^{K-s}\bar \phi^{L-s}f^{i_1}\cdots f^{i_s}\bar f_{i_1}\cdots \bar f_{i_s}\right)\,,
\label{eq:PJd}
\end{equation}
where the $\SU(2)_R$ indices are contracted and the trace is the quiver trace introduced in Section~\ref{Sec:HarmonicOrbifold}. We can confirm that every term satisfies the orbifold invariance condition~\eqref{eq: condition} and thus the quiver traces close. Every term contains $(K-s)+(L-s)+2s=K+L$ scalar fields and has the same $\U(1)_r$ charge $K-L$. We will show below that the relative coefficients $c_s$ can be fixed to
\begin{equation}
c_s=(-1)^s\frac{K!L!}{(K-s)!(L-s)!\,s!(s+1)!}\,.
\label{eq:cr-coeff}
\end{equation}
In particular, the first two components of the descendant are
\begin{equation}
\tilde U_{K,L} = \mathcal N_{K,L}\bigg[ \Tr_{\rm sym}\left( \phi^{K}\bar \phi^{L} \right) - \frac{KL}{2} \Tr_{\rm sym}\left( \phi^{K-1} \bar \phi^{L-1} f^i\bar f_i \right)+\ldots \bigg]\,.
\label{eq:BPS-first-components}
\end{equation}
For the supergraph calculation in Section~\ref{Sec:non-extremal}, we have to distinguish the coefficient
multiplying the symmetrised trace from that of a fixed planar ordering. For
the $s=0$ term a fixed ordered word occurs $ K!
L!$ times in the permutation sum in~\eqref{eq:symmetrised-trace}.
For the $s=1$ term, a fixed ordered word with a specified
$f^i\tilde f_i$ component occurs
$ (K-1)!
(L-1)!$ times. The common factor $1/(K+L)!$
cancels in their ratio, and we find the relative factor between individual orderings at $s=0$ and $s=1$ to take the value
\begin{equation}
\frac{c_1(K-1)!(L-1)!}
{c_0\,K!L!}
=-\frac12\,.
\label{eq:BPS-fixed-word-factor}
\end{equation}
This factor appeared in~\eqref{eq:schematic desc} and was used in the superspace calculation.

\paragraph{Determination of the coefficients} We will now derive~\eqref{eq:cr-coeff}. The coefficients $c_s$ in~\eqref{eq:PJd} implement the projection onto the traceless half-BPS $\SO(6)$ representation. To determine them, we may temporarily neglect the ordering inside the symmetrised trace and represent the parent $\SO(6)$ tensor structure by commuting variables
\begin{equation}
P_{K,L} = \sum_{s=0}^{\min(K,L)} c_s\, \phi^{K-s}\bar \phi^{L-s} (f^i\bar f_i)^s\,.
\end{equation}
This homogeneous polynomial representative has to be harmonic with respect to the $\SO(6)$ trace operator. In the $\mathcal N=2$ decomposition relevant here, and up to an irrelevant overall normalisation, this operator takes the form
\begin{equation}
\Delta_{\rm scal} = \partial_\phi \partial_{\bar \phi} + \partial_{f^i}\partial_{\bar f_i}\,.
\label{eq:scalar-trace-operator}
\end{equation}
and we have to enforce
\begin{equation}
\Delta_{\rm scal}P_{K,L}=0\,.
\label{eq:BPS-harmonic-condition}
\end{equation}

Let us now consider the coefficient of $\phi^{K-s-1}\bar \phi^{L-s-1} (f^i\bar f_i)^s$ in~\eqref{eq:BPS-harmonic-condition}. There are two contributions. The first from acting with the $\partial_\phi \partial_{\bar \phi}$ part of the trace operator on the term with
$s$ hypermultiplet pairs, which gives
\begin{equation}
\partial_\phi\partial_{\bar \phi} \left[ \phi^{K-s}\bar \phi^{L-s} (f^i\bar f_i)^s \right] = \left(K-s\right)\left(L-s\right) \phi^{K-s-1}\bar \phi^{L-s-1} (f^i\bar f_i)^s\,.
\label{eq:ZbarZ-trace}
\end{equation}
The combinatorial factors simply count the possible $\phi$ and $\bar \phi$ fields that can be contracted. The second contribution comes from the term containing $s+1$ hypermultiplet pairs. Writing $F=f^i\bar f_i$, we first have
\begin{equation}
\partial_{\bar f_i}F^{s+1} = (s+1)f^iF^s\,.
\end{equation}
Acting with the remaining derivative and summing over the $\SU(2)_R$ index gives
\begin{equation}
\begin{aligned}
\partial_{f^i}\partial_{\bar f_i}F^{s+1} &= (s+1)\partial_{f^i}\left(f^iF^s\right) \\
&=(s+1)\left[ \delta_i^{\,i}F^s + s\,f^i\bar f_i F^{s-1} \right] \\
&= (s+1)(s+2)F^s\,,
\end{aligned}
\label{eq:qtildeq-trace}
\end{equation}
where $\delta_i^{\,i}=2$ for $\SU(2)_R$ representations.

Both~\eqref{eq:ZbarZ-trace} and~\eqref{eq:qtildeq-trace} produce the same monomial. Requiring its coefficient to vanish in $\Delta_{\rm scal}P_{K,L}$ therefore gives
\begin{equation}
c_s\left(K-s\right)\left(L-s\right) + c_{s+1}(s+1)(s+2) = 0\,,
\end{equation}
or equivalently
\begin{equation}
c_{s+1} = -\frac{\left(K-s\right)\left(L-s\right)}{(s+1)(s+2)} c_s\,.
\label{eq:descendant-recursion}
\end{equation}
Choosing $c_0=1$, the recursion is solved by~\eqref{eq:cr-coeff}.

\bibliography{refs}
\end{document}